\documentclass[10pt,journal,compsoc]{IEEEtran}

\ifCLASSOPTIONcompsoc
  \usepackage[nocompress]{cite}
\else
  \usepackage{cite}
\fi

\usepackage[utf8]{inputenc}
\usepackage[english]{babel}
\usepackage{amsmath}

\usepackage{graphicx}
\usepackage{caption}
\usepackage{listings}
\usepackage{float}
\usepackage{algorithm2e}
\usepackage{caption}
\usepackage{subcaption}
\usepackage{array}
\usepackage{booktabs}

\usepackage{hyperref}
\usepackage{makecell}

\usepackage[table]{xcolor}

\usepackage{cardEstMacrosTikz}
\usepackage{cardEstMacros}

\begin{document}
\title{A General Cardinality Estimation Framework for Subgraph Matching in Property Graphs}
\author{Wilco~van~Leeuwen, George~Fletcher, Nikolay~Yakovets 
\IEEEcompsocitemizethanks{\IEEEcompsocthanksitem W.J.~van~Leeuwen, G.H.L.~Fletcher and N.~Yakovets are with TU Eindhoven. E-mail: \{w.j.v.leeuwen, g.h.l.fletcher, n.yakovets\}@tue.nl}
\thanks{This work has been submitted to the IEEE for possible publication. Copyright may be transferred without notice, after which this version may no longer be accessible.}
}

\markboth{}%
{}
%

\IEEEtitleabstractindextext{%
\begin{abstract}
Many techniques have been developed for the cardinality estimation problem in data management systems. 
In this document, we introduce a framework for cardinality estimation of \queryPatterns{} over property graph databases, which makes it possible to analyze, compare and combine different cardinality estimation approaches.
This framework consists of three phases: obtaining a set of estimates for some subqueries, extending this set and finally combining the set into a single cardinality estimate for the query.
We show that (parts of) many of the existing cardinality estimation approaches can be used as techniques in one of the phases from our framework.
The three phases are loosely coupled, this makes it possible to combine (parts of) current cardinality estimation approaches.
We create a graph version of the Join Order Benchmark to perform experiments with different combinations of techniques.
The results show that \queryPatterns{} \emph{without} property constraints can be accurately estimated using synopses for small patterns. 
Accurate estimation of \queryPatterns{} \emph{with} property constraints require new estimation techniques to be developed that capture correlations between the property constraints and the topology in graph databases.
\end{abstract}

}

\maketitle

\IEEEdisplaynontitleabstractindextext

%
\IEEEpeerreviewmaketitle


\IEEEraisesectionheading{\section{Introduction}\label{sec:relWorkIntro}}

\IEEEPARstart{C}{}\textit{ardinality estimation} can be defined as the task of estimating the
number of results returned by a given query over a given database instance.
This is a fundamental data management problem, important
across many practical scenarios, e.g., in query planning, approximate query
evaluation, and query execution progress estimation.  The basic evaluation criteria
for cardinality estimation techniques are: estimation accuracy,
estimation time, memory cost, and preparation time.  Different applications
might value these dimensions differently, leading to trade-offs, e.g., between
estimation accuracy and memory footprint. It has been shown that poor query plans
are regularly produced by the query optimizers of current database management
systems and that cardinality estimation errors are the main reason for
these bad query plans \cite{leis2015good,leis2018query}.  Hence, while a
classical topic heavily studied for decades, cardinality estimation remains a
significant challenge in practice.

In this work, we survey cardinality estimation solutions for \queryPatterns{} on property
graph databases.  Property graphs are a popular data model in industry, part of
the upcoming ISO graph database query language standard \cite{GQLstandard}.  Essentially, a
property graph is a graph where both nodes and edges can have labels (e.g.,
Person) and have a set of associated key-value pairs, which are called the properties, where the property key specifies the meaning of the property value (e.g., MemberSince = 2008-11-10). 
A \queryPattern{} (also known as a conjunctive graph query, or a subgraph matching query \cite{bonifati2018querying})
specifies a graph pattern of interest (e.g., Persons and the Movies they ActIn)
and also possibly constraints on attributes (e.g., only those Persons who
became a member after 1990-12-08).  \QueryPatterns{} are fundamental in both
theory and practice, forming the backbone of virtually all queries expressed in
contemporary graph database query languages
\cite{bonifati2018querying,BonifatiMT20,GQLstandard}.  Cardinality estimation for
\queryPatterns{} over property graph databases gives rise to new challenges due
to the schemaless design of graph databases and the correlation between the
data (property values and labels) and the topology (connectivity) of the graph.
The focus of our survey allows us to shed deep light on an important, timely,
highly challenging, and broadly applicable subclass of the cardinality
estimation problem.  Further, many of the concepts of our survey generalize to other
typical settings, for example conjunctive queries on relational databases.

There is a large and rich literature on cardinality estimation for queries on database instances which can be used for \queryPatterns{} on property graphs. 
Each technique requires specialized statistics or
indexes and uses a set of simplifying assumptions to overcome missing
information or to improve on criteria of interest (i.e., accuracy, time, space, etc.).  
Hence, it is very difficult for  practitioners, systems designers, and researchers to navigate the state of the art and understand which solutions work or not for a given application scenario.
The most recent and complete survey here is the groundbreaking G-CARE framework \cite{ParkKBKHH20}.
G-CARE performs extensive experiments on different cardinality estimation approaches 
with queries of varying sizes, result sizes and topologies, 
on both real and synthetic data sets.
%
This study addresses two limitations of G-CARE. 
First, G-CARE studies cardinality estimation techniques for 
\queryPatterns{} without predicates on the properties.
This study includes extensive experiments on property predicates.
Our experiments show that properly handling property predicates 
is essential for accurate cardinality estimation.
Second, the focus of G-CARE is on comparing different cardinality estimation approaches. 
In this study, we also focus on combining parts of different cardinality estimation approaches with the aim of producing superior estimation techniques.

Our contribution in this survey is to introduce a framework that
makes it possible to explain the state of the art and systematically analyze, compare and combine the current
cardinality estimation techniques for \queryPatterns{} on property graphs.  We start by viewing a \queryPattern{} as a specification of a set of constraints (Section \ref{sec:prelim}). 
We then highlight the basic ingredients of contemporary estimators and how they together are used to define concrete cardinality estimation solutions in a general framework (Section \ref{sec:framework}).
The first phase of the framework consists of estimation techniques for individual constraints (Section \ref{sec:est_ic}) and of estimation techniques for multiple constraints (Section \ref{sec:est_mc}).
The second phase consists of
methods for extending partial estimates (Section \ref{sec:extend_PES}).
Finally, the third phase consists of
methods for combining partial estimates in order to obtain a final estimate for the complete query (Section \ref{sec:est_all}).

We leverage our framework to undertake a comprehensive comparative analysis of the state of the art cardinality estimators, shedding fresh light on their trade-offs and relative strengths and weaknesses (Sections \ref{sec:exp_Setup}-\ref{sec:exp}).  Given these new deep insights, we conclude the survey with recommendations for practitioners and researchers (Section \ref{sec:concusionFutureWork}). With this work we aim to help move the field forward towards more effective cardinality estimation solutions in practical graph data management.


 

\section{Preliminaries}
\label{sec:prelim}


\subsection{Basic Property Graph Query Pattern}
\label{sec:qgp}
A \textit{basic property graph query pattern}, or shortly \textit{\queryPattern{}}, consists of a set of vertices and a set of edges. Vertices and edges have unique identifiers. Each edge is associated with exactly one source vertex and one target vertex. Vertices and edges can have a (possibly empty) set of label and a (possibly empty) set of key-value properties. 

A \queryPattern{} $Q$ is defined as the tuple (\edges{}, \vertices{}, \idSet{}, \edgeGreek{}, \labelsGreek{}, \propGreek{}, $\Theta$, \labelSet{}, \propKeySet{}, \propValSet{}).\footnote{We use the dot notation to refer to an element in $Q$ or $G$, e.g. $Q.\propGreek{}$} 
The set of edge identifiers \edges{} and the set of vertex identifiers \vertices{} are disjoint, and $\idSet{} = \edges{} \cup \vertices{}$.
The total function \edgeGreek{}: \edges{} $\rightarrow$ (\vertices{} $\times$ \vertices{}) maps each edge identifier to a pair of source-target vertex identifiers. 
The total function $\labelsGreek{}: \idSet{} \rightarrow$ $\mathcal{P}(\labelSet{})$, where $\labelSet{}$ is a finite set of labels and $\mathcal{P}(\labelSet{})$ is the powerset of \labelSet{}, maps edge and vertex identifiers to a (possible empty) set of labels.
The partial function $\propGreek{}: (\idSet{} \times \propKeySet{} \times \Theta) \rightarrow \propValSet{}$
defines the key-property constraints of vertices and edges, 
where \propKeySet{} is a finite set of properties, \propValSet{} is a set of values 
and $\Theta$ is a set of binary predicates, 
i.e. set of functions $\theta:(\propValSet{} \times \propValSet{}) \rightarrow \{True, False\}$.
Table \ref{tab:notationQueryAndGraph} summarizes our notation for \queryPatterns{}.

\begin{example}
\label{ex:formal_def_job18a}
The formal description of the \queryPattern{} in Figure \ref{fig:queryJOB18a} is $Q_{example}$ = (\edges{ex}, \vertices{ex}, \edgeGreek{ex}, \labelsGreek{ex}, \propGreek{ex}, \labelSet{ex}, \propKeySet{ex}, \propValSet{ex}, $\Theta_{ex}$), where:\\ \\
{\scriptsize 
\ttfamily	
$\edges{}_{ex}$ = \{id1, id3, id5, id7\}\\
$\vertices{}_{ex}$ = \{id0, id2, id4, id6, id8\}\\
$\idSet{}_{ex}$ = \{id0, id1, $\dots$, id8\}
\vspace{2mm}\\
\edgeDef{ex}{id1}{id0}{id2}, $\dots$\\
\labelDef{ex}{id0}{'title'},\labelDef{ex}{id1}{'budget'}, $\dots$\\
\propDef{ex}{id6}{note}{\text{IN}}{\{'(producer)','(executive producer)'\}},\\
\propDef{ex}{id8}{gender}{=}{m},\\
\propDef{ex}{id8}{name}{\text{CONTAINS}}{Tim}
\vspace{2mm}\\
Lab$_{ex}$ = \{'title', 'votes', $\dots$\}\\
Prop$_{ex}$ = \{'note', 'gender', 'name'\}\\
Val$_{ex}$ = \{"\{'(producer)','(executive producer)'\}", 'm', 'Tim'\}\\
$\Theta_{ex}$=\{=, IN, CONTAINS\}
}

\end{example}

\subsection{Property Graph Data Model}
\label{sec:propGraphModel}
A property graph instance is defined as a tuple $G$ = (\edges{}, \vertices{}, \idSet{}, \edgeGreek{}, \labelsGreek{}, \propGreek{}, \labelSet{}, \propKeySet{}, \propValSet{}). 
Only the equality operator is allowed for $\Theta$ in \propGreek{} in this definition, therefore $\Theta$ is irrelevant.
%
%
%
%
Let $\ell_e(G) \subseteq G.$\labelSet{} be the set of labels that occur on edges in graph $G$ and $\ell_v(G) \subseteq G.$\labelSet{} be the set of all labels that occur on vertices in graph $G$.
Table \ref{tab:notationQueryAndGraph} summarizes our notation.

\begin{table}
\centering
\caption{Notation for \QueryPatterns{} and Property Graph Instances}
\label{tab:notationQueryAndGraph}
{\tiny
\begin{tabular}{| l | l || l | l | }
\hline
$Q$ & \QueryPattern & 
$G$ & Property graph instance\\ \hline
$Q.$\labelSet{} & Set of labels in $Q$ &
$G.$\labelSet{} & Set of labels in $G$\\ \hline
$Q.$\propKeySet{} & Set of property keys in $Q$ &
$G.$\propKeySet{} & Set of property keys in $G$\\ \hline
$Q.$\propValSet{} & Set of property values in $Q$ &
$G.$\propValSet{} & Set of property values in $G$\\ \hline
$Q.\Theta$ & Set of binary predicates used in $Q$ &
&\\ \hline
$Q.$\edges{} & Set of edge ids in $Q$ &
$G.$\edges{} & Set of edge ids in $G$\\ \hline
$Q.$\vertices{} & Set of vertex ids in $Q$ &
$G.$\vertices{} & Set of vertex ids in $G$\\ \hline
$Q.\idSet{}$ & $Q.\edges{} \cup Q.\vertices{}$ &
$G.\idSet{}$ & $G.\edges{} \cup G.\vertices{}$\\ \hline
$Q.$\edgeGreek{} & $Q.\edges{} \rightarrow (Q.\vertices{} \times Q.\vertices{})$ &
$G.$\edgeGreek{} & $G.\edges{} \rightarrow (G.\vertices{} \times G.\vertices{})$\\ \hline
$Q.$\labelsGreek{} & $Q.\idSet{} \rightarrow \mathcal{P}(Q.\labelSet{})$ &
$G.$\labelsGreek{} & $G.\idSet{} \rightarrow \mathcal{P}(G.\labelSet{})$\\ \hline
$Q.\propGreek{}$ & $(Q.\idSet{} \times Q.\propKeySet{} \times Q.\Theta) \rightarrow Q.\propValSet{}$ &
$G.\propGreek{}$ & $(G.\idSet{} \times G.\propKeySet{}) \rightarrow G.\propValSet{}$\\ \hline
& & 
$G.\ell_e$ & $\cup_{e \in G.E} G.\labelsGreek{}(e)$\\ \hline
& &
$G.\ell_v$ & $\cup_{v \in G.V} G.\labelsGreek{}(v)$ \\ \hline
\end{tabular}
}
\end{table}

\begin{table}
\centering
\caption{Notation (section where notation is introduced)}
\label{tab:notationInDoc}
{\scriptsize
\begin{tabular}{| l || l |}
\hline
$\mapping{D \rightarrow R}$ & \makecell{A mapping that maps each element in the \\domain $D$ to an element in range $R$ (\ref{intro:mapping})} \\ \hline
$\mathcal{M}_{D \rightarrow R}$ & Set of all possible mappings $\mapping{D \rightarrow R}$ (\ref{intro:mappings})\\ \hline
$|S|$ & Cardinality of set $S$ (\ref{intro:card})\\ \hline
$\constr{}$ & A constraint (\ref{intro:constraint})\\ \hline
$\constrSet{}$ & A set of constraints (\ref{intro:constraintSet})\\ \hline
$\constrSet{}(Q)$ & Set of constraints defined by $Q$ (\ref{intro:queryConstraints}) \\ \hline
$\constrSet{}.\idSet{}$ &  Set of query ids contained in \constrSet{} (\ref{intro:constrIds})\\ \hline
$\SAT{\constr{}}{D}$ & \makecell{Set of all mappings in $\mappingSet{}_{\xToGraphId{D}}$ \\that satisfy constraint $\constr{}$ (\ref{intro:SATc})}\\ \hline
\makecell{$\Pr[\SAT{\constr{}}{D}]$\\
$\Pr[\Sat{\constr{}}]$}
& \makecell{Fraction of all mappings in $\mappingSet{}_{\xToGraphId{D}}$ \\that satisfy constraint $\constr{}$ (\ref{intro:probSATc})}\\ \hline
$\Pr[\Sat{\constrSet{}}]$ 
& \makecell{Fraction of all mappings in $\mappingSet{}_{\xToGraphId{D}}$ that \\satisfy all constraints in \constrSet{} (\ref{intro:probSATc})}\\ \hline
$PE$ & Partial Estimate (\ref{intro:PE})\\ \hline
$PE.s$ & Selectivity value of $PE$ (\ref{intro:PE_s})\\ \hline
$PE.\constrSet{}$ & Set of constraints in $PE$ (\ref{intro:PE_C})\\ \hline
$PET$ & Partial Estimation Technique (\ref{intro:PET})\\ \hline
$PES$ & Partial Estimate Set (\ref{intro:PES})\\ \hline
$EPEST$ & Extend PES Technique (\ref{intro:EPEST})\\ \hline
$EPES$ & Extended PES (\ref{intro:EPES})\\ \hline
$CPES$ & Complete PES (\ref{intro:CPES})\\ \hline
CT & Combine Technique (\ref{intro:CT})\\ \hline
\end{tabular}
}
\end{table}

\subsection{Semantics of \QueryPatterns{}}
\label{sec:qgp_semantics}
Consider a \queryPattern{} $Q$ and a property graph instance $G$.
\label{intro:mapping}Let mapping $\mapping{Q.\idSet{}\rightarrow G.\idSet{}}$ be a total function from the edge and vertex identifiers in $Q$ to edge and vertex identifiers in $G$, i.e. \mapping{}: $Q.\idSet{}\rightarrow G.\idSet{}$. 
\label{intro:mappings}Let $\mathcal{M}_{Q.\idSet{}\rightarrow G.\idSet{}}$ be the set of all possible mappings from $Q$ to $G$. Table \ref{tab:notationInDoc} shows a summary of the notation used in this document.

A mapping \mapping{} is called a \textit{homomorphic match}\footnote{Some systems require isomorphic matches, which can be obtained by adding the requirement MR5: For each $i,j \in Q.\idSet{}$: if $i \neq j$, then \m{$i$} $\neq$ \m{j}. See \cite{angles2017foundations} for more advanced isomorphic-based semantics, like no-repeated-node and no-repeated-edge semantics.} from $Q$ on $G$ if and only if the following \textit{match requirements} are satisfied:
\begin{enumerate}
\item [MR1:] For each $v \in Q.$\vertices{}: \m{$v$} $\in G$.\vertices{};
\item [MR2:] For each $e \in Q$.\edges{}: if $Q.$\edgeDef{}{$e$}{$s$}{$t$}, \\then $G.$\edgeDef{}{\m{$e$}}{\m{$s$}}{\m{$t$}};
\item [MR3:] For each $i \in Q.\idSet{}$ and $l \in Q.\labelSet{}$:\\if \labelConstrCtxVar{$l$}{$i$}{Q.}, then \labelConstrCtxVar{$l$}{\m{$i$}}{$G.$};
\item [MR4:] For each $i \in Q.\idSet{}, k \in Q.\propKeySet{}, \theta \in Q.\Theta$ and $v \in Q.$\propValSet{}: if $Q.$\propDefVar{}{$i$}{$k$}{$\theta$}{$v$}, then \\$G.$\dataPropDefVar{}{\m{$i$}}{$k$}{$w$} and $w \theta v$\footnote{Here, we use the infix notation of the binary predicate $\theta$} holds for some $w \in G.$\propValSet{}.
\end{enumerate}

Here, MR1 and MR2 require that the mapping agrees with the topology defined in the query. Notice that MR1 is required when the query contains disconnected vertices.
%
MR3 defines constraints on the existence of labels and MR4 defines constraints on the existence of properties.

Finding all matches for a \queryPattern{} is also called \textit{subgraph pattern matching} and is at the core of many query languages\cite{angles2017foundations}, e.g. SPARQL\footnote{https://www.w3.org/TR/rdf-sparql-query/}, Cypher\footnote{https://neo4j.com/developer/cypher/}, PGQL\cite{van2016pgql}, G-CORE\cite{angles2018g}.

The total number of mappings is $|\mathcal{M}_{Q,G}|$, where \label{intro:card}$|S|$ denotes the cardinality of set $S$.
The number of these mappings that are matches is called the \textit{cardinality} of $Q$ on $G$ and the fraction is called the \textit{selectivity}.


\subsection{Query Constraints}
\label{sec:queryConstraints}
Given a \queryPattern{} $Q$, the set of constraints of $Q$, $\constrSet{}(Q)$\label{intro:queryConstraints}, is defined as follows:
\begin{enumerate}
\item For each vertex id $v \in Q.\vertices{}$: \vertexConstr{$v$} $\in \constrSet{}(Q)$
\item For each edge id $e \in Q.\edges{}$: \edgeConstr{$e$} $\in \constrSet{}(Q)$
\item For each edge id $e \in Q.\edges{}$: \\
if $Q.$\edgeDef{}{$e$}{$s$}{$t$}, \\
then \srcConstr{s}{e} $ \in \constrSet{}(Q)$ and \trgConstr{t}{e} $ \in \constrSet{}(Q)$
\item For each $i \in Q.\idSet{}$ and $l \in Q.\labelSet{}$: \\
If $l \in Q.\labelsGreek{}(i)$, \\
then \hasLabelConstr{$i$}{$l$} $\in \constrSet{}(Q)$
\item For each $i \in Q.\idSet{}, k \in Q.\propKeySet{}$, $\theta \in Q.\Theta$ and $v \in Q.\propValSet{}$: \\
If $Q.\propGreek{}(i,k,\theta)=v$, \\
then \hasPropKeyConstr{$i$}{$k$} $\in \constrSet{}(Q)$ and\\ 
\hasPropConstr{$i$}{$k$}{$\theta$}{$v$} $\in \constrSet{}(Q)$
\item $\constrSet{}(Q)$ does not contain any other constraints
\end{enumerate}

\label{intro:constraint}A single constraint is denoted by $\constr{}$, 
\label{intro:constraintSet}a set of constraints by $\constrSet{}$ and 
the set of all constraints of $Q$ by $\constrSet{}(Q)$.
A constraint contains one or two identifiers from a query $Q$. 
\label{intro:constrIds}Let $\constrSet{}.\idSet{}$ represent the set of query identifiers contained in the constraints $\constrSet{}$.

Given a $\constr{}$ constraint, a mapping $\mapping{\{\constr{}\}.\idSet{} \rightarrow G.\idSet{}}$ and a graph $G$, it is possible to verify if $\mapping{}$ satisfies $\constr{}$ in $G$:
\begin{itemize}
\item If $\constr{} = $\vertexConstr{$i$}, then verify if $\m{i} \in G.\vertices{}$
\item If $\constr{} = $\edgeConstr{$i$}, then verify if $\m{i} \in G.\edges{}$
\item If $\constr{} = $\srcConstr{$v$}{$e$}, then verify if \\
$G.$\edgeDef{}{$\m{e}$}{$\m{v}$}{$v'$} for some $v' \in G.\vertices{}$
\item If $\constr{} = $\trgConstr{$v$}{$e$}, then verify if \\
$G.$\edgeDef{}{$\m{e}$}{$v'$}{$\m{v}$} for some $v' \in G.\vertices{}$
\item If $\constr{} = $\hasLabelConstr{$i$}{$l$}, then verify if \\
$l \in \labelsGreek{}(\m{i})$
\item If $\constr{} = $\hasPropKeyConstr{$i$}{$k$}, then verify if \\
$G.\propGreek{}(\m{i}, k) = w$ is defined for some $w \in G.\propValSet{}$
\item If $\constr{} = $\hasPropConstr{$i$}{$k$}{$\theta$}{$v$}, then verify if \\
$G.\propGreek{}(\m{i}, k) = w$ and $w \theta v$ holds for some $w \in G.\propValSet{}$
\end{itemize}

A mapping $\mapping{Q.\idSet{} \rightarrow G.\idSet{}}$ satisfies all constraints $\constrSet{}(Q)$ of query $Q$ in graph $G$, if and only if $\mapping{Q.\idSet{} \rightarrow G.\idSet{}}$ is a homomorphic match from $Q$ to $G$.
Table \ref{tab:setsOfConstraintsNaming} shows the categorization of constraints based on what they primarily deal with: \emph{topology} or \emph{data}. 

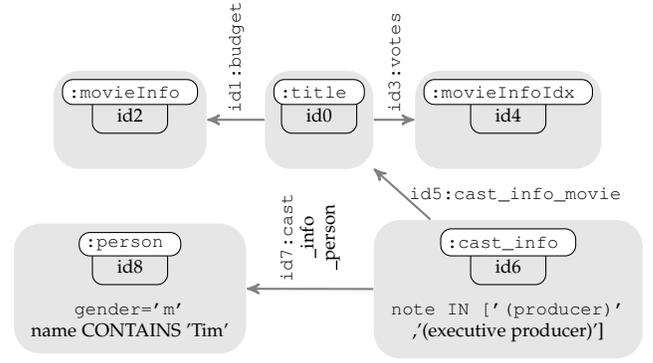
\begin{figure}
	\centering
	
\begin{tikzpicture}
	[
		scale=0.5,
		propNode/.append style={node distance=1.5cm and 1.5cm}
	]	
	
	\propVertex{0}{0}{p}{id8}{:person}{gender='m' \\ name CONTAINS 'Tim'}
	\propVertexLocRef{above}{idp}{mi}{id2}{:movieInfo}{}	
	\propVertexLocRef{right}{idmi}{t}{id0}{:title}{}	
	\propVertexLocRef{right}{idt}{miIdx}{id4}{:movieInfoIdx}{}	
	\propVertexLocRef{below}{idmiIdx}{ci}{id6}{:cast\_info}{note IN ['(producer)'\\,'(executive producer)']}

	\draw [propEdge] (ci) to node [color=black, align=center, anchor=west, rotate=90] {\imageText{id7:cast\\ \_info\\ \_person}} (p);
	\draw [propEdge] (ci) to node [color=black, align=center, anchor=west] {\imageText{id5:cast\_info\_movie}} (t);
	\draw [propEdge] (t) to node [color=black, align=center, anchor=west, rotate=90] {\imageText{id1:budget}} (mi);
	\draw [propEdge] (t) to node [color=black, align=center, anchor=west, rotate=90] {\imageText{id3:votes}} (miIdx);
\end{tikzpicture}
	\caption[caption]{JOB query 18a visualized as a \queryPattern{}}
	\label{fig:queryJOB18a}
\end{figure}

\vspace{2mm}
\begin{example}
\label{ex:runningExampleQuery_constraints}
%
The constraints $\constrSet{}(Q)$ of the \queryPattern{} $Q$ in Figure \ref{fig:queryJOB18a} are given as follows:\\ \\
{ \scriptsize
$\constr{0}$: \vertexConstr{id0},\dots,
$\constr{4}$: \vertexConstr{id8},\\
$\constr{5}$: \edgeConstr{id1},\dots,
$\constr{8}$: \edgeConstr{id7},\\
$\constr{9}$: \srcConstr{id0}{id1},\dots,
$\constr{16}$: \trgConstr{id8}{id7},\\
$\constr{17}$: \hasLabelConstr{id0}{title},\dots,
$\constr{25}$: \hasLabelConstr{id8}{person},\\
$\constr{26}$: \hasPropConstr{id6}{note}{IN}{['(producer)',' \\(executive producer)']},\\
$\constr{27}$: \hasPropConstr{id8}{gender}{=}{'m'},\\
$\constr{28}$: \hasPropConstr{id8}{name}{CONTAINS}{'Tim'},\\
$\constr{29}$: \hasPropKeyConstr{id6}{note},\\
$\constr{30}$: \hasPropKeyConstr{id8}{gender},\\
$\constr{31}$: \hasPropKeyConstr{id8}{name}.
}
%
\end{example}

\begin{table*}
\centering
\caption{Categorization of constraints}
\label{tab:setsOfConstraintsNaming}
\begin{tabular}{| c | c | c | c | c | c | c |}
\hline
\multicolumn{7}{|c|}{Query Pattern Constraints} \\ \hline
\multicolumn{6}{|c|}{Query Pattern Schema Constraints} & \multicolumn{1}{c|}{Property-value Constraints} \\ \hline
\multicolumn{5}{|c|}{Labeled Topological Constraints} & \multicolumn{2}{c|}{Property Constraints} \\ \hline
\multicolumn{4}{|c|}{Topological Constraints} & \multicolumn{3}{c|}{Data Constraints} \\ \hline \hline
\vertexConstr{$i$} & \edgeConstr{$i$} & \srcConstr{$v$}{$e$} & \trgConstr{$v$}{$e$} & \hasLabelConstr{$i$}{$l$} &\hasPropKeyConstr{$i$}{$k$} & \hasPropConstr{$i$}{$k$}{$\theta$}{$v$}  \\ \hline 
\end{tabular}
\end{table*}

\subsection{Probability, Selectivity and Cardinality}
\label{sec:selEst}
Consider obtaining a random mapping $\mapping{Q.\idSet{}\rightarrow G.\idSet{}}$ from $\mathcal{M}_{Q.\idSet{}\rightarrow G.\idSet{}}$ by throwing a fair $|G.\idSet{}|$-sides dice for every $i \in Q.\idSet{}$ to obtain $\m{i}$. This gives a \textit{sample space} of $G.\idSet{}_1 \times \dots \times G.\idSet{}_{|Q.\idSet{}|}$. 
%
%
%
Then, the probability of obtaining mapping $\mapping{}=(\omega_1, \dots, \omega_{|Q.\idSet{}|})$ is 
\begin{align*}
\Pr[(\omega_1, \dots, \omega_{|Q.\idSet{}|})]
&= \Pr[\omega_1] \cdot ... \cdot \Pr[\omega_{|Q.\idSet{}|}]
= (1/|G.\idSet{}|)^{|Q.\idSet{}|},
\end{align*}
because the outcomes of the dice throws are mutually independent.

Subsets of the sample space are called \textit{events}. 
\label{intro:SATc}Let $\SAT{\constr{}}{Q.\idSet{}}$ be the event that contains all mappings in $\mappingSet_{\queryIdToGraphId{}}$ that satisfy the constraint $\constr{}$:
\begin{align*}
\SAT{\constr{}}{Q.\idSet{}}
&= \{\mapping{} \in \mappingSet_{\queryIdToGraphId{}} \mid \mapping{} \text{ satisfies constr } \constr{}\}.
\end{align*}

Since each element from our sample set has the same probability, \label{intro:probSATc}the probability of event $\SAT{\constr{}}{Q.\idSet{}}$ is defined as:
\begin{align*}
\Pr[\SAT{\constr{}}{Q.\idSet{}}] &= \frac{|\SAT{\constr{}}{Q.\idSet{}}|}{|\mappingSet_{\queryIdToGraphId{}}|} = \frac{|\SAT{\constr{}}{\constr{}.\idSet{}}|}{|\mappingSet_{\xToGraphId{\constr{}.\idSet{}}}|}
\end{align*}

\begin{figure}
	\centering
	
\begin{tikzpicture}
	\draw (1,1.2) node {$Q$};
	\node at (0,0) [propNode](idq1)  {\imageText{q1}};
	\node at (2,0) [propNode](idq3)  {\imageText{q3}};
	\draw [propEdge] (idq1) to node [color=black, align=center, anchor=south] {\imageText{q2}} (idq3);
	
	\draw [dashed] (3,-1) -- (3,1.2);
	
	\draw (5,1.2) node {$G$};	
	\draw (5,0.8) node [node font=\footnotesize] {$\Omega_{G}=\{g1,g2,g3,g4\}$};
	
	\node at (4,0) [propNode](idg1)  {\imageText{g1}};
	\node at (6,0) [propNode](idg3)  {\imageText{g3}};
	\draw [propEdge] (idg1) to [color=black, align=center, anchor=south, bend left=20] node {\imageText{g2}} (idg3);	
	\draw [propEdge] (idg3) to [color=black, align=center, anchor=south, bend left=20] node {\imageText{g4}} (idg1);	
	
	\draw [->, dashed, color=red](0,-0.25) -- (0.8,-1.2);
	\draw [->, dashed, color=red](0.85,-1.2) -- (4,-0.25);
	\draw [->, dashed, color=red](1,0) -- (1.3,-1.2);
	\draw [->, dashed, color=red](1.35,-1.2) -- (4,-0.25);	
	\draw [->, dashed, color=red](2,-0.25) -- (1.8,-1.2);
	\draw [->, dashed, color=red](1.85,-1.2) -- (4,-0.25);	
	
	\draw [->, color=green](0,-0.25) -- (2.95,-1.2);
	\draw [->, color=green](3,-1.2) -- (6,-0.25);
	\draw [->, color=green](1,0) -- (3.4,-1.2);
	\draw [->, color=green](3.45,-1.2) -- (5,-0.3);	
	\draw [->, color=green](2,-0.25) -- (3.8,-1.2);
	\draw [->, color=green](3.85,-1.2) to[bend right=55]  (4.1,-0.25);	

	\draw (3,-1.5) node [node font=\footnotesize]{$\mappingSet{}=\{(g1,g1,g1),\dots,(g3,g4,g1),\dots,(g4,g4,g4)\}$};
	\draw (3.5,-2.6) node [node font=\footnotesize]{
\begin{tabular}{l l}
$\Sat{\text{\vertexConstr{q1}}}$ & $=\{(g1,*,*), (g3,*,*)\}$\\
$\Sat{\text{\vertexConstr{q3}}}$ & $=\{(*,*,g1), (*,*,g3)\}$\\
$\Sat{\text{\edgeConstr{q2}}}$ & $=\{(*,g2,*), (*,g4,*)\}$\\
$\Sat{\text{\srcConstr{q1}{g2}}}$ & $=\{(g1,g2,*), (g3,g4,*)\}$\\
$\Sat{\text{\trgConstr{q3}{q2}}}$ & $=\{(*,g2,g3), (*,g4,g1)\}$
\end{tabular}
	};
	
	\draw (3,-3.7) node [node font=\footnotesize]{where $(x,*,*)$ expands to $(x,g1,g1),(x,g1,g2),\dots,(x,g4,g4)$};
	\draw (3,-4.6) node [node font=\footnotesize]{
\begin{tabular}{r l}
$\Sat{\constrSet{}(Q)}$ 
& $= \bigcap_{\constr{} \in \constrSet{}(Q)}\Sat{\constr{}}$\\
& $= \Sat{\text{\vertexConstr{q1}}} \cap \dots \cap \Sat{\text{\trgConstr{q3}{q2}}}$\\
& $=\{(g1,g2,g3), (g3,g4,g1)\}$
\end{tabular}	
};

\end{tikzpicture}
	\caption[caption]{The set $\Sat{\constrSet{}(Q)}$ can be obtained by taking the intersection of the events $\Sat{\constr{}}$ for $\constr{} \in \constrSet{}(Q)$.}
	\label{fig:ProbNotationExample}
\end{figure}
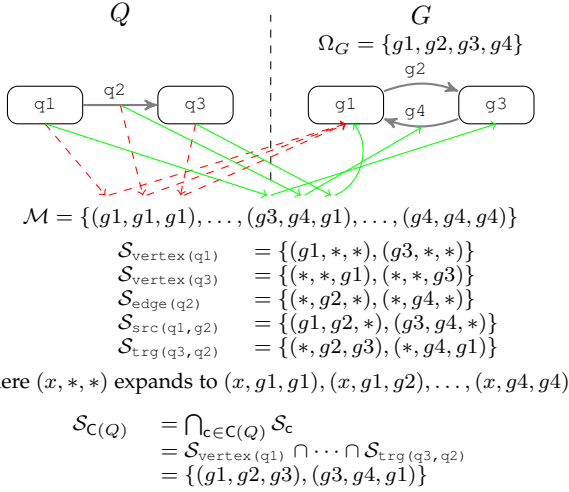

The event ``$\mapping{}$ satisfies all constraints in $\constrSet{}$'' can be expressed as: $\bigcap_{\constr{} \in \constrSet{}}\Sat{\constr{}}$\footnote{Whenever $Q$ and $G$ are clear from the context or not relevant for the discussion, we write $\Sat{\constr{}}$ instead of $\SAT{\constr{}}{Q.\idSet{}}$.}. This will be written shortly as $\Sat{\constrSet{}}$.
Notice that the event $\Sat{\constrSet{}_i \cup \constrSet{}_j}$ represents the set of mappings that satisfy all constraints in $\constrSet{}_i$ \textit{and} all constraints in $\constrSet{}_j$. This corresponds to $\Sat{\constrSet{}_i} \cap \Sat{\constrSet{}_j}$.
Whereas the event $\Sat{\constrSet{}_i} \cup \Sat{\constrSet{}_j}$ represents the set of mappings that satisfies all constraints in $\constrSet{}_i$ \textit{or} all constraints in $\constrSet{}_j$.
Figure \ref{fig:ProbNotationExample} shows an example with the notation introduced in this section.

Within the database literature, the term ``selectivity of $\constr{}$'' is usually used to represent the probability $\Pr[\Sat{\constr{}}]$. 
In this document, we will use both terms interchangeably, since they mean the same thing.
The \emph{goal} in this document is to estimate the selectivity $\Pr[\Sat{\constrSet{}(Q)}]$, since multiplying this with $|\mappingSet_{\queryIdToGraphId{}}|$ (which is equal to $|G.\idSet{}|^{|Q.\idSet{}|}$) gives an estimate for the cardinality $|\Sat{\constrSet{}(Q)}|$. 


\section{Selectivity Estimation Framework}
\label{sec:framework}

The task of obtaining an estimate for the selectivity of a query can be simplified by first obtaining estimates for \emph{subsets} of the query constraints and then \emph{combining} those estimates.
%

An estimate for a subset of the query constraints is called a \textit{partial estimate} (PE)\label{intro:PE}. 
A PE contains a set of constraints (PE.\constrSet{})\label{intro:PE_C} and a selectivity estimate (PE.$s$)\label{intro:PE_s} for that set of constraints, such that $\Pr[\Sat{PE.\constrSet{}}]$ is approximated by $PE.s$.
The set of all PEs for a query will be called its \textit{partial estimate set} (PES)\label{intro:PES}.

Techniques that are able to obtain PEs are called Partial Estimation Techniques (PETs)\label{intro:PET}. A PET requires specific statistics or indexes (\textit{prerequisites}) in order to obtain a PE for a specific set of constraints (\textit{targeted set of constraints}) using a specialized procedure (\textit{estimation procedure}).
For each PET, these three elements will be described (summarized in Table \ref{tab:PETs_mc}).
Techniques that rely on the existence of other partial estimates are called Extend PES Techniques (EPESTs\label{intro:EPEST}). They add PEs to the PES by combining several PEs using specialized assumptions.
The resulting set of PEs is called an Extended PES (EPES\label{intro:EPES}).
Finally, a selectivity estimate for a \queryPattern{} is obtained by combining all PEs of a \textit{complete} PES (CPES\label{intro:CPES}) using a general combine technique (CT\label{intro:CT}).

Notice that a PES\footnote{We will refer to EPES and CPES usually as PES when it will not lead to ambiguity.} should be \textit{complete} before it can be combined. A PES is complete w.r.t. a query $Q$, when every constraint from $\constrSet{}(Q)$ occurs is at least one partial estimate from the PES. Therefore, it must hold that $\bigcup_{PE \in PES_Q} PE.\constrSet{} = \constrSet{}(Q)$.
A PES can be made complete by adding a PE of each missing constraint with a default selectivity value.

\begin{figure}
\centering
\includegraphics[scale=0.55]{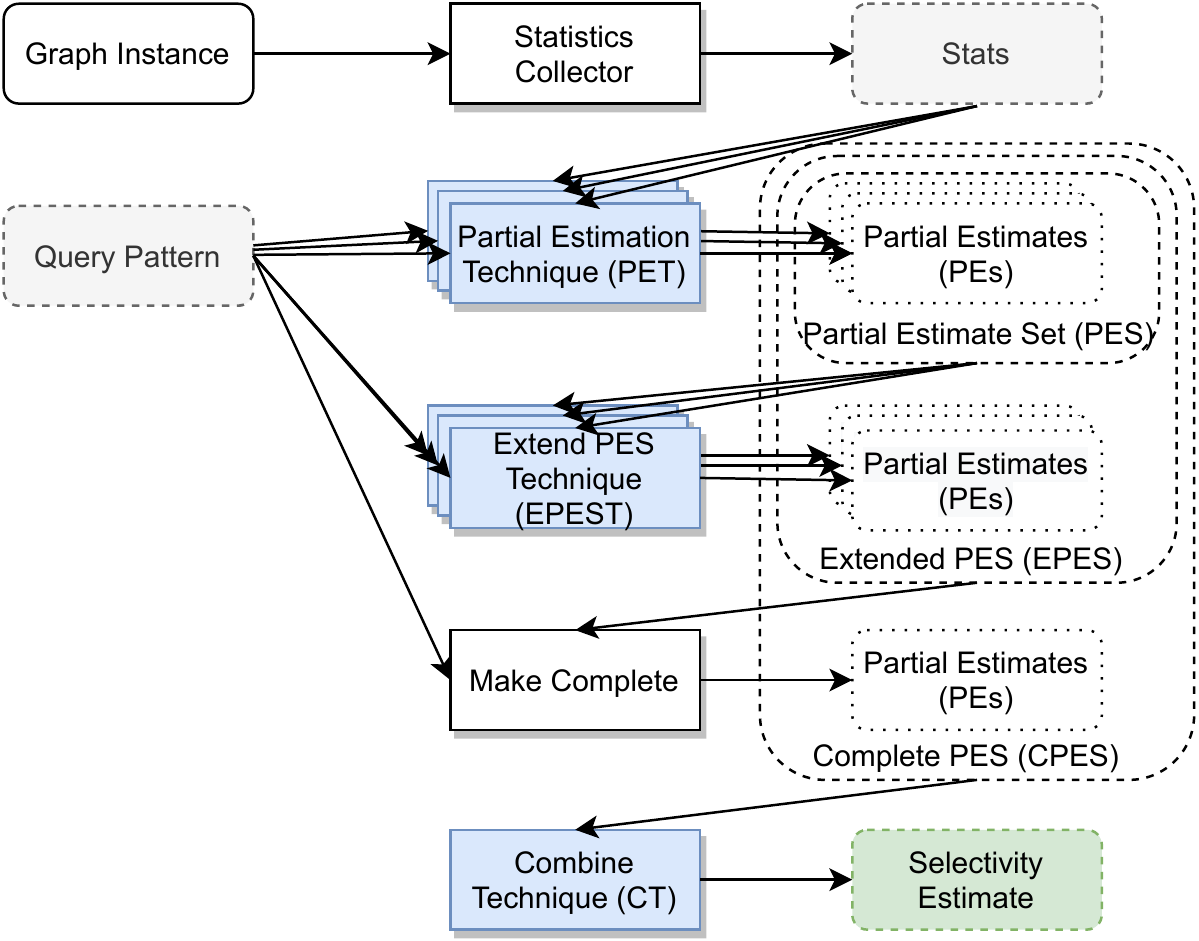}
\caption{Selectivity Estimation Framework}
\label{fig:general_estimation_process}
\end{figure}

The general framework is illustrated in Figure \ref{fig:general_estimation_process}, where the gray dashed boxes are considered as input and the green dashed box as output.
\footnote{The \textit{statistics collection procedure}, i.e. the step from graph instance to statistics, is separate from the estimation procedure. It can be executed after batch loading of a graph instance, at user's request, at low system utilization, at periodic intervals or at demand. Some statistics can also be obtained using a feedback loop after executing queries.}
Processes are shown as boxes with sharp corners and input and output are shown as boxes with rounded corners. The blue boxes represent the three phases of the framework, for which multiple techniques are available. Partially overlapping blue boxes mean that zero or more options can be chosen.
Algorithm \ref{alg:generalSelEstProcess} shows the whole framework in more detail.

\begin{algorithm}
\SetAlgoLined
\KwData{$Q$: a \queryPattern{}}
\KwData{$stats_{G}$: available statistics for instance $G$}
\KwData{$PETs$: A set of partial estimation techniques}
\KwData{$EPESTs$: A set of extend PES techniques}
\KwData{$CT$: Combination technique}
\KwResult{Selectivity estimate for $Q$ on graph instance $G$, i.e. an estimate for $\Pr_G[\Sat{\constrSet{}(Q)}]$,}
$PES_Q$ = \{\}\;
\For{each partial estimation technique $PET$ in $PETs$}{
	$cSets$ $\gets$ find all sets of constraints (subsets of $\constrSet{}(Q)$) that belong to the targeted set of constraints of $PET$\;
	\For{each $cSet$ in $cSets$}{
		\uIf{required stats are available in $stats_G$ to obtain an estimate for $cSet$ using $PET$}{
			$selEst$ $\gets$ obtain selectivity estimate for $cSet$ using statistics and estimation procedure from $PET$\;
			add $(cSet, selEst)$ to $PES_Q$\;
		}
	}
}
$EPES_Q$ = \{\}\;
$EPES_Q$.addAll($PES_Q$)\;
\For{each Extend PES Technique $EPEST$ in $EPESTs$}{
	$EPEs \gets$ obtain extended partial estimates using $EPEST(PES_Q, Q)$\;
	$EPES_Q$.addAll($EPEs$)\;
}
$CPES_Q$ = makeComplete($EPES_Q, Q$)\;
$selEst \gets$ combine all PEs in $CPES_Q$ using $CT$\;
\textbf{return} $selEst$\;
\caption{Selectivity Estimation Framework}
\label{alg:generalSelEstProcess}
\end{algorithm}

\section{PETs for Individual Constraints}
\label{sec:est_ic}

This section deals with PETs for obtaining estimates for individual constraints. 
The \textit{targeted set of constraints} consists of individual constraints of a specific type. 
The following subsections will describe the \textit{estimation procedures} and \textit{prerequisites} for each type.

\vspace{1mm}

\noindent \textbf{Vertex constraints.}
The constraint \vertexConstr{$i$} is satisfied for every mapping that maps query id $i$ to an element in $G.\vertices{}$. All other query ids can be mapped to any element in $G.\idSet{}$. This gives a total of $|G.\vertices{}| \cdot |G.\idSet{}|^{|Q.\idSet{}|-1}$ mappings in event $\SAT{\text{\vertexConstr{$i$}}}{Q.\idSet{}}$.
Therefore, $\Pr[\Sat{\text{\vertexConstr{$i$}}}] = \frac{|G.\vertices{}| \cdot |G.\idSet{}|^{|Q.\idSet{}|-1}}{|G.\idSet{}|^{|Q.\idSet{}|}} = \frac{|G.\vertices{}|}{|G.\idSet{}|}$. The required prerequisites are stored values for $|G.\vertices{}|$ and $|G.\idSet{}|$.

\vspace{1mm}

\noindent \textbf{Edge constraints.}
The selectivity of \edgeConstr{$i$} constraints, where $i \in Q.\idSet{}$, is $|G.\edges{}|/|G.\idSet{}|$. Required prerequisites are stored values for $|G.\edges{}|$ and $|G.\idSet{}|$.

\vspace{1mm}
\noindent \textbf{Src/Trg constraints.}
The constraint \srcConstr{$i_1$}{$i_2$} is satisfied if $i_1$ is mapped to an element $v \in G.\vertices{}$ and $i_2$ is mapped an element in $e \in G.\edges{}$ such that $G.$\edgeDef{}{$e$}{$v$}{$v'$} for some $v' \in G.\vertices{}$. 
Each edge has exactly one source vertex, therefore $i_2$ can be mapped to any element in $G.\edges{}$, then there remains only one option for $i_1$. All other query ids can be mapped to any element in $G.\idSet{}$.  
This gives a total of $|G.\edges{}| \cdot 1 \cdot |G.\idSet{}|^{|Q.\idSet{}|-2}$ mappings in \SAT{\text{\srcConstr{$i_1$}{$i_2$}}}{Q.\idSet{}}.
Therefore, $\Pr[\Sat{\text{\srcConstr{$i_1$}{$i_2$}}}] = 
\frac{|G.\edges{}| \cdot |G.\idSet{}|^{|Q.\idSet{}|-2}}{|G.\idSet{}|^{|Q.\idSet{}|}} = \frac{|G.\edges{}|}{|G.\idSet{}|^2}$. Required prerequisites are $|G.\edges{}|$ and $|G.\idSet{}|$.
The selectivity of \trgConstr{$i_1$}{$i_2$} is the same, because each edge also has exactly one target vertex.

\vspace{1mm}

\noindent \textbf{Label constraints.}
When the number of different labels ($|G.\labelSet{}|$) is small, $\Pr[\Sat{\text{\hasLabelConstr{$i$}{$l$}}}]$ can be precomputed for every $l \in G.\labelSet{}$ (prerequisites).
%
Otherwise, it is possible to treat them as property values with a special property key, e.g. 'hasLabel', and use one of the selectivity estimation methods for property-value constraints.

\vspace{1mm}

\noindent \textbf{Property-key constraints.}
When the number of different property keys ($|G.\propKeySet{}|$) is small, $\Pr[\Sat{\text{\hasPropKeyConstr{$i$}{$k$}}}]$ can be precomputed for every $k \in G.\propKeySet{}$ (prerequisites).
Otherwise, it is possible to treat them as property values with a special property key, e.g. 'hasKey', and use one of the selectivity estimation methods for property-value constraints.

\vspace{1mm}

\noindent \textbf{Property-value constraints.}
The remaining constraints are the property-value constraints, e.g. \hasPropConstr{$i$}{birthyear}{$\leq$}{2000}. 
Storing the exact selectivity of every property constraint in a synopsis (prerequisites) requires only a synopsis lookup (estimation procedure) to obtain the exact selectivity. 
However, storing exact selectivities can be prohibitively expensive as the set $G.$\propValSet{} can be very large in property graph instances. 
This requires specialized techniques to obtain estimates for the selectivity of those constraints.

One way is to assign default selectivity values based on the operator used, i.e. property constraints with equality operator get a selectivity estimate of $1/10$, with inequality get $9/10$ and with other operators get $1/3$\cite{selinger1979access}. We will call this technique ``\textit{Default values}''.
This technique has no prerequisites and the estimation procedure consist of applying simple rules, i.e. if operator is '=', then selectivity is $1/10$.

\label{sec:ip_sampling}
Another way is to store a small sample of the whole data and compute the selectivity of a constraint on that sample when needed. Use that selectivity as an approximation of the selectivity on the whole data.
Prerequisite for this technique is a sample of $G.\idSet{}$.
Sampling is very diverse, but is known to have problem with highly selective constraints and requires large estimation times \cite{cormode2012synopses}.

\label{sec:ip_histograms}
Property constraints with a comparison operator, i.e. $=,\neq,<,\leq,>,\geq$, can be approximated using histograms. A histogram summarizes the distribution of a collection of element by partitioning them into buckets. Different partitioning schemes have been introduced\cite{poosala1996improved, ioannidis2003history} which makes it possible to choose between construction complexity and estimation accuracy. 
Assumptions are used to approximate the values and their frequencies within the buckets.
Prerequisites are histograms for the property keys.

Property constraints with a substring operator, i.e. 'CONTAINS' or 'LIKE' in SQL, can be approximated using pruned suffix trees (PSTs) \cite{jagadish1999substring}.

A summary of the different techniques is given in Table \ref{tab:est_ip}. 
The different approaches all have their own advantages and disadvantages, which makes them useful in different scenarios. 
Also, multiple techniques can be used. Most frequently occurring query constraints can be precomputed. 
Histograms can be constructed for property keys that occur in many constraints with comparison operators. 
While the remaining constraints can be estimated using sampling or using default values.

\begin{table}
\centering
\caption{Obtain estimates for individual constraints}
\label{tab:est_ip}
{\tiny
\begin{tabular}{| l || l | c | c | c | c |}
 \hline
\textbf{Technique} 
& \textbf{Prerequisites}
& \textbf{\criteria{Acc}} 
& \textbf{\criteria{EstTime}} 
& \textbf{\criteria{Mem}} 
& \textbf{\criteria{PrepTime}}\\ \hline \hline
Synopsis Lookup 
& \makecell{Exact selectivity of\\ all indiv. constraints} 
& \darkgreenCell{+ +} 
& \darkgreenCell{+ +} 
& \darkredCell{- -}
& \darkredCell{- -} \\ \hline
Default values 
& None 
& \darkredCell{- -}
& \darkgreenCell{+ +} 
& \darkgreenCell{++} 
& \darkgreenCell{+ +} \\ \hline
Sampling & \makecell{Sample of $G.$\idSet{}} 
& \greyCell{+/-}
& \redCell{-}
& \greenCell{+} 
& \greenCell{+} \\ \hline
Histograms/PSTs & \makecell{Histogram/PST for\\ every property key} 
& \greyCell{+/-} 
& \greenCell{+}  
& \redCell{-}
& \redCell{-} \\  \hline
\end{tabular}
}
\end{table}

\section{PETs for Multiple Constraints}
\label{sec:est_mc}

Previous section showed techniques for estimating the selectivity of every individual constraint. 
%
%
%
%
The techniques described in this section are able to obtain estimates for multiple constraints (Table \ref{tab:PETs_mc} shows a summary), but are not applicable in general to obtain an estimate for \emph{any} possible \queryPattern{}.
Section \ref{sec:est_all} shows how all estimates obtained for multiple constraints (in this section) and for each individual constraint (in the previous section) can be combined to obtain a final estimate for \textit{any} \queryPattern{}.

When the estimation procedure obtains a cardinality estimate $c$ for the set of constraints \constrSet{}, then the selectivity estimate $\Pr[\Sat{\constrSet{}}]$ is $c/|G.\idSet{}|^{|\constrSet{}.\idSet{}|}$.

\begin{table*}
\caption{Summary of different PETs for obtaining estimates for multiple constraints.}
\label{tab:PETs_mc}
\begin{tabular}{|p{0.15\textwidth}|p{0.18\textwidth}|p{0.25\textwidth}|p{0.32\textwidth}|}
    \hline
    \textbf{Technique} & \textbf{Targeted Set of Constraints} & \textbf{Prerequisites} & \textbf{Estimation Procedure} \\ \hline \hline
    Topological Synopsis                
    & Topological Patterns 
    & Topological Synopsis
    & Synopsis Lookup\\ \hline
    Labeled Topological Synopses~\cite{aboulnaga2001estimating}        
    & Labeled Topological Patterns 
    & Labeled Topological Synopsis 
    & Synopsis Lookup\\ \hline
    System R's Join Size Estimation~\cite{selinger1979access}     
    & Labeled Star Patterns 
    & Cardinality and the number of distinct source 
    and target vertices of every labeled edge pattern
    & Apply inclusion and uniform distribution assumptions\\ \hline
    BoundSketch~\cite{cai2019pessimistic}                         
    & Any Labeled Topological Pattern 
    & Cardinality and maximum degree for 
    every vertex in every partition
    & Use upper bound as cardinality estimate. Obtain upper bound by using degree statistics and partitioning.\\ \hline
    Characteristic Sets~\cite{neumann2011characteristic}                 
    & Source/Target Labeled Topological Patterns 
    & For each CS in the data: 
    the number of vertices that have that CS
    and the count for each label in that CS
    & Find the CSs in the data
    that are supersets of the query CS
    and sum their contributions.\\ \hline
    Multidimensional Histograms~\cite{poosala1997selectivity, bruno2001stholes}        
    & Property-value constraints to same query id 
    & The multidimensional histograms
    & Sum the fractions of the buckets in the histogram that satisfy the constraints.\\ \hline
    Sampling                            
    & Any set of \queryPattern{} constraints on a specific pattern type 
    & A random sample of all patterns in the data that belong to a specific pattern type
    & Use the fraction of the sample that satisfies the constraints as selectivity estimate.\\ \hline
    Wander Join~\cite{li2016wander}                        
    & Any set of constraints such that a valid walk plan (ordering of the query edges) exists
    & Indexes to efficiently perform a random walk
    & Perform random walks. If the walk satisfies the constraints, then add 1/(probability of obtaining that walk) to the result. Divide the result by the number of random walks performed to obtain the cardinality estimate \\ \hline
\end{tabular}
\end{table*}

\subsection{Estimation for Topological Constraints}
The topology of a graph refers to its vertices, edges and their connectivity.
Different topological patterns can be defined, which are subclasses of general \queryPatterns{}:

\begin{itemize}

%
%
%
\item 
An \textit{edge pattern}, \edgePattern{$i_1$}{$i_2$}{$i_3$}{1}, consists of a single edge together with its source and target vertices.
\item 
A \textit{chain pattern}, 
\tikz{
\node[scale=0.7] at (0,0) [draw, ellipse, inner sep=1pt](src)  {$i_1$};
\node[scale=0.7] at (0.7,0) [draw, ellipse, inner sep=1pt](trg)  {$i_3$};
\node[scale=0.7] at (1.4,0) [ellipse, inner sep=1pt](dots)  {$\dots$};
\node[scale=0.7] at (2.4,0) [draw, ellipse, inner sep=1pt](nMin2)  {$i_{x-2}$};
\node[scale=0.7] at (3.4,0) [draw, ellipse, inner sep=1pt](n)  {$i_{x}$};
\draw [->](src) to node [color=black, align=center, anchor=south, scale=0.6] {$i_2$} (trg);
\draw [->](trg) to node [color=black, align=center, anchor=south, scale=0.6] {$i_4$} (dots);
\draw [->](dots) to node [color=black, align=center, anchor=south, scale=0.6] {$i_{x-3}$} (nMin2);
\draw [->](nMin2) to node [color=black, align=center, anchor=south, scale=0.6] {$i_{x-1}$} (n);
},
 is a chain of edge patterns in one direction. Let $ep_1, ep_2, \dots ep_n$ be edge patterns. Then, $ep_1/ep_2/\dots/ep_n$ is a chain pattern when the target vertex of $ep_j$ is the source vertex of $ep_{j+1}$ for all $1 \leq j \leq n-1$.
\item 
A \textit{star pattern} is a set of edge patterns that all have a common vertex. 
When all edge patterns have a common source vertex, then the pattern is also called a \textit{source star pattern},
\tikz{
\node[scale=0.7] at (0,0) [draw, ellipse, inner sep=1pt](center)  {$i_1$};

\node[scale=0.7] at (-1,0.15) [draw, ellipse, inner sep=1pt, minimum height=10pt](leftDots1)  {$\dots$};
\node[scale=0.7] at (-1,0) [draw, ellipse, inner sep=1pt, minimum height=10pt](leftDots2)  {$\dots$};
\node[scale=0.7] at (-1,-0.15) [draw, ellipse, inner sep=1pt, minimum height=10pt](leftDots3)  {$\dots$};


\draw [->, dashed](center) to node [color=black, align=center, anchor=south, scale=0.6] {} (leftDots1);
\draw [->, dashed](center) to node [color=black, align=center, anchor=south, scale=0.6] {} (leftDots2);
\draw [->, dashed](center) to node [color=black, align=center, anchor=south, scale=0.6] {} (leftDots3);

}
.
When all edge patterns have a common target vertex, then the pattern is also called a \textit{target star pattern},
\tikz{
\node[scale=0.7] at (0,0) [draw, ellipse, inner sep=1pt](center)  {$i_1$};

\node[scale=0.7] at (-1,0.15) [draw, ellipse, inner sep=1pt, minimum height=10pt](leftDots1)  {$\dots$};
\node[scale=0.7] at (-1,0) [draw, ellipse, inner sep=1pt, minimum height=10pt](leftDots2)  {$\dots$};
\node[scale=0.7] at (-1,-0.15) [draw, ellipse, inner sep=1pt, minimum height=10pt](leftDots3)  {$\dots$};


\draw [->, dashed](leftDots1) to node [color=black, align=center, anchor=south, scale=0.6] {} (center);
\draw [->, dashed](leftDots2) to node [color=black, align=center, anchor=south, scale=0.6] {} (center);
\draw [->, dashed](leftDots3) to node [color=black, align=center, anchor=south, scale=0.6] {} (center);

}
.
\end{itemize}

Other frequently occurring patterns in the query workload can be used, e.g. triangles, cycles, trees, etc. See Bonifati et al. \cite{BonifatiMT20} for a study of frequently occurring query patterns.

The size of a topological pattern is defined as the number of edge patterns involved in that pattern.


\begin{example}
The \queryPattern{} in Figure \ref{fig:queryJOB18a} contains, among others, four subpatterns that are edge patterns, two subpatterns that are size-2 chain patterns and two subpatterns that are size-2 source star patterns.
\end{example}

\subsubsection{Topological Synopsis Lookup}
A topological synopsis contains the cardinalities of topological patterns.
For example, it can contain the cardinalities of a chain pattern of size $2$ and source star patterns of size $2$ and $3$.
%
The cardinalities can be precomputed and stored at a small cost, when the patterns are small.   

\vspace{1mm}
\noindent \textbf{Targeted Set of Constraints.}
All topological subpatterns of the \queryPattern{} for which the synopsis stores its cardinality.

\vspace{1mm}
\noindent \textbf{Prerequisites.}
The topological synopsis.

\vspace{1mm}
\noindent \textbf{Estimation Procedure.}
The cardinality can be obtained using a lookup in the synopsis. 

\subsection{Estimation for Labeled Topological Constraints}
Vertices and edges can have labels, which can have a large impact on the cardinality of a query.
%
%

Labeled topological patterns are an extension of topological patterns, where each query id is allowed to have at most one label e.g., 
a \textit{labeled edge pattern}, \textit{labeled chain pattern}, etc.


\subsubsection{Labeled Topological Synopsis Lookup}
\label{sec:structSyn}
A labeled topological synopsis contains cardinalities of labeled topological patterns belonging to a specific class, e.g. labeled edge patterns or labeled chain patterns of size $2$.


The specific pattern class and its size are considered the parameters of a labeled topological synopsis. 
For example, a labeled topological synopsis can be specified by labTopSyn(chain, $2$), which represents a synopsis which stores the cardinality of all labeled chain patterns of size $2$.

Aboulnage et al.~\cite{aboulnaga2001estimating} introduced a synopsis for labeled chain pattern up to size $n$ and used that synopsis to obtain estimates for labeled chain patterns of size $m > n$ using Markov Chains:
{ \footnotesize
\begin{align*}
\Pr[\Sat{\constrSet{}(t_1/\dots/t_m)}] = &\Pr[\Sat{\constrSet{}(t_1/\dots/t_n)}] 
 \cdot \prod_{i=1}^{m-n} \frac{\Pr[\Sat{\constrSet{}(t_{1+i}/\dots/t_{n+i})}]}{\Pr[\Sat{\constrSet{}(t_{1+i}/\dots/t_{n+i-1})}]}
\end{align*}
}
where $t_i/\dots/t_j$ represents a chain of labeled edge patterns $t_x$ for $i \leq x \leq j$.
This formula is a special case of the combination technique based on conditional independence (with sorting strategy based on maximum overlap), which is explained in detail in Section \ref{sec:combProc_indep}.

%

\vspace{1mm}
\noindent \textbf{Targeted Set of Constraints.}
All subpatterns of the \queryPattern{} belonging to the pattern class of the labeled topological synopsis.

\vspace{1mm}
\noindent \textbf{Prerequisites.}
The labeled topological synopsis.

\vspace{1mm}
\noindent \textbf{Estimation Procedure.}
The cardinality can be obtained using a lookup in the synopsis.

\subsubsection{\SystemR{}'s Join Size Estimation using Inclusion and Uniform Distribution Assumptions}
\label{sec:tb_join_size_est}
Join size estimation in relational databases has its origin from \textit{\SystemR{}}~\cite{selinger1979access}. 
%
Considering each labeled edge pattern as a relation, with an attribute for the source vertex id, edge id and target vertex id, allows us to apply \SystemR{}'s join size estimation method in the context of this document. 

\SystemR{} uses the cardinality and column cardinality of each relation (i.e. number of edges and number of distinct source and target vertices for each labeled edge pattern), and three assumptions to estimate the cardinality of any sequence of equality joins (i.e. any labeled topological pattern). 
The three assumptions are 1) the inclusion assumption (all values in the joining attribute with the lowest number of distinct values are included in the set of values of the other joining attribute), 
2) the uniform distribution assumption (all distinct values of an attribute have the same frequency) and 
3) the preservation of value set assumption (all distinct values in the non-joining attributes are preserved after the join).

\vspace{1mm}

\noindent \textbf{Targeted Set of Constraints.}
Using all three assumptions, it is possible to obtain cardinality estimates for any labeled topological pattern. 
Using only the first two assumptions, it is possible to obtain estimates for any star pattern, which will be the focus of this section.


\vspace{1mm}

\noindent \textbf{Prerequisites.}
Let $S_G(ep)$ and $T_G(ep)$ be the multiset of source and target vertices on the edges in $G$ that match labeled edge pattern $ep$.
Let $S'_G(ep)$ and $T'_G(ep)$ be the set versions of $S_G(ep)$ and $T_G(ep)$, where duplicates are eliminated.

This approach can be applied to star pattern $S$ when the following statistics are available for every edge pattern $ep$
 in $S$ for graph instance $G$:
\begin{itemize}
\item $n_G(ep)$: cardinality of $ep$ in $G$
\item $|S'_G(ep)|$: the number of distinct source vertices of the edges in $G$ that match $ep$
\item $|T'_G(ep)|$: the number of distinct target vertices of the edges in $G$ that match $ep$
\end{itemize}

%

\vspace{1mm}

\noindent \textbf{Estimation Procedure.}
%
Let $f(M,v)$ represent the frequency of vertex $v$ in multiset $M$.
Consider a \textit{labeled star pattern} that consists of labeled edge patterns $\{ep_1, \dots, ep_n\}$ and a center vertex $v$. Let $X_G(ep_i)$ be $S_G(ep_i)$ if $v$ is the source vertex of $ep_i$ and $T_G(ep_i)$ if $v$ is the target vertex of $ep_i$.

In general, the cardinality of the star pattern is
\begin{align*}
&\sum_{v \in X'_G(ep_1) \cap \dots \cap X'_G(ep_n)} \prod_{1 \leq i \leq n}f(X_G(ep_i), v)
\end{align*}
Assuming a uniform distribution gives:
\begin{align*}
&\approx \sum_{v \in X'_G(ep_1) \cap \dots \cap X'_G(ep_n)} \prod_{1 \leq i \leq n}\frac{n_G(ep_i)}{|X'_G(ep_i)|}
\end{align*}
The sum is independent of the value $v$:
\begin{align*}
= |X'_G(ep_1) \cap \dots \cap X'_G(ep_n)| \prod_{1 \leq i \leq n}\frac{n_G(ep_i)}{|X'_G(ep_i)|}
\end{align*}
Now, use the inclusion assumption to estimate the size of the intersection:
\begin{align*}
\approx min(X'_G(ep_1), \dots, X'_G(ep_n)) \prod_{1 \leq i \leq n}\frac{n_G(ep_i)}{|X'_G(ep_i)|}
\end{align*}

This makes it possible to estimate the cardinality of a star pattern $S$ using only $n_G(ep), |S'_G(ep)|$ and $|T'_G(ep)|$ for every edge pattern $ep$ in $S$.

\subsubsection{Bound Sketch}
Consider a size-2 labeled chain pattern $ep_i/ep_j$ = \chainTwoPattern{$i_1:l_1$}{$i_2:l_2$}{$i_3:l_3$}{$i_4:l_4$}{$i_5:l_5$}{1.5}{3} where $ep_i$ = \edgePattern{$i_1:l_1$}{$i_2:l_2$}{$i_3:l_3$}{1.5} and $ep_j$ = \edgePattern{$i_3:l_3$}{$i_4:l_4$}{$i_5:l_5$}{1.5} are labeled edge patterns, $i_1, i_2, \dots$ are ids and $l_1, l_2, \dots$ are labels.

If the cardinalities of $ep_i$ and $ep_j$ are known, then the cardinality of $ep_i/ep_j$ is at most $|ep_i| \cdot |ep_j|$. 
This upper bound occurs if every mapping from $ep_i$ matches with every mapping from $ep_j$, which can be visualized as 
\tikz{
\node[scale=0.7] at (0,0) [draw, ellipse, inner sep=1pt](center)  {$y$};

\node[scale=0.7] at (-1,0.15) [draw, ellipse, inner sep=1pt, minimum height=10pt](leftDots1)  {$\dots$};
\node[scale=0.7] at (-1,0) [draw, ellipse, inner sep=1pt, minimum height=10pt](leftDots2)  {$\dots$};
\node[scale=0.7] at (-1,-0.15) [draw, ellipse, inner sep=1pt, minimum height=10pt](leftDots3)  {$\dots$};

\node[scale=0.7] at (1,0.15) [draw, ellipse, inner sep=1pt, minimum height=10pt](rightDots1)  {$\dots$};
\node[scale=0.7] at (1,0) [draw, ellipse, inner sep=1pt, minimum height=10pt](rightDots2)  {$\dots$};
\node[scale=0.7] at (1,-0.15) [draw, ellipse, inner sep=1pt, minimum height=10pt](rightDots3)  {$\dots$};

\draw [->](leftDots1) to node [color=black, align=center, anchor=south, scale=0.6] {$:l_2$} (center);
\draw [->](leftDots2) to node [color=black, align=center, anchor=south, scale=0.6] {} (center);
\draw [->](leftDots3) to node [color=black, align=center, anchor=south, scale=0.6] {} (center);

\draw [->](center) to node [color=black, align=center, anchor=south, scale=0.6] {$:l_4$} (rightDots1);
\draw [->](center) to node [color=black, align=center, anchor=south, scale=0.6] {} (rightDots2);
\draw [->](center) to node [color=black, align=center, anchor=south, scale=0.6] {} (rightDots3);
}.

Let $d^{v}_{ep}$ be the maximum degree of vertices in $\m{v}$ in the mappings from edge pattern $ep$.
Above is a special case where $d^{i_3}_{ep_i} = |ep_i|$ and where $d^{i_3}_{ep_j} = |ep_j|$.

When $d^{i_3}_{ep_j}$ is known, then each mapping from $ep_i$ can be matched with at most $d^{i_3}_{ep_j}$ mappings from $ep_j$, which leads to the upper bound $|ep_i| \cdot d^{i_3}_{ep_j}$ for the cardinality of $ep_i/ep_j$.
Similar, when $d^{i_3}_{ep_i}$ is known, then this leads to the upper bound $d^{i_3}_{ep_i} \cdot |ep_j|$ for the cardinality of $ep_i/ep_j$.
When both $d^{i_3}_{ep_j}$ and  $d^{i_3}_{ep_i}$ are known, the upper bound can be obtained by taking the minimum of both upper bounds.

Cai et al.~\cite{cai2019pessimistic} improve upon this kind of upper bounds, by partitioning the data of the join attributes. Their data summary is called a Bound Sketch.
For example, above upper bound can be improved by partitioning the mappings of $ep_i$ and $ep_j$ on their values for $\m{i_3}$, and storing the cardinality and maximum degree for each partition. 
Then, the cardinality estimate of the pattern $ep_i/ep_j$ is the sum of the upper bound obtained from each partition.
The upper bound for the cardinality can be used as an estimate for the cardinality.

\vspace{1mm}
\noindent \textbf{Targeted Set of Constraints.}
Any labeled topological pattern.

\vspace{1mm}
\noindent \textbf{Prerequisites.}
Cardinality and degree statistics for each partition.

\vspace{1mm}
\noindent \textbf{Estimation Procedure.}
Obtain upper bounds for each bounding formula. Use the lowest one as cardinality estimate. Obtain an upper bound by applying a bounding formula to each partition and sum the results.

%
%


\subsection{Estimation for Query Pattern Schema Constraints}
The class of schema constraints extends the class of labeled topological constraints with 'has property key' constraints.

\subsubsection{Characteristic Sets}
\label{sec:char_set}
Frequently, in graph instances, many vertices have the same property keys and the same labels on adjacent edges. This insight was exploited by Neumann and Moerkotte \cite{neumann2011characteristic} and will be discussed in this section.

The set of labels on all outgoing edges of a vertex $v$ is called the Characteristic Set (CS) of $v$, denoted by $CS(v)$ \cite{neumann2011characteristic}. 
This definition was defined on the RDF data model\cite{world2014rdf,schreiber2014rdf}. 
In the RDF data model, properties are encoded as edges, with the property key on the edge and the property value on the target vertex. 
Since we are focusing on property graphs, we extend the definition of $CS(v)$ to also include all property keys of vertex $v$.

\vspace{1mm}

\noindent \textbf{Targeted Set of Constraints.}
For each labeled source star pattern, $P_{SS}$, this approach can produce an estimate for the set of constraints that define $P_{SS}$, i.e. $\constrSet{}(P_{SS})$, together with all 'has property key' constraints on its center vertex.

\vspace{1mm}

\noindent \textbf{Prerequisites.}
Let $CS(G)$ be the set of all different CSs that occur in the graph instance $G$. In theory, each vertex can have a different CS, but, in practice, the number of CSs in an instance is usually much smaller\cite{neumann2011characteristic}. 

For each CS $CS' \in CS(G)$ in the graph instance $G$, store:
\begin{itemize}
\item $|CS'|$: The number of vertices $v$ in the data, such that the CS of $v$ is equal to $CS'$ (i.e. $CS(v)=CS'$)
\item For each edge label $l \in CS'$:
	\begin{itemize}
	\item $CS'.count(l)$: The total number of outgoing edges with label $l$ from the vertices with CS $CS'$
	\end{itemize}
\end{itemize}

Each query id can have only one value for a property key in our model. Therefore, for each property key $k \in CS'$, $CS'.count(k) = |CS'|$. Therefore, this does not need to be stored explicitly for every property key in a CS.
 

A large number of different characteristic sets in a graph instance will lead to slow estimation time, since finding all supersets of the CS of the query will lead to a noticeable overhead (see Estimation Procedure). 
Neumann and Moerkotte \cite{neumann2011characteristic} propose to merge a CS into a CS that is a superset to limit to total number of different CSs. 
They suggest to keep the $10000$ most frequent CSs and merge the remaining CSs. 
Merging is first attempted directly by finding the smallest possible superset (use the most frequent one as tiebreaker). 
When no superset can be found, the set is first split (largest possible set that is a subset of another CS, and 'the rest') and then merging both parts individually.

\vspace{1mm}

\noindent \textbf{Estimation Procedure.}
Let $P'_{SS}$ denote the a labeled source star pattern extended with the 'has property key' constraints on its center vertex.
Let $CS(P'_{SS})$ be the CS of the (sub)query pattern $P'_{SS}$.
Estimate the cardinality of $P'_{SS}$ as:
\begin{align*}
&cardEst(P'_{SS}) \approx \\
&\sum_{\{CS' \in CS(G) \mid CS' \supseteq CS(P'_{SS})\}} \left( |CS'| \cdot  \prod_{e \in CS(P'_{SS})} \frac{CS'.count(e)}{|CS'|} \right)
\end{align*}

Here, CSs in $CS(G)$ that includes statistics of interest to estimate of the cardinality of $P'_{SS}$ are those that are supersets of $CS(P'_{SS})$. 
The first part $|CS'|$ gives the number of vertices with CS $CS'$. 
Each vertex with CS $CS'$ can have multiple edges with the label $l \in CS(P'_{SS})$. 
A \textit{uniform distribution} is assumed, i.e. each vertex with CS $CS'$ has approximately $\frac{CS'.count(l)}{|CS'|}$ number of edges with label $l$.
If $e$ is a property key, then $CS'.count(e)=|CS'|$.


\begin{example}
Our example query from Figure \ref{fig:queryJOB18a} has three CSs with at least two elements: $C_{Q1}=\{budget, votes\}$, $C_{Q2}=\{cast\_info\_person, cast\_info\_movie, note\}$ and $C_{Q3}=\{gender, name\}$. 

\end{example}

With minor differences, this technique can also be used for labeled target star patterns.

\subsection{Estimation for Property-value Constraints}
\label{sec:mc_propConstrs}
All previous approaches focused on obtaining estimates for the topology of the query and extensions that also include labels and 'has property key' constraints. 
Combination of property-value constraints will be the focus of this section.

\subsubsection{Multidimensional Histogram}
Section \ref{sec:ip_histograms} showed that the distribution of property values for a specific property key can be approximated using a histogram. 
Extension are developed to approximate joint distributions of values from more than one property key \cite{poosala1997selectivity, bruno2001stholes}. 

A difficult question is for which set of property keys to construct a multidimensional histogram.
Bruno and Chaudhuri \cite{bruno2002exploiting} follow a workload driven approach to find the combinations of constraints that will benefit most from additional statistics.

\vspace{1mm}
\noindent \textbf{Targeted Set of Constraints.}
A set of property-value constraints referring to the same query id.

\vspace{1mm}
\noindent \textbf{Prerequisites.}
A multidimensional histogram for a combination of all property keys of the property-value constraints.

\vspace{1mm}
\noindent \textbf{Estimation Procedure.}
Find the buckets in the multidimensional histogram that contain elements that satisfy the constraints and estimate the fraction of the element within those buckets that satisfy the constraint. The sum of those values will be the cardinality estimate. 
Intra bucket estimation is done using assumptions like the uniform spread and uniform distribution assumption.

\subsection{Estimation for Query Pattern Constraints}
\label{mc_topologyAndData}
This section will focus on estimation techniques for sets of constraints that can include any combination of \queryPattern{} constraints.

\subsubsection{Sampling}
\label{sec:mcSampling}
A sample of all graph instance ids can be used to obtain a cardinality estimate for any \queryPattern{} by evaluating the \queryPattern{} on the sample and scaling up the number of results obtained. 
However, evaluating a \queryPattern{} over a sample can make the estimation time large, especially when the sample size needs to be large in order to obtain accurate estimates (e.g. for queries containing very selective constraints).
Therefore, we focus on using sampling to obtain estimates for small patterns together with all data constraints on those patterns.

Each graph pattern in the graph instance that belongs to a specific pattern type is sampled with a specific probability. 
The pattern type and probability are considered as the parameters of this technique, i.e. $S(pt, pr)$. 
As pattern type, one could think of: id, (labeled) vertex pattern, (labeled) edge pattern, etc.
The probability defines a trade-off between estimation time and estimation accuracy. A larger probability, is expected to lead to a larger sample size, a larger estimation time and a better estimation accuracy.

\vspace{1mm}
\noindent \textbf{Targeted Set of Constraints.}
All subpatterns of the \queryPattern{} that belong to the pattern type $pt$ together with all the data constraints on those subpatterns.

\vspace{1mm}
\noindent \textbf{Prerequisites.}
A random sample of all patterns in the graph instance of the type $pt$, where each pattern is chosen with a probability of $pr$.

\vspace{1mm}
\noindent \textbf{Estimation Procedure.}
The selectivity can be estimated by the fraction of the patterns in the sample that satisfy all the data constraints. 
This approach assumes that the sample is representative for the whole collection of patterns, i.e. the fraction of patterns that satisfy the constraints is the same in the sample and in the whole collection.

\begin{example}
$S(id, 0.001)$ can obtain an estimate for the constraints \vertexConstr{id8}, \hasLabelConstr{id8}{person}, \hasPropConstr{id8}{gender}{=}{m}, \hasPropConstr{id8}{name}{CONTAINS}{Tim} from Figure \ref{fig:queryJOB18a}, because they all refer to the same query id, namely $id8$.
\end{example}

\subsubsection{Wander Join}
Wander Join~\cite{li2016wander} is a sampling technique that obtains samples of the results of a (sub)query by performing random walks over the topology of the (sub)query.

\vspace{1mm}
\noindent \textbf{Targeted Set of Constraints.}
Any set of constraints on the subpatterns of the \queryPattern{}, such that there exists an ordering of the query edges ($e_1,e_2,\dots,e_n$) where $e_i$ has a common vertex with some $e_j$ where $j < i$ and $e_i$ has an index on that vertex.

\vspace{1mm}
\noindent \textbf{Prerequisites.}
Indexes to efficiently obtain all edges with a specific source or target vertex, in order to perform efficient random walks.

\vspace{1mm}
\noindent \textbf{Estimation Procedure.}
Perform $x$ random walks according to the valid order, where $x$ is a parameter that defines the trade-off between estimation time and estimation accuracy. 
Each random walk returns a value. For cardinality estimation this would be either $1$ or $0$ (i.e. the random walk satisfies all constraints or it does not). 
Divide the value by the probability of occurrence of the walk that was obtained. 
Finally, take the average over the values obtain by the $x$ random walks to obtain a cardinality estimate.

\section{Extend PES Techniques}
\label{sec:extend_PES}

Techniques that can obtain partial estimates for \queryPatterns{} that rely on the existence of other partial estimates are covered in this section. 


\subsection{Implied Constraints}
If the constraint \srcConstr{$i_1$}{$i_2$} from \queryPattern{} $Q$ is satisfied for a mapping $\mapping{}$ to graph $G$, then it must be the case that the constraints \vertexConstr{$i_1$} and \edgeConstr{$i_2$} are also satisfied. 
This implication holds for any $i_1, i_2 \in Q.\idSet{}$, due to the definition of the source constraint. 
Namely, if \srcConstr{$i_1$}{$i_2$} is satisfied in $G$, then $G.$\edgeDef{}{$i_2$}{$i_1$}{$v'$} for some $v' \in G.\vertices{}$. 
Since \edgeGreek{} is defined as $\edges{} \rightarrow (\vertices{} \times \vertices{})$, it must hold that $i_2 \in \edges{}$ and $i_1 \in \vertices{}$, which is specified by the constraints \edgeConstr{$i_2$} and \vertexConstr{$i_1$}.

If the PES contains an estimate for \srcConstr{$i_1$}{$i_2$}, for any $i_1, i_2 \in Q.\idSet{}$, then a new PE will be added to the PES consisting of the set of constraints \{\srcConstr{$i_1$}{$i_2$}, \vertexConstr{$i_1$}, \edgeConstr{$i_2$}\} with the selectivity of the PE for \srcConstr{$i_1$}{$i_2$}.
More precisely, $\Pr[\Sat{\{\text{\srcConstr{$i_1$}{$i_2$}, \vertexConstr{$i_1$}, \edgeConstr{$i_2$}}\}}] = \Pr[\Sat{\text{\srcConstr{$i_1$}{$i_2$}}}]$, for any $i_1, i_2 \in Q.\idSet{}$.

Similarly, \trgConstr{$i_1$}{$i_2$} implies \vertexConstr{$i_1$} and \edgeConstr{$i_2$} for any $i_1, i_2 \in Q.\idSet{}$.
Also, \hasPropConstr{$i$}{$k$}{$\theta$}{$v$} implies \hasPropKeyConstr{$i$}{$k$}, for any $i \in Q.\idSet{}, k \in Q.\propKeySet{}, \theta \in Q.\Theta, v \in Q.\propValSet{}$.

\subsection{Implied Constraints Assumptions}
Each query id can have many property constraints associated with it. Neumann and Moerkotte \cite{neumann2011characteristic} noticed that usually one property-value constraint is extremely selective, which (nearly) implies the other property-value constraints. 
This means that, when the most selective property-value constraint on a query id is satisfied, then all other property-value constraints on that query id are (usually) satisfied too.

Whenever detailed statistics about combinations of those constraints are not available, instead of assuming independence between those constraints, assuming an \emph{implication} from the most selective one to the others might be more appropriate.


\begin{example}
Consider a graph $G$ to which the query from Figure \ref{fig:queryJOB18a} is executed.
If $G$ has less ids that have the property key ``name'' that contains the word ``Tim'' (\constr{28} from Example \ref{ex:runningExampleQuery_constraints}) than ids that have the property key ``gender'' with the value ``m'' (\constr{27}), i.e. $\Pr[\Sat{\constr{28}}] < \Pr[\Sat{\constr{27}}]$, then constraint $\constr{28}$ is assumed, by this approach, to imply $\constr{27}$. 
This gives $\Pr[\Sat{\{\constr{27},\constr{28}\}}] = \Pr[\Sat{\constr{28}}]$.
\end{example}

This idea can be extended to a more general form, which requires as input a pattern class and a class of constraints. Above description has 'Id' as pattern class and 'Property-value constraints' as class of constraints, which will be identified as \statShort{IP(Id, PropValueConstrs)}.

The generalized procedure works as follows. 
Let the defined pattern class be $x$ and the class of constraints be $y$.
For every pattern $x_i$ in the \queryPattern{} that belongs to the class $x$, find $PE(x_i, y)$.
The set $PE(x_i, y)$ consists of all PEs in the PES where all constraints belong to the class $y$ and where the query ids in the constraints of the PE is a subset of the query ids in the constraints of $x_i$, i.e. $(PE.\constrSet{}).\idSet{} \subseteq \constrSet{}(x_i).\idSet{}$. 
An implication assumption is applied from the constraints in the PE in $PE(x_i, y)$ with the lowest selectivity estimate (PE.$s$) to the union of all constraints in the PEs in $PE(x_i, y)$. More precisely, $PE_{min}(x_i, y).\constrSet{}$ is assumed to imply $\bigcup_{PE \in PE(x_i, y)}PE.\constrSet{}$, where $PE_{min}(x_i, y)$ is the PE in $PE(x_i, y)$ that has the lowest selectivity.


\section{Combine Techniques}
\label{sec:est_all}

This section will show different techniques to combine the partial estimates of complete PESs into a single selectivity estimate.
The different approaches that will be discussed are summarized in Table \ref{tab:combining_pk}. 
Notice that these techniques do not depend on the availability of specific statistics or indexes, therefore they can be applied in all cases (on PESs that are \emph{complete}). 
Statistical assumptions are used to combine PEs. 
This part of the estimation procedure can introduce a large estimation error. 
Therefore, it is important to make the best out of the available partial estimates, since they are obtained using available statistics instead of rough assumptions.
\begin{table}[H]
\centering
\caption{General Combine Techniques}
\label{tab:combining_pk}
\begin{tabular}{| l | l |}
\hline
\textbf{Technique} & \textbf{Assumptions made} \\ \hline \hline
\ctShort{CondIndep} & \makecell{Conditional Independence\\between sets of constraints}\\ \hline
\ctShort{MaxEnt} & \makecell{Most uniform probability mass function that\\satisfies all constraints from the $PES$} \\ \hline
\ctShort{Upper Bound} & Worst case scenario (from comp. perspective)\\ \hline
\ctShort{Lower Bound} & Best case scenario (from comp. perspective)\\ \hline
\end{tabular}
\end{table}

\subsection{Conditional Independence Assumptions}
\label{sec:combProc_indep}
The most simple way of combining all partial estimates for the individual constraints (using techniques from Section \ref{sec:est_ic}) is by assuming independence between the different constraints. This makes it possible to obtain an estimate by multiplying the individual selectivity estimates: 
\begin{equation}
\label{eq:combine_indep}
\Pr[\Sat{\constrSet{}(Q)}] = \prod_{\constr{} \in \constrSet{}(Q)} \Pr[\Sat{\constr{}}]
\end{equation}

This formula implies independence between all constraints and does not make use of the partial estimates obtained for multiple constraint using techniques from Section \ref{sec:est_mc}.

If a partial estimate set $PES_Q$ is complete, w.r.t. query $Q$, then the intersection of all events $\Sat{PE.\constrSet{}}$ for $PE \in PES_Q$ gives the event $\Sat{\constrSet{}(Q)}$.
Therefore, the probability of all query constraints can be defined in terms of a complete partial estimate set $PES_Q$:
$$
\Pr[\Sat{\constrSet{}(Q)}] = \Pr[\bigcap_{PE \in PES_Q} \Sat{PE.\constrSet{}}].
$$

Assuming some order of the partial estimates in $PES_Q$, makes it possible to apply the chain rule of probability:
\begin{align*}
\Pr[\bigcap_{PE \in PES_Q} \Sat{PE.\constrSet{}}] &= 
\prod_{i \in [1,..,|PES_Q|]} \Pr[\Sat{PE_i.\constrSet{}} \mid \bigcap_{j =1}^{i-1} \Sat{PE_j.\constrSet{}}]\\
&= \prod_{i \in [1,..,|PES_Q|]}\Pr[\Sat{PE_i.\constrSet{}} \mid \Sat{\bigcup_{j=1}^{i-1} PE_j.\constrSet{}}].
\end{align*}

This formula performs a product over all PEs in the PES. 
The values of the product consist of the probability that the set of constraints (of the current PE) hold, given that all previously considered constraints hold.
If the PES is complete, then all constraints are considered in the end.

We are limited to the PEs that are available in the PES. Therefore, the probability will be approximated using conditional independence where necessary. 
This gives:
\begin{align*}
&\Pr[\Sat{\constrSet{}(Q)}] \\
&\approx \prod_{i \in [1,..,|PES_Q|]}\Pr[\Sat{PE_i.\constrSet{}''} \mid \Sat{(\bigcup_{j=1}^{i-1} PE_j.\constrSet{})''  \cap PE_i.\constrSet{}''}]\\
&= \prod_{i \in [1,..,|PES_Q|]} PE_i.s / \Pr[\Sat{(\bigcup_{j=1}^{i-1} PE_j.\constrSet{})''  \cap PE_i.\constrSet{}''}]\\
\end{align*}
where $S''$ represents the set of all constraints in $\constrSet{}(Q)$ implied by $S$, which can be obtain from the set $S$ as follows:
\begin{itemize}
\item add all elements from $S$ to $S''$;
\item for each \srcConstr{$i_1$}{$i_2$} (or \trgConstr{$i_1$}{$i_2$}) where $i_1, i_2 \in Q.\idSet{}$, add \vertexConstr{$i_1$} and \edgeConstr{$i_2$} to $S''$;
\item for each \propConstrVars{$i_1$}{$k$}{$\theta$}{$v$} where $i_1 \in Q.\idSet{}$, $k \in Q.\propKeySet{}$, $\theta \in Q.\Theta$, $v \in Q.\propValSet{}$, add \hasPropKeyConstr{$i_1$}{$k$}.
\end{itemize}

The probability of a subset of the constraints in a partial estimate might be needed due to the intersection in the 'given' part of the conditional probability. This selectivity might not be available. 
Therefore, the general selectivity estimation process is called recursively on this subproblem.
The lower bound for this selectivity is set to $\Pr[\Sat{PE_i.\constrSet{}}]$, because any subset of $PE_i.\constrSet{}$ imposes no more constraints than $PE_i.\constrSet{}$ does.
%

Notice that the resulting formula leads to (\ref{eq:combine_indep}) when the PES includes only PEs for individual constraints and no constraints imply other constraints.

Different orderings of the PES can lead to different selectivity estimates. Some possible options that can be used as sorting strategy are:
\begin{enumerate}
\item Primary sort on the number of constraints in the partial estimates (descending). Secondary sort on the selectivity value (ascending).
\item Primary sort on the deviation from the independence assumption (descending). Deviation from independence for partial estimate $PE$ is defined as: $max(PE.s / \prod_{\constr{} \in PE.\constrSet{}}\Pr[\Sat{\constr{}}],$ $\prod_{\constr{} \in PE.\constrSet{}}\Pr[\Sat{\constr{}}] / PE.s)$. This gives the value $1$ if the constraints in $PE$ are mutually independent and a higher value the more dependent they are. 
Secondary sort on the selectivity value (ascending).
\item Primary sort on the overlap between its constraints and the previously considered constraints (descending).
Secondary sort on the deviation from independence (descending).
\end{enumerate}

Algorithm \ref{alg:combine_indep} shows the general procedure of combining all partial estimates using the chain rule of probability and conditional independence assumptions. 
The algorithm uses \textit{sort()}, \textit{impl()} (short for implied) and \textit{selEst()} as subroutines. 
The first two are explained in the text above and the last one is the general selectivity estimation process, which is shown in Algorithm \ref{alg:generalSelEstProcess}. 
The parameters involving the (sets of) techniques of choice and the available graph instance statistics are shown as dots to keep the description clean.

\begin{algorithm}
\SetAlgoLined
\KwData{CPES: a complete partial estimate set}
\KwResult{Selectivity estimate for the conjunction of all constraints in the PEs of CPES}
sort(CPES)\;
constraintsDone = \{\}\;
currentEst = 1.0\;
\For{each partial estimate $PE$ in CPES}{
	cDoneAndImplied = impl(constraintsDone)\;
	cImpl = impl($PE.\constrSet{}$)\;
	intersection = cDoneAndImplied $\cap$ cImpl\;
	\uIf{intersection is empty}{
		currentEst *= $PE.s$\;
	}
	\uElseIf{intersection != cImpl}{
		currentEst *= $\frac{PE.s}{max(PE.s,\text{ } selEst(\text{intersection},\dots))}$\;
	}		
	constraintsDone = constraintsDone $\cup PE.\constrSet{}$\;
}
\textbf{return} currentEst\;
\caption{combineIndep(CPES, \dots)}
\label{alg:combine_indep}
\end{algorithm}


\subsection{The Maximum Entropy approach}
\label{sec:maxEnt}
Each constraint $\constr{i} \in \constrSet{}(Q)$ can be modeled as a Bernoulli variable, $X_i$, where $\Pr[X_i]=\Pr[\Sat{\constr{i}}]$.
Then, $\Pr[\Sat{\constrSet{}(Q)}]$ corresponds to the joint probability $\Pr[X_{1}=True, X_{2}=True, \dots, X_{n}=True]$, where $n=|\constrSet{}(Q)|$.

Instead of estimating only the joint probability $\Pr[X_{1}=True, X_{2}=True, \dots, X_{n}=True]$, Markl et al. \cite{markl2007consistent} approximate the whole joint probability mass function $\Pr[X_1=x_1, \dots, X_n=x_n]$ where $x_i$ is $True$ or $False$ for $1 \leq i \leq n$.
The idea is to assign the probabilities such that:
\begin{itemize}
\item each probability is $\geq 0$ and $\leq 1$;
\item the sum of all probabilities adds up to $1$;
\item for each partial estimate $PE$ it holds that:
the sum of all items, where $X_i=True$ for all $i$ such that $\constr{i} \in PE.\constrSet{}$, is equal to $PE.s$.
\end{itemize}

Since many probability mass functions might satisfy above requirements, the one which maximizes the Entropy function will be chosen. 
This is the most uniform probability mass function (i.e. the one with the largest uncertainty) that is consistent with the information from the complete partial estimate set (CPES).

\subsubsection{Constraint Optimization Problem}
Let $S \subseteq \constrSet{}(Q)$, then $\Pr[S \wedge \neg (\constrSet{}(Q)-S)]$ is the probability that all variables $X_i$ are $True$ for all $i$ such that $\constr{i} \in S$ and all variables $X_j$ are false for all $j$ such that $\constr{j} \in (\constrSet{}(Q)-S)$.

The task is now to find the joint probability $\Pr[S \wedge \neg (\constrSet{}(Q)-S)]$, for each $S$ where $S \subseteq \constrSet{}(Q)$, such that
$$
\sum_{\{S \mid S \subseteq \constrSet{}(Q)\}} \Pr[S \wedge \neg (\constrSet{}(Q)-S)] = 1
$$
and for each partial estimate $PE$
$$
\sum_{\{S \mid PE.\constrSet{} \subseteq S \subseteq \constrSet{}(Q)\}} \Pr[S \wedge \neg (\constrSet{}(Q)-S)] = PE.s
$$
that maximizes the entropy function:
$$
-\sum_{\{S \mid S \subseteq \constrSet{}(Q)\}} \Pr[S \wedge \neg (\constrSet{}(Q)-S)] \text{ log } \Pr[S \wedge \neg (\constrSet{}(Q)-S)]
$$

The resulting joint probability mass function makes it possible to estimate the selectivity of the query $Q$ and all of its subqueries. The selectivity of a subquery that contains only the constraints $V' \subseteq \constrSet{}(Q)$ can be estimated by
$$
\sum_{\{S \mid V' \subseteq S \subseteq \constrSet{}(Q)\}} \Pr[S \wedge \neg (\constrSet{}(Q)-S)]
$$
which sums all probabilities in the joint probability mass function where all variables $X_i$ are $True$ for all $i$ such that $\constr{i} \in V'$. The remaining variable can be either $True$ or $False$.

The selectivity of query $Q$ is the probability in the estimated joint probability mass function where all variable are true.

Markl et al. \cite{markl2007consistent} solve this constraint optimization problem using an iterative scaling algorithm based on Lagrange multipliers.

\subsection{Obtaining Cardinality Bounds}
\label{sec:card_bounds}
The above methods all construct a single value as an estimate for the cardinality of a query. 
The real cardinality can be very different from this estimate. 
Therefore, this section will introduce a technique to obtain a lower and an upper bound such that the real cardinality is likely to fall within that range, depending on the error guarantees of the PEs in the PES.

\subsubsection{Finding the Upper Bound}
The selectivity of all constraints in $\constrSet{}(Q)$ is less or equal than the selectivity of a subset of $\constrSet{}(Q)$.
The PES of query $Q$ ($PES_Q$) stores estimates for subsets of $\constrSet{}(Q)$. Therefore, an upper bound can be obtained directly from $PES_Q$:
\begin{align}
\label{eq:ub_ks_basic}
Pr[\Sat{\constrSet{}(Q)}] &\leq min_{PE \in PES_Q}PE.s
\end{align}

Now consider a special case: Let $PE_k$ and $PE_l$ be elements from $PES_Q$. 
If $(PE_k.\constrSet{}).\idSet{}$ and $(PE_l.\constrSet{}).\idSet{}$ are disjoint, then the events $\Sat{PE_i.\constrSet{}}$ and $\Sat{PE_j.\constrSet{}}$ are independent.
This makes it possible to obtain the selectivity of the intersection of both events as follows: $\Pr[\Sat{PE_i.\constrSet{}} \cap \Sat{PE_j.\constrSet{}{}}] = \Pr[\Sat{PE_i.\constrSet{}}] \cdot \Pr[\Sat{PE_j.\constrSet{}}] = PE_i.s \cdot PE_j.s$.

This can be generalized to any subset $S$ of $PES_Q$, such that the query ids in the constraints of all pairs of PEs from $S$ are disjoint. Let $indep(PES_Q)$ represent all those sets. This gives
\begin{align}
\label{eq:ub_ks_indep}
Pr[\Sat{\constrSet{}(Q)}] &\leq min_{I \in indep(PES_Q)}\prod_{PE \in I}PE.s
\end{align}

Note that (\ref{eq:ub_ks_indep}) is an improvement upon (\ref{eq:ub_ks_basic}).\footnote{Each set that consists of a single $PE$ from $PES_Q$ is by definition an element of $indep(PES_Q)$. Also $PE.s \leq 1$. Therefore, (\ref{eq:ub_ks_indep}) cannot lead to a larger upper bound than (\ref{eq:ub_ks_basic})}

\subsubsection{Finding the Lower Bound}
For most PESs, there is not enough evidence that a query return at least one result. This means that a lower bound of zero for $\Pr[\Sat{\constrSet{}(Q)}]$ is common.
Only when the selectivity values of the PEs are large, it might be possible to obtain a lower bound larger than $0$. 
\begin{example}
For example, $PES_Q=\{PE_1.s=0.8, PE_2.s=0.7, PE_3.s=0.9\}$ gives a lower bound $0.4$ for the selectivity $\Pr[\Sat{\PK{}_1.\constrSet{} \cup \PK{}_2.\constrSet{} \cup \PK{}_3.\constrSet{}}]$. 
%
%
%
\end{example}

By making the sets of mappings that satisfy the different PEs as disjoint as possible, it is possible to obtain the smallest possible set of mappings that must satisfy all partial estimates. This can be computed using the formula $lbSel(PES_Q) = 1 - \sum_{PE \in PES_Q}(1-PE.s)$, where $(1-PE.s)$ represent the fraction of mappings that do not satisfy the constraints in $PE.\constrSet{}$.
%
The final lower bound is $max(0, lbSel(PES_Q))$.
A lower bound larger than $0$ will only be obtained when $\sum_{PE \in PES_Q}PE.s > |PES_Q|-1$, where $|PES_Q|$ is the number of PEs in the PES.

Notice that, in order to obtain the best lower bound, all PEs $PE_i$ need to be removed for which there exists another PE $PE_j$ where $PE_i.\constrSet{} \subseteq PE_j.\constrSet{}$.

\section{Experimental Setup}
\label{sec:exp_Setup}

The Join Order Benchmark (JOB) was introduced by Leis et al. \cite{leis2015good, leis2018query}. It consists of a fixed database instance and a fixed query workload. The database instance is a real instance of the \textit{Internet Movie Data Base} (IMDB), which is full of correlations and non-uniform distributions. The query workload consists of 113 analytical SQL queries that were manually constructed to represent questions that could have been asked by a movie enthusiast. The queries contain 33 structures, each with several variants that differ in selections (i.e. data constraints) only.

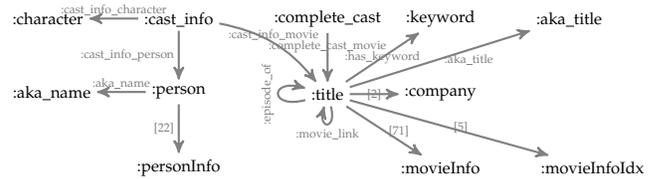
\begin{figure}
	\centering
	
\begin{tikzpicture}[thick, scale=0.65, every node/.style={transform shape}]
	
	\node at (0,0) [](t){:title};
	\node [right=of t](c){:company};
	\node [above=of c](kw){:keyword};
	\node [right=of kw](akat){:aka\_title};
	\node [below=of c](mi){:movieInfo};
	\node [right=of mi](miIdx){:movieInfoIdx};
	
	\node [above=of t](cc){:complete\_cast};
	
	\node [left=of cc](ci){:cast\_info};
	\node [left=of ci](char){:character};
	\node [below=of ci](p){:person};
	\node [below=of p](pi){:personInfo};
	\node [left=of p](akan){:aka\_name};
	
	\draw [propEdge] (t) to node {[2]} (c);
	\draw [propEdge] (t) to node {:has\_keyword} (kw);
	\draw [propEdge] (t) to node [anchor=west] {:aka\_title} (akat);
	\draw [propEdge] (t) to node [anchor=west] {[71]} (mi);
	\draw [propEdge] (t) to node [anchor=west] {[5]} (miIdx);

	\draw [propEdge] (cc) to node [anchor=south] {:complete\_cast\_movie} (t);
	
	\draw [propEdge] (ci) to [bend left=20] node [anchor=south] {:cast\_info\_movie} (t);
	\draw [propEdge] (ci) to node [anchor=south] {:cast\_info\_character} (char);
	\draw [propEdge] (ci) to node [anchor=east] {:cast\_info\_person} (p);
	
	\draw [propEdge] (p) to node [anchor=east] {[22]} (pi);
	\draw [propEdge] (p) to node [anchor=south] {:aka\_name} (akan);
	
	\draw [propEdge] (t) to [loop left] node [anchor=south, rotate=90] {:episode\_of} (t);
	
	\draw [propEdge] (t) to [loop below] node {:movie\_link} (t);
\end{tikzpicture}
\caption[caption]{Graph schema for the data from the Join Order Benchmark. The edges with label $[x]$, e.g. from \textit{:person} to \textit{:personInfo}, can have one of $x$ different labels, e.g. \textit{birth date}, \textit{death date}, \textit{height}, \textit{mini biography}, etc.}
\label{fig:JOB_graph_schema}
\end{figure}

We have translated the relational database instance into a property graph database instance and each SQL query in the workload into an openCypher query. 
See Figure \ref{fig:JOB_graph_schema} for the graph schema.
Table \ref{tab:abbreviations} shows the abbreviations that will be used in the experiment results and descriptions.\footnote{A sorting strategy will be abbreviated by a combination like \sortStratShort{SaNd}, where \sortStratShort{Sa} defines the primary sorting criteria and \sortStratShort{Nd} the secondary sorting criteria.}

\subsection{Translation from relational to graph instance}
\label{sec:JOB_translate_to_graph_instance}
For the IMDB schema, we refer to Figure 2 in \cite{leis2018query}.
We partitioned the relation into three sets $I, E$ and $V$, i.e. \textbf{I}dentifying relations, \textbf{E}dge relations and \textbf{V}ertex relations.
A relation belongs to $I$ if it is only used to identify a string value, i.e. it only has two attributes, the primary key and the string value.
A relation belongs to $E$ if it contains exactly two foreign key attributes to relations not in $I$ and it does not have any references from other relations to its primary key:
\textit{movie\_companies} to \textit{title} and \textit{company},
\textit{movie\_link} twice to \textit{title}
and  \textit{movie\_keyword to \textit{title} and \textit{keyword}}.

The remaining relations belong to $V$.

\textit{I = \{link\_type, role\_type, complete\_cast\_type, company\_type, kind\_type, info\_type\}}

\textit{E = \{movie\_companies, movie\_link, movie\_keyword\}}

\textit{V = \{title, complete\_cast, keyword, aka\_title, company\_name, movie\_info, movie\_info\_idx, cast\_info, char\_name, name, aka\_name, person\_info\}}

We created a vertex for each tuple in the relations in $V$. We assigned the name of the relation as a label to the vertex. If a relation in $V$ contains a foreign key (e.g. \textit{movie\_id} in \textit{aka\_title}), then an edge will be created between the vertices associated with the tuple in the relation and the tuple to which it refers. 

For each tuple in the relations in $E$, we created an edge between the vertices associated with the tuples to which both foreign keys refer. If the relation contains only one attribute that is not a foreign key or primary key, then the value of that attribute is used as edge label (e.g. \textit{company\_type} for \textit{movie\_companies}). Otherwise, a label based on the name of the relation is assigned, e.g. \textit{has\_keyword} for \textit{movie\_keyword}.

The remaining attributes are added as  key-value properties to the vertices and edges, where the key is the attribute name.

\subsection{Basic Query and Data Statistics}
\label{exp:subpattern_overview}
Several statistics from Section \ref{sec:est_mc} are tailored to specific patterns. If those patterns do not occur frequently in the data and the queries, then those statistics lose their benefit. Table \ref{tab:JOB_patterns} shows the number of occurrences of several patterns in the JOB query workload. There are much more source star patterns than chain patterns and much more chain patterns than target star pattern. Therefore, statistics for source star patterns have the potential to improve the estimation accuracy for many queries.
Table \ref{tab:JOB_data_stats} shows some basic statistics of the graph version of the JOB dataset.\\

\noindent
\begin{minipage}[t]{0.64\columnwidth}
\begin{table}[H]
\centering
\caption{Types of patterns in the OpenCypher version of the JOB queries and the number of times they occur.}
\label{tab:JOB_patterns}
{\scriptsize
\begin{tabular}{| l | l |}
\hline 
\textbf{Pattern} & \textbf{Occ.} \\ \hline \hline
Query edges & 464\\ \hline
Chains length $2$ & 201\\ \hline
Chains length $>2$ & 0\\ \hline
Source stars size $2$ & 356\\ \hline
Source stars size $3$ & 130\\ \hline
Source stars size $4$ & 20\\ \hline
Source stars size $>4$ & 0\\ \hline
Target stars size $2$ & 15\\ \hline
Target stars size $>2$ & 0\\ \hline
\end{tabular}
}
\end{table}
\end{minipage}
\begin{minipage}[t]{0.35\columnwidth}
\begin{table}[H]
\centering
\caption{Basic statistics of the graph version of the JOB dataset.}
\label{tab:JOB_data_stats}
{\scriptsize
\begin{tabular}{| l | r |}
\hline
\multicolumn{2}{|c|}{\textbf{JOB data stats}}\\ \hline\hline
$|G.\edges{}|$ & $119343754$\\ \hline 
$|G.\vertices{}|$ & $52639796$\\ \hline
$|G.\idSet{}|$ & $171983550$\\ \hline
$|G.\ell_v|$ & $12$\\ \hline 
$|G.\ell_e|$ & $106$\\ \hline 
\end{tabular}
}
\end{table}
\end{minipage}

\vspace{1mm}

\begin{table}
\centering
{ \scriptsize
\caption{Abbreviations used in the experiments.}
\label{tab:abbreviations}
\begin{tabular}{| l | l |}
\hline
\multicolumn{2}{|c|}{\textbf{PETs for multiple constraints}}\\ \hline
\statShort{EP} & Labeled topological synopsis for edge patterns\\ \hline
\statShort{SysR} & \SystemR{}'s join size estimation\\ \hline
\statShort{cX} & \makecell{Labeled topological synopses\\
for chain patterns up to size $X$}\\ \hline
\statShort{sX} & \makecell{Labeled topological synopses\\
for source star patterns up to size $X$}\\ \hline
\statShort{tX} & \makecell{Labeled topological synopses\\
for target star patterns up to size $X$}\\ \hline
\statShort{CS} & Characteristic Sets\\ \hline
\statShort{S(pa,pr)} &Sampling with pattern \statShort{pa} and probability \statShort{pr}\\ \hline
\multicolumn{2}{|c|}{\textbf{EPESTs}}\\ \hline
\statShort{IP(p,c)} & \makecell{Implied constraints assumptions\\
with pattern \statShort{p} and constraint type \statShort{c}}\\ \hline
\statShort{id} & Pattern: Query identifier\\ \hline
\statShort{ep} & Pattern: Edge pattern\\ \hline
\statShort{p} & Constraint type: Property constraints\\ \hline
\statShort{pv} & Constraint type: Property-value constraints\\ \hline
\statShort{a} & Constraint type: All \queryPattern{} constraints\\ \hline
\hline
\multicolumn{2}{|c|}{\textbf{Combination Techniques (CTs)}}\\ \hline
\ctShort{condIndep} & Conditional independence assumptions\\ \hline
\ctShort{maxEnt} & Maximum Entropy\\ \hline
\ctShort{bounds} & Upper and lower bounds\\ \hline
\hline
\multicolumn{2}{|c|}{\textbf{Sorting Strategies}}\\ \hline
\sortStratShort{a} & ascending\\ \hline
\sortStratShort{d} & descending\\ \hline
\sortStratShort{S} & Selectivity value\\ \hline
\sortStratShort{N} & Number of conjuncts\\ \hline
\sortStratShort{Di} & Deviation from Independence assumption\\ \hline
\sortStratShort{Mo} & Maximum Overlap\\ \hline
\end{tabular}
}
\end{table}

\subsection{Notes on Disjunctions in JOB Queries}
Some queries in the JOB contain disjunctions, e.g.  [\hasLabelConstr{e1}{'production companies'} OR \hasLabelConstr{e1}{'distributors'}]. The way we handled them is by generating a set of queries without disjunctions and finally unifying their results, by assuming that the sets of results are disjoint.

\section{Example of the Estimation Process}
\label{sec:exampleEstProcess}

\begin{figure*}
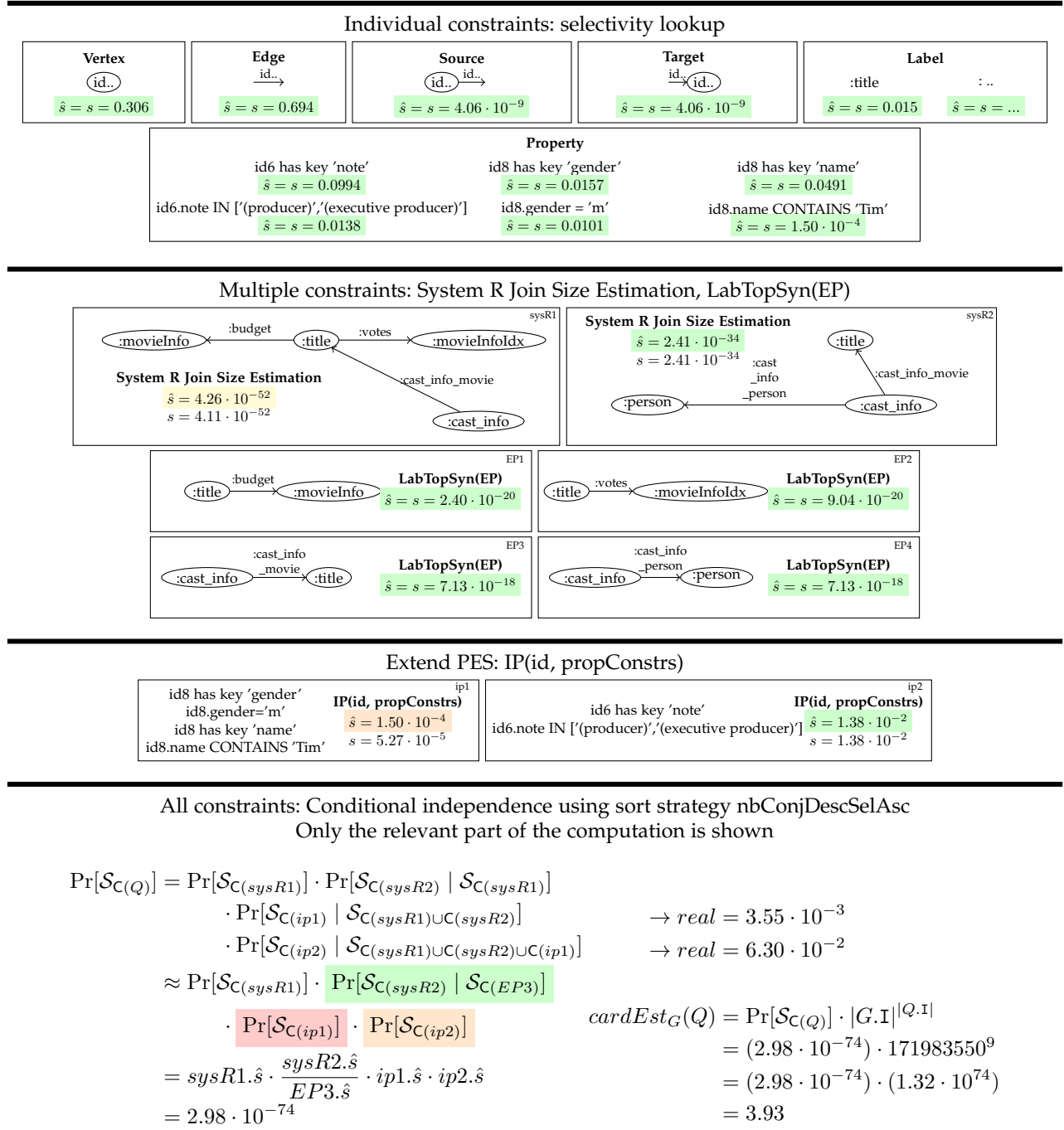

\centering
\tikz[scale=1.3]{	
	\draw (0,0) [fill] rectangle (13, 0.05);
}
Individual constraints: selectivity lookup\\
\vspace{2pt}
\tikz[scale=1.3]{	
	
	\draw (-0.5,-0.8) rectangle (1.5,0.2);
	
	\node[scale=0.7] at (0.5, 0) [align=center] (vertex){\textbf{Vertex}};
	
	
	\node[scale=0.7] at (0.5, -0.3) [draw, ellipse, inner sep=1pt, align=center] (id){id..};
	\node[scale=0.7] at (0.5, -0.6) [align=center] (s){\colorbox{green!20}{$\hat{s}=s = 0.306$}};
}
\tikz[scale=1.3]{	
	
	\draw (-0.2,-0.8) rectangle (1.7,0.2);
	
	\node[scale=0.7] at (0.75, 0) [align=center] (edge){\textbf{Edge}};
	
	
	\node[scale=0.7] at (0.5, -0.3) (src){};
	\node[scale=0.7] at (1, -0.3) (trg){};
	\draw[->] (src) to node [color=black, align=center, anchor=south, scale=0.6] {id..} (trg);
	\node[scale=0.7] at (0.75, -0.6) [align=center] (s){\colorbox{green!20}{$\hat{s}=s = 0.694$}};
}
\tikz[scale=1.3]{	
	
	\draw (-0.4,-0.8) rectangle (2.3,0.2);
	
	\node[scale=0.7] at (0.95, 0) [align=center] (src){\textbf{Source}};
	
	
	\node[scale=0.7] at (0.7, -0.3) [draw, ellipse, inner sep=1pt] (src){id..};
	\node[scale=0.7] at (1.3, -0.3) (trg){};
	\draw[->] (src) to node [color=black, align=center, anchor=south, scale=0.6] {id..} (trg);
	\node[scale=0.7] at (0.95, -0.6) [align=center] (s){\colorbox{green!20}{$\hat{s}=s = 4.06 \cdot 10^{-9}$}};
}
\tikz[scale=1.3]{	
	
	\draw (-0.4,-0.8) rectangle (2.3,0.2);
	
	\node[scale=0.7] at (0.9, 0) [align=center] (trg){\textbf{Target}};
	
	
	\node[scale=0.7] at (0.65, -0.3) (src){};
	\node[scale=0.7] at (1.15, -0.3) [draw, ellipse, inner sep=1pt, inner sep=1pt] (trg){id..};
	\draw[->] (src) to node [color=black, align=center, anchor=south, scale=0.6] {id..} (trg);
	\node[scale=0.7] at (0.9, -0.6) [align=center] (s){\colorbox{green!20}{$\hat{s}=s = 4.06 \cdot 10^{-9}$}};
}
\tikz[scale=1.3]{	
	
	\draw (-0.5,-0.8) rectangle (2.5,0.2);
	
	\node[scale=0.7] at (1, 0) [align=center] (label){\textbf{Label}};
	
	\node[scale=0.7] at (0.25, -0.3) (title){:title};
	\node[scale=0.7] at (0.35, -0.6) [align=center] (s){\colorbox{green!20}{$\hat{s}=s = 0.015$}};
	
	\node[scale=0.7] at (1.75, -0.3) (other){: ..};
	\node[scale=0.7] at (1.75, -0.6) [align=center] (s){\colorbox{green!20}{$\hat{s}=s = ...$}};
}

\vspace{2pt}
\tikz[scale=1.3]{	
	
	\draw (-2,-1.2) rectangle (7.5,0.2);
	
	\node[scale=0.7] at (3, 0) [align=center] (prop){\textbf{Property}};
	
	\node[scale=0.7] at (0, -0.3) (keyNote){id6 has key 'note'};
	\node[scale=0.7] at (0, -0.5) [align=center] (sKeyNote){\colorbox{green!20}{$\hat{s}=s = 0.0994$}};
	\node[scale=0.7] at (0, -0.8) [align=center](note){id6.note IN ['(producer)','(executive producer)']};
	\node[scale=0.7] at (0, -1) (sNote){\colorbox{green!20}{$\hat{s}=s = 0.0138$}};
	
	\node[scale=0.7] at (3, -0.3) (keyGender){id8 has key 'gender'};
	\node[scale=0.7] at (3, -0.5) [align=center] (sKeyGender){\colorbox{green!20}{$\hat{s}=s = 0.0157$}};
	\node[scale=0.7] at (3, -0.8) [align=center](gender){id8.gender = 'm'};
	\node[scale=0.7] at (3, -1) (sGender){\colorbox{green!20}{$\hat{s}=s = 0.0101$}};
	
	\node[scale=0.7] at (6, -0.3) (keyName){id8 has key 'name'};
	\node[scale=0.7] at (6, -0.5) [align=center] (sKeyName){\colorbox{green!20}{$\hat{s}=s = 0.0491$}};
	\node[scale=0.7] at (6, -0.8) [align=center](name){id8.name CONTAINS 'Tim'};
	\node[scale=0.7] at (6, -1) (sName){\colorbox{green!20}{$\hat{s}=s = 1.50 \cdot 10^{-4}$}};
	
}

\vspace{2pt}
\tikz[scale=1.3]{	
	\draw (0,0) [fill] rectangle (13, 0.05);
}

Multiple constraints: System R Join Size Estimation, LabTopSyn(EP)\\
\vspace{2pt}
\tikz[scale=1.3]{	
	
	\draw (-1, -0.3) rectangle (5, 1.4);
	
	\node[scale=0.5] at (4.8, 1.3) (id){sysR1};
	
	\node[scale=0.7] at (0, 1) [draw, ellipse, inner sep=1pt, align=center] (mi){:movieInfo};
	\node[scale=0.7] at (2, 1) [draw, ellipse, inner sep=1pt] (t){:title};
	\node[scale=0.7] at (4, 1) [draw, ellipse, inner sep=1pt, align=center] (miIdx){:movieInfoIdx};
	\node[scale=0.7] at (4, 0) [draw, ellipse, inner sep=1pt] (ci){:cast\_info};
	\node[scale=0.7] at (0.8,0.3) [align=center] (estText) {\textbf{System R Join Size Estimation}\\ \colorbox{yellow!20}{$\hat{s} = 4.26 \cdot 10^{-52}$}\\$s = 4.11 \cdot 10^{-52}$};

	\draw[->] (ci) to node [color=black, align=center, anchor=west, scale=0.6] {:cast\_info\_movie} (t);
	\draw[->] (t) to node [color=black, align=center, anchor=south, scale=0.6] {:budget} (mi);
	\draw[->] (t) to node [color=black, align=center, anchor=south, scale=0.6] {:votes} (miIdx);
}
\tikz[scale=1.3]{	
	\draw (-1,-0.5) rectangle (4.3,1.2);
	
	\node[scale=0.5] at (4.1, 1.1) (id){sysR2};
	
	\node[scale=0.7] at (0, 0) [draw, ellipse, inner sep=1pt] (p){:person};
	\node[scale=0.7] at (2.5, 0.8) [draw, ellipse, inner sep=1pt] (t){:title};
	\node[scale=0.7] at (3, 0) [draw, ellipse, inner sep=1pt] (ci){:cast\_info};	
	\node[scale=0.7] at (0.5,0.8) [align=center] (estText) {\textbf{System R Join Size Estimation}\\ \colorbox{green!20}{$\hat{s} = 2.41 \cdot 10^{-34}$}\\ $s = 2.41 \cdot 10^{-34}$};

	\draw[->] (ci) to node [color=black, align=center, anchor=south, scale=0.6] {:cast\\ \_info\\ \_person} (p);
	\draw[->] (ci) to node [color=black, align=center, anchor=west, scale=0.6] {:cast\_info\_movie} (t);
}

\vspace{2pt}
\tikz[scale=1.3]{
	\draw (-0.7,-0.5) rectangle (4,0.5);
	\node[scale=0.5] at (3.8, 0.4) (id){EP1};
	\node[scale=0.7] at (0,0) [draw, ellipse, inner sep=1pt](src)  {:title};
	\node[scale=0.7] at (1.5,0) [draw, ellipse, inner sep=1pt](trg)  {:movieInfo};
	\draw [->](src) to node [color=black, align=center, anchor=south, scale=0.6] {:budget} (trg);
	
	\node[scale=0.7] at (3,0) [align=center] (estText) {\textbf{LabTopSyn(EP)}\\ \colorbox{green!20}{$\hat{s}=s = 2.40 \cdot 10^{-20}$}};
}
\tikz[scale=1.3]{
	\draw (-0.7,-0.5) rectangle (4,0.5);
	\node[scale=0.5] at (3.8, 0.4) (id){EP2};
	\node[scale=0.7] at (-0.35,0) [draw, ellipse, inner sep=1pt](src)  {:title};
	\node[scale=0.7] at (1.3,0) [draw, ellipse, inner sep=1pt](trg)  {:movieInfoIdx};
	\draw[->](src) to node [color=black, align=center, anchor=south, scale=0.6] {:votes} (trg);
	
	\node[scale=0.7] at (3,0) [align=center] (estText) {\textbf{LabTopSyn(EP)}\\ \colorbox{green!20}{$\hat{s}=s = 9.04 \cdot 10^{-20}$}};
}

\vspace{2pt}
\tikz[scale=1.3]{
	\draw (-0.7,-0.5) rectangle (4,0.5);
	\node[scale=0.5] at (3.8, 0.4) (id){EP3};
	\node[scale=0.7] at (0,0) [draw, ellipse, inner sep=1pt](src)  {:cast\_info};
	\node[scale=0.7] at (1.5,0) [draw, ellipse, inner sep=1pt](trg)  {:title};
	\draw[->](src) to node [color=black, align=center, anchor=south, scale=0.6] {:cast\_info\\ \_movie} (trg);
	
	\node[scale=0.7] at (3,0) [align=center] (estText) {\textbf{LabTopSyn(EP)}\\ \colorbox{green!20}{$\hat{s}=s = 7.13 \cdot 10^{-18}$}};
}
\tikz[scale=1.3]{
	\draw (-0.7,-0.5) rectangle (4,0.5);
	\node[scale=0.5] at (3.8, 0.4) (id){EP4};
	\node[scale=0.7] at (0,0) [draw, ellipse, inner sep=1pt](src)  {:cast\_info};
	\node[scale=0.7] at (1.5,0) [draw, ellipse, inner sep=1pt](trg)  {:person};
	\draw [->](src) to node [color=black, align=center, anchor=south, scale=0.6] {:cast\_info\\ \_person} (trg);
	
	\node[scale=0.7] at (3,0) [align=center] (estText) {\textbf{LabTopSyn(EP)}\\ \colorbox{green!20}{$\hat{s}=s = 7.13 \cdot 10^{-18}$}};
}

\tikz[scale=1.3]{	
	\draw (0,0) [fill] rectangle (13, 0.05);
}

Extend PES: IP(id, propConstrs)

\vspace{2pt}
\tikz[scale=1.3]{
	\draw (-1.2,-0.5) rectangle (3,0.5);
	\node[scale=0.5] at (2.8, 0.4) (id){ip1};
	\node[scale=0.7] at (0,0) [align=center](constrs)  {id8 has key 'gender'\\ id8.gender='m'\\id8 has key 'name'\\ id8.name CONTAINS 'Tim'};

	\node[scale=0.7] at (2,0) [align=center] (estText) {\textbf{IP(id, propConstrs)}\\ \colorbox{orange!20}{$\hat{s} = 1.50 \cdot 10^{-4}$}\\$s = 5.27 \cdot 10^{-5}$};
}
\tikz[scale=1.3]{
	\draw (-2,-0.5) rectangle (3.5,0.5);
	\node[scale=0.5] at (3.3, 0.4) (id){ip2};
	\node[scale=0.7] at (0,0) [align=center](constrs)  {id6 has key 'note'\\ id6.note IN ['(producer)','(executive producer)']};
	
	\node[scale=0.7] at (2.6,0) [align=center] (estText) {\textbf{IP(id, propConstrs)}\\ \colorbox{green!20}{$\hat{s} = 1.38 \cdot 10^{-2}$}\\$s = 1.38 \cdot 10^{-2}$};
}

\tikz[scale=1.3]{	
	\draw (0,0) [fill] rectangle (13, 0.05);
}

All constraints: Conditional independence using sort strategy nbConjDescSelAsc\\
Only the relevant part of the computation is shown

\vspace{2pt}
\begin{minipage}[c]{0.65\columnwidth}
\begin{align*}
\Pr[\Sat{\constrSet{}(Q)}] &= \Pr[\Sat{\constrSet{}(sysR1)}] \cdot \Pr[\Sat{\constrSet{}(sysR2)} \mid \Sat{\constrSet{}(sysR1)}] \\
&\hspace{1cm}
\cdot \Pr[\Sat{\constrSet{}(ip1)} \mid \Sat{\constrSet{}(sysR1) \cup \constrSet{}(sysR2)}]\\ 
&\hspace{1cm}
\cdot \Pr[\Sat{\constrSet{}(ip2)}  \mid \Sat{\constrSet{}(sysR1) \cup \constrSet{}(sysR2) \cup \constrSet{}(ip1)}]\\
&\approx \Pr[\Sat{\constrSet{}(sysR1)}] \cdot \colorbox{green!20}{$\Pr[\Sat{\constrSet{}(sysR2)} \mid \Sat{\constrSet{}(EP3)}]$} \\
&\hspace{1cm}
 \cdot \colorbox{red!20}{$\Pr[\Sat{\constrSet{}(ip1)}]$} \cdot \colorbox{orange!20}{$\Pr[\Sat{\constrSet{}(ip2)}]$}\\
&= sysR1.\hat{s} \cdot \frac{sysR2.\hat{s}}{EP3.\hat{s}} \cdot ip1.\hat{s} \cdot ip2.\hat{s}\\
&= 2.98 \cdot 10^{-74}
\end{align*}
\end{minipage}
\begin{minipage}[c]{0.34\columnwidth}
\begin{align*}
&\\
\rightarrow real &= 3.55 \cdot 10^{-3}\\
\rightarrow real &= 6.30 \cdot 10^{-2}\\
&\\
cardEst_G(Q) &= \Pr[\Sat{\constrSet{}(Q)}] \cdot |G.\idSet{}|^{|Q.\idSet{}|}\\
&= (2.98 \cdot 10^{-74}) \cdot 171983550^{9}\\
&= (2.98 \cdot 10^{-74}) \cdot (1.32 \cdot 10^{74})\\
&= 3.93
\end{align*}
\end{minipage}

\tikz[scale=1.3]{	
	\draw (0,0) [fill] rectangle (13, 0.05);
}

	\caption[caption]{Example of an estimation process of the query from Figure \ref{fig:queryJOB18a}.}
	\label{fig:bigExample}
\end{figure*}

Figure \ref{fig:bigExample} shows an example of the estimation process for the query from Figure \ref{fig:queryJOB18a} on the property graph version of the JOB dataset. 

The PET for the individual constraints that is used is synopsis lookup, where the synopsis contains the precomputed selectivities of the required individual constraints.
The PETs for multiple constraints that are used are: \statShort{SysR} and \statShort{EP}.
The EPEST \statShort{IP[id,p]} is used, which assumes implied constraints between property constraint referring to the same query id.

Using those PETs, EPEST and the combination technique \ctShort{condIndep}\sortStratShort{(NdSa)}, the cardinality of the query from Figure \ref{fig:queryJOB18a} is estimated to be almost $4$ in the JOB dataset. 
The real cardinality of this query in the JOB dataset is $410$.

Each PE contains a selectivity estimate $\hat{s}$. 
For analysis, we also added the exact selectivity values $s$ for each pattern for which a PE is available.
All estimates in green do not introduce any error.
Estimates in yellow introduce a very small errors (q-errors less than $2$).
Estimates in orange introduce some errors (q-errors between $2$ and $10$).
Finally, estimates in red introduce large errors (q-errors larger than $10$).

The errors introduced during the estimation process will be investigated in the following subsections.

\vspace{1mm}
\noindent \textbf{System R's assumptions.}
The inclusion assumption and uniform distribution assumption caused a very small overestimation error in the PE sysR1. 
The PE sysR2 does not introduce any error, since each \textit{cast\_info} vertex (i.e. vertex with the label \textit{cast\_info}) has exactly one outgoing edge with label \textit{cast\_info\_movie} to a \textit{title} vertex and exactly one outgoing edge with label \textit{cast\_info\_person} to a \textit{person} vertex. 
Therefore, inclusion and uniform distribution assumptions hold.

\vspace{1mm}
\noindent \textbf{Implied constraints assumptions.}
The implied constraints assumption (an element whose name contains 'Tim' determines that the element has gender 'm') in ip1 caused a small overestimation error.

\vspace{1mm}
\noindent \textbf{Independence between labeled topological patterns.}
$\Pr[\Sat{\constrSet{}(sysR2)} \mid \Sat{\constrSet{}(sysR1)}]$ is approximated by $\Pr[\Sat{\constrSet{}(sysR2)} \mid \Sat{\constrSet{}(EP3)}]$.
This assumes conditional independence between the pattern $P_{bv}:$\srcStarTwoPattern{:movieInfo}{:budget}{:title}{:votes}{:movieIndoIdx}{2}{4} and $P_{p}:$\edgePattern{:cast\_info}{:cast\_info\\ \_person}{:person}{2} given $P_{m}:$\edgePattern{:cast\_info}{:cast\_info\\ \_movie}{:title}{2}. This did not introduce an estimation error. The reason for this is that every \textit{cast\_info} vertex in the dataset has exactly one outgoing edge with label \textit{cast\_info\_person} to a \textit{person} vertex and exactly one outgoing edge with label \textit{cast\_info\_movie} to a \textit{title} vertex. Therefore $P_p$ only depends on $P_m$ and not on $P_{bv}$.

\vspace{1mm}
\noindent \textbf{Independence between property constraints and labeled topological pattern.}
$\Pr[\Sat{\constrSet{}(ip1)} \mid \Sat{\constrSet{}(sysR1) \cup \constrSet{}(sysR2)}]$ (which has real $s=3.55 \cdot 10^{-3}$) is approximated by $\Pr[\Sat{\constrSet{}(ip1)}]$ (which has real $s=5.27 \cdot 10^{-5}$). The leads to a large q-error of $67.4$.
Notice that $\Pr[\Sat{\constrSet{}(ip1)}]$ itself is approximated using implied constraints assumptions. This approximation causes a small overestimation, which cancels out a part of the underestimation for using $\Pr[\Sat{\constrSet{}(ip1)}]$ as an approximation for $\Pr[\Sat{\constrSet{}(ip1)} \mid \Sat{\constrSet{}(sysR1) \cup \constrSet{}(sysR2)}]$. This eventually leads to a q-error of $23.7$.

Approximating 
$\Pr[\Sat{\constrSet{}(ip1)} \mid \Sat{\constrSet{}(sysR1) \cup \constrSet{}(sysR2)}]$ 
by
$\Pr[\Sat{\constrSet{}(ip1)} \mid \Sat{\{\text{\vertexConstr{id8}}\}}]$ (which has real $s=1.72 \cdot 10^{-4}$) would improve estimation accuracy (q-error of $20.6$).
However, approximating it by
$\Pr[\Sat{\constrSet{}(ip1)} \mid \Sat{\{\text{\vertexConstr{id8}, \hasLabelConstr{id8}{person}}\}}]$ (which has real $s=2.17 \cdot 10^{-3}$) would lead to a very accurate estimate (q-error of $1.6$).
This shows a strong positive correlation between property and label constraints.

\vspace{1mm}
\noindent \textbf{Independence between property constraints with labeled topological pattern and other property constraints.}
$\Pr[\Sat{\constrSet{}(ip2)}  \mid \Sat{\constrSet{}(sysR1) \cup \constrSet{}(sysR2) \cup \constrSet{}(ip1)}]$ (which has real $s=6.30 \cdot 10^{-2}$)
is approximated by
$\Pr[\Sat{\constrSet{}(ip2)}]$ (which has real $s=1.38 \cdot 10^{-2}$).
The leads to a q-error of $4.6$.

Approximating it using $\Pr[\Sat{\constrSet{}(ip2)} \mid \Sat{\{\text{\vertexConstr{id6}}\}}]$ (which has real $s=4.52 \cdot 10^{-2}$) improves estimation accuracy (q-error of $1.4$).
%
Approximating it using $\Pr[\Sat{\constrSet{}(ip2)} \mid \Sat{\{\text{\vertexConstr{id6}, \hasLabelConstr{id6}{cast\_info}}\}}]$ (which has real $s=6.56 \cdot 10^{-2}$) would lead to an almost perfect estimate (q-error of $1.0$).
This, again, shows a strong positive correlation between property and label constraints.

\vspace{1mm}
\noindent \textbf{Summary.}
The illustrated estimation process in Figure \ref{fig:bigExample} introduces the largest errors by assuming independence between each property constraint and the labeled topological constraints. Therefore, the selectivity estimate can be improved by adding partial estimates that contain combinations of property constraints, edge or vertex constraints and label constraints. 

\section{Experiments}
\label{sec:exp}
Section \ref{sec:exampleEstProcess} identified different source of errors that can be introduced within the estimation process. 
In the following sections, these error sources are investigated further. 
All experiments are performed on the graph version of the JOB dataset and using the OpenCypher version of the JOB query workload. 

For all experiments, a synopsis lookup for individual constraints is used. This allows us to focus on the estimation errors introduced by PETs for multiple constraints, EPESTs and CTs. 
In a real world scenario, individual constraints might be estimated using histograms or sampling.

Several experimental results are visualized using boxplots. The boxplot is drawn, such that 50\% of the values fall within the box, $90\%$ within the whiskers. The remaining $10\%$ is shown as dots (top $5\%$ and bottom $5\%$). The median is drawn as an orange line within the box.

Section \ref{sec:experiments_PET_error} shows experiments regarding the errors introduced by some (extend) partial estimation techniques.
Experiments related to the errors that are introduced by the combination techniques are shown in Section \ref{sec:experiments_CT_error}.

\vspace{1mm}
\noindent \textbf{Key observations from the experiments:}
\begin{itemize}
\item for labeled star patterns, a combination of a labeled edge pattern synopsis (PET \statShort{EP}) and conditional independence assumptions (CT \ctShort{condIndep}) requires less statistics than System~R's join size estimation (PET \statShort{SysR}) and its largest Q-error is $7\times$ lower. The median Q-errors are similar. (Section \ref{exp:sysR_accuracy});
\item for labeled source star patterns with property-key constraints, the PET \statShort{CS} has better estimation accuracy than structural synopses combined with \ctShort{condIndep}. Its median Q-error is $60\times$ lower. (Section \ref{sec:CS_exp});
\item using implication assumptions instead of independence assumptions leads to a median Q-error that is $20\times$ lower for estimating multiple property-value constraints referring to the same query id. (Section \ref{sec:IP_exp})\\
Section \ref{sec:IP_exp_withPropConsts} also supports this claim for general \queryPatterns{}, independence assumptions can lead to Q-errors that are $10^{24}\times$ larger than using implication assumptions on the JOB subqueries with less than 7 edges; 
\item sorting strategies (for CT  \ctShort{condIndep}) with primary sort \sortStratShort{Sd} or \sortStratShort{Na} should be avoided (Section \ref{sec:condIndep_topOnly});
\item \ctShort{condIndep} outperforms \ctShort{maxEnt} and \ctShort{bounds} (Section \ref{sec:condIndep_topOnly});
\item in the absence of property constraints, labeled structural synopses with CT \ctShort{condIndep}\sortStratShort{(MoDi)} can obtain accurate estimates efficiently Section \ref{sec:condIndep_topOnly}); 
\item in the presence of property constraints, labeled structural synopses for patterns that are larger than edge patterns do not improve estimation accuracy when correlations between property and labeled topological constraints are not captured (Section \ref{sec:labTopSyn_withPropConsts});
\item sampling improves estimation accuracy (median Q-error is $16\times$ lower), but increases estimation time (median estimation time is $135\times$ higher), compared to the best alternative using implication assumptions, on the JOB subqueries with less than 7 edges (Section \ref{sec:sampling_exp_withPropConsts}).
\end{itemize}

\subsection{Estimation Error Experiments for PETs and EPESTs}
\label{sec:experiments_PET_error}

Partial estimation techniques like \statShort{c2} can obtain partial estimates that are exact. However, PETs like \statShort{CS} or \statShort{SysR} can obtain partial estimates that are not exact. Those techniques rely on a combination of specialized statistics and assumptions. The assumptions can introduce errors in the partial estimates. This section investigates on the estimation errors introduced by PETs and EPESTs.

\subsubsection{System R Join Size Estimation}
\label{exp:sysR_accuracy}
The PET \statShort{SysR} (Section \ref{sec:tb_join_size_est}) can obtain estimates for labeled star patterns, using labeled edge pattern statistics, inclusion assumptions and uniform distribution assumptions.
This section focuses on the accuracy of estimating labeled star patterns, using \statShort{SysR} and alternative approaches.

\vspace{1mm}
\noindent \textbf{Theoretical Analysis}
The full estimation error introduced by \statShort{SysR} on star patterns is due to the inclusion assumptions and uniform distribution assumptions. 
The inclusion assumption can \emph{only} lead to an overestimate. 
The uniform distribution assumption can lead to both over- and underestimations.

The PET \statShort{SysR} underestimates in the case that 1) the degree distributions are skewed, 2) vertices with a relative high number of edges in one edge pattern also have a relatively high number of edges in other edge patterns and 3) the inclusion assumption is approximately satisfied.

The PET \statShort{SysR} overestimates in the case that in inclusion assumption does not hold (for example when $|T'_G(ep1) \cap S'_G(ep2)|$ is much smaller than $min(|T'_G(ep1), S'_G(ep2)|)$ in a pattern $ep1/ep2$), or when 1) the degree distributions are skewed and 2) vertices with a relative high number of edges in one edge pattern have a relatively low number of edges in other edge patterns.

\begin{figure}
\centering
\includegraphics[width=\linewidth]{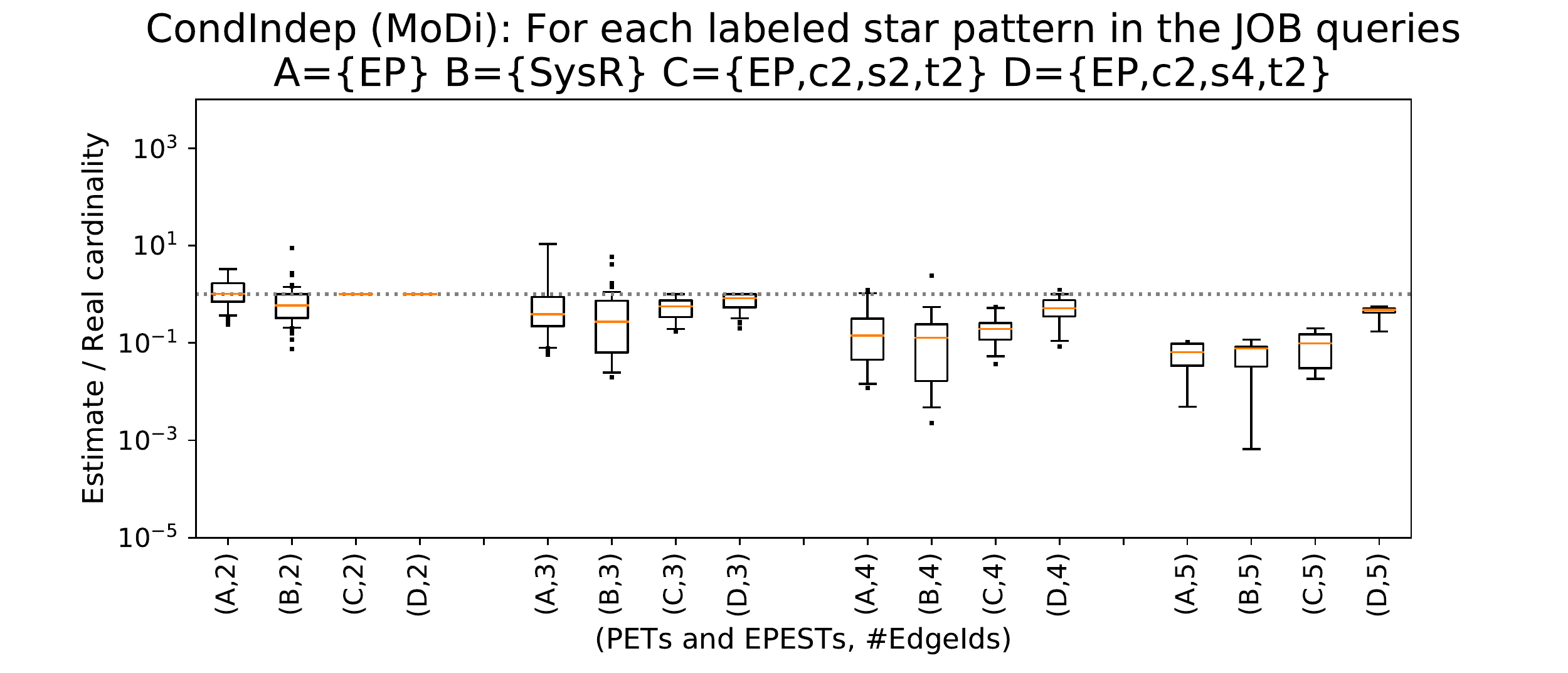}
\caption{Estimation error for all star patterns in the JOB queries, partitioned on the size of the star patterns, using different PETs. The CT \ctShort{condIndep}\sortStratShort{(MoDi)} is used when the PETs are not able to produce an estimate for the whole star pattern.}
\label{fig:sysR_analysis}
\end{figure}

\vspace{1mm}
\noindent \textbf{Analysis on Star Patterns in JOB Queries.}
For each labeled star pattern in the JOB queries with size $\geq 2$, the exact selectivity and estimates using \statShort{SysR} and alternative approaches are obtained. 
Figure \ref{fig:sysR_analysis} shows a summary of these results by plotting the values estimate/real in boxplots partitioned by the size of the star patterns. 
%
%
In order to compare the errors obtained using the PET \statShort{SysR}, alternative approaches (labeled topological synopses) are also illustrated in Figure \ref{fig:sysR_analysis}. 
The PES will be combined using \ctShort{condIndep} with sort strategy \sortStratShort{MoDi}.
For the alternative approaches, all PEs are exact. 
Therefore, the full estimation error is due to (conditional) independence assumptions that are made during combining the PES using CT \ctShort{condIndep}.
For the PET \statShort{SysR}, combing the PES is not necessary, since \statShort{SysR} can obtain an estimate for every star pattern.
Therefore, the full estimation error is due to the assumptions made by \statShort{SysR}.


From Figure \ref{fig:sysR_analysis}, it can be seen that
\begin{itemize}
\item PEs obtained using \statShort{SysR} will introduce estimation errors up to $3$ orders of magnitude for the JOB queries on the JOB dataset;
\item larger star patterns lead to larger underestimations of the selectivity;
\item when statistics about labeled topological pattern (larger than edge patterns) are available, then it is better to use PETs that use those statistics (e.g. \statShort{c2,s2,s3}) and combine using conditional independence assumptions, instead of using \statShort{SysR};
\item the approach \statShort{EP} is performing comparable to \statShort{SysR} (and in many cases, slightly better), while it requires less statistics than \statShort{SysR}.
\end{itemize}

The PET \statShort{EP} with CT \ctShort{condIndep}\sortStratShort{(MoDi)} gives a median and max Q-error of $2.0$ and $205$, where \statShort{SysR} gives $2.1$ and $1515$, on the star patterns in the JOB query workload. 
The max Q-error is $7\times$ larger for \statShort{SysR}, while it requires more statistics than \statShort{EP} with \ctShort{condIndep}.
As a result, PET \statShort{SysR} is not recommended.

\subsubsection{Characteristic Sets}
\label{sec:CS_exp}
The PET \statShort{CS} (Section \ref{sec:char_set}) can obtain estimates for labeled source star patterns with 'has property key' constraints on their center vertices.
A source star pattern with exactly one label on each edge together with the 'has property key' constraints on its center vertex will be called a \textit{csPattern}.

\begin{figure}
\centering
\includegraphics[scale=0.38]{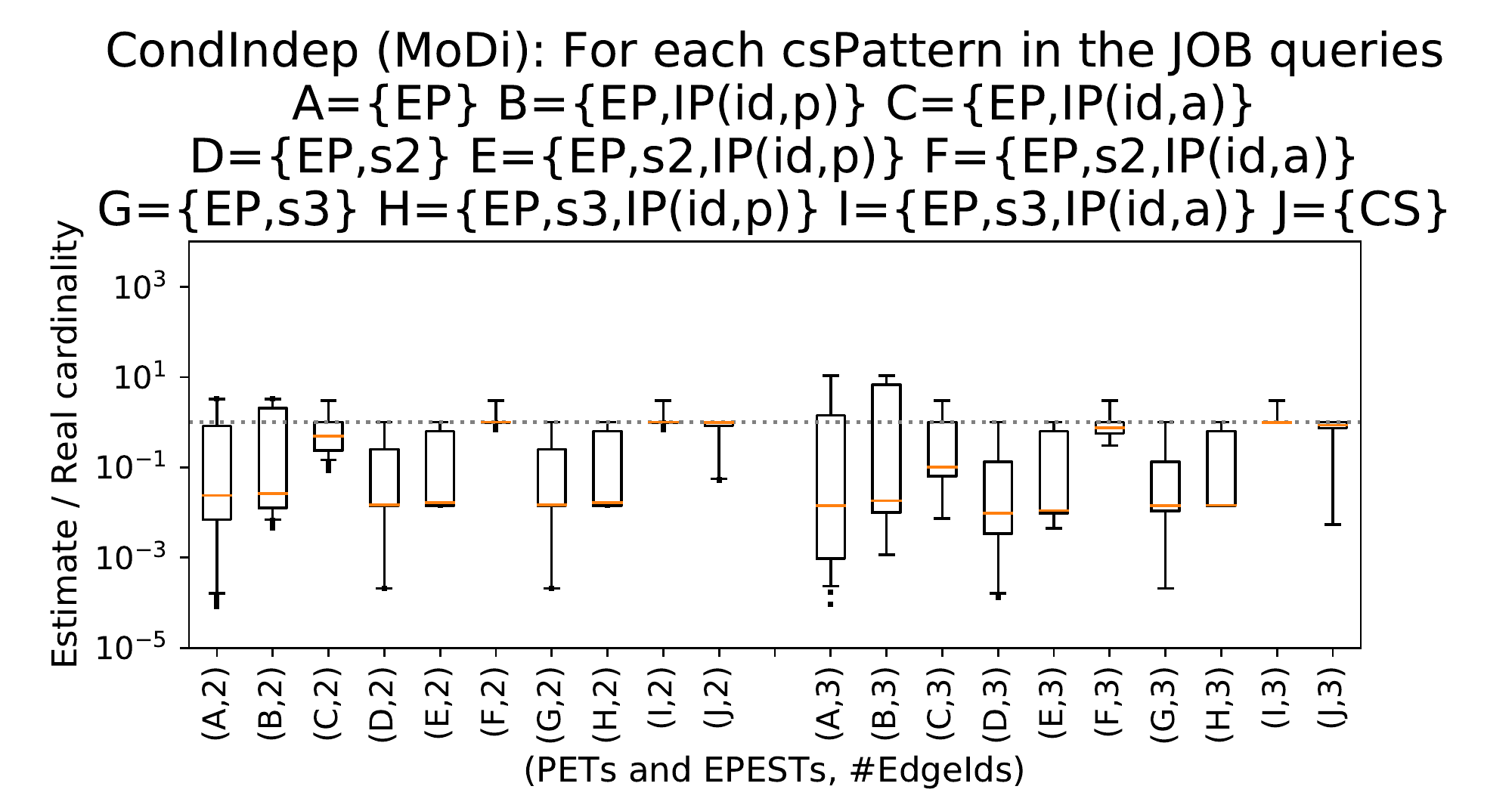}
\caption{Estimation error for all csPatterns in the JOB queries, with at least two edge pattern, partitioned on the size of the csPatterns, using different PETs. The CT \ctShort{condIndep}\sortStratShort{(MoDi)} is used when the PETs are not able to produce an estimate for the whole csPattern.}
\label{fig:cs_analysis}
\end{figure}

For each csPattern in the JOB queries, which contains at least two edge patterns, the exact selectivity and an estimate using \statShort{CS} are obtained. 
Figure \ref{fig:cs_analysis} shows a summary of these results together with alternative techniques, similar as was done in Section \ref{exp:sysR_accuracy}.


Figure \ref{fig:cs_analysis} shows that \statShort{CS} performs very well compared to \{\statShort{EP,s2}\} and \{\statShort{EP,s3}\}. The approaches \{\statShort{EP,s2}\} and \{\statShort{EP,s3}\} have to assume independence between the labeled topological pattern and the 'has property key' constraints, also between all of the 'has property key' constraints.
Those independence assumptions lead to many large underestimations.
\statShort{CS} gives median and max Q-error of $1.1$ and $184$, where \{\statShort{EP}, \statShort{s3}\} gives  $68.0$ and $4762$. Therefore, the median Q-error of \statShort{CS} is $60\times$ lower than for \{\statShort{EP}, \statShort{s3}\} on all csPatterns in the JOB query workload.

Instead of assuming independence between all 'has property key' constraints, it is possible to assume implications using \statShort{IP(id,p)}.
Figure \ref{fig:cs_analysis} shows that the estimates using implied 'has property key' constraints perform better than assuming independence between the 'has property key' constraints. 
However, the selectivity of most csPatterns is still underestimated by more than an order of magnitude. 
Another source of underestimation is the independence assumption between the 'has property key' constraints and the labeled topological constraints.
This source of underestimation can be removed by assuming implications using \statShort{IP(id,a)}.
Figure \ref{fig:cs_analysis} shows that most csPatterns are estimated very accurately with techniques \statShort{EP} and \statShort{IP(id,a)}.
The addition of \statShort{s2} and \statShort{s3} helps to further improve the selectivity estimation accuracy.

The advantage for the \statShort{CS} approach compared to the approaches based on implication assumptions is that \statShort{CS} is based on real statistics. Therefore, it is expected to perform well even in cases where implication and independence assumptions do not hold. 
However, the price that \statShort{CS} pays is an increase in memory cost, preparation time and estimation time.

\subsubsection{Implied Constraints Assumptions}
\label{sec:IP_exp}
The EPEST \statShort{IP(id,pv)} can obtain an estimate for all property-value constraints that refer to the same query id.
We obtained for each query id in each JOB query all property-value constraints.
%
\begin{figure}
\begin{minipage}[c]{0.6\columnwidth}
\includegraphics[scale=0.4]{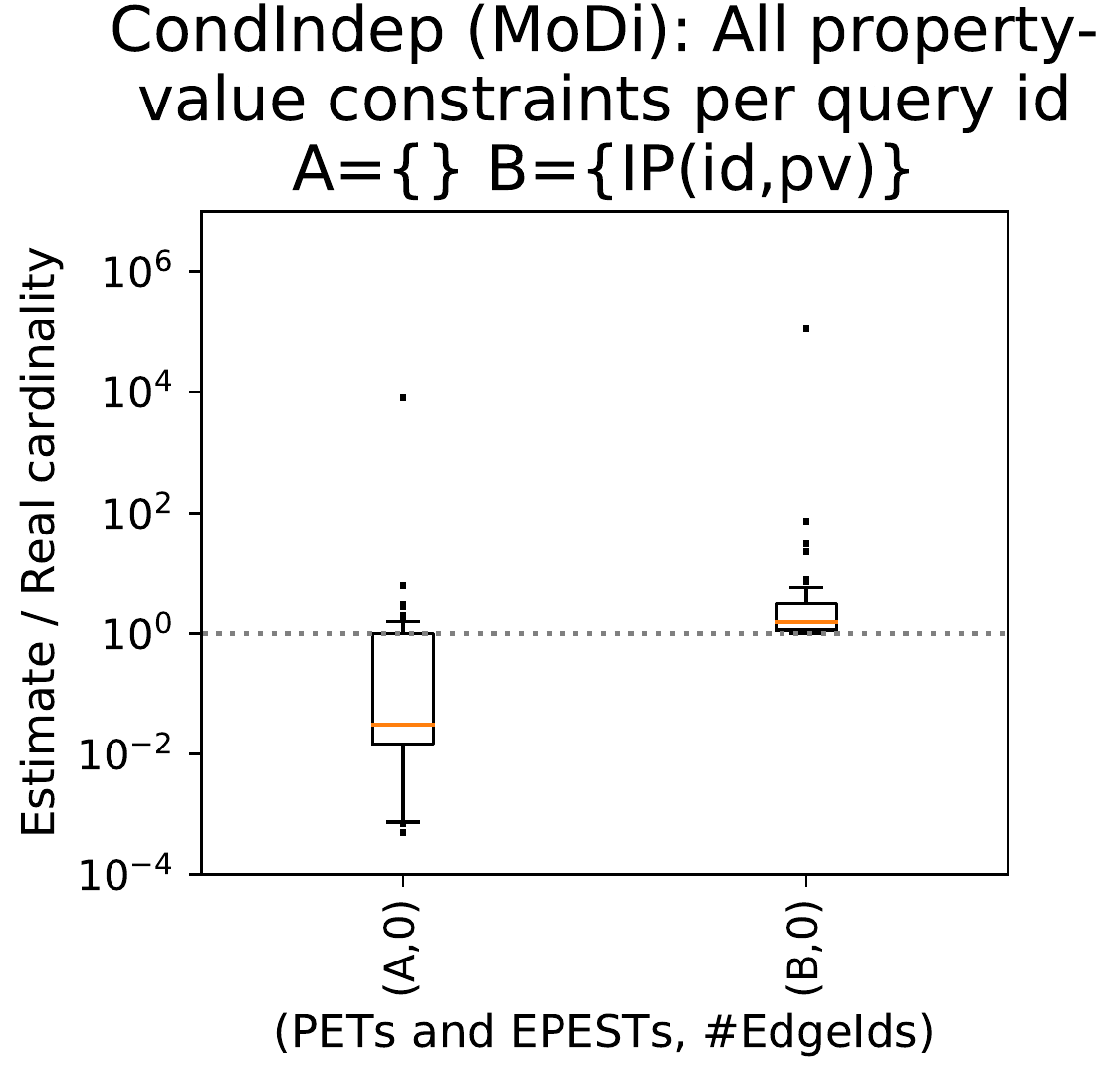}
\end{minipage}
\begin{minipage}[c]{0.38\columnwidth}
\caption{Estimation error for the set of all property-value constraints that refer to the same query id, for each query id in the JOB queries.}
\label{fig:ip_prop_id_analysis}
\end{minipage}
\vspace{-5mm}
\end{figure}
With only the exact selectivities of the individual constraints (no additional PETs and EPESTs), the full estimation error is caused by the CT.
Figure \ref{fig:ip_prop_id_analysis} shows this using CT \ctShort{condIndep}(\sortStratShort{MoDi}).
From the boxplot above \statShort{(\_,0)}, it can be seen that assuming independence mostly leads to underestimations. However, it can also lead a large overestimation.

With the EPEST \statShort{IP(id,pv)}, an estimate for all constraints can be obtained. Therefore, the full estimation error is caused by the implication assumptions within \statShort{IP(id,pv)}.
Clearly, \statShort{IP(id,pv)} cannot lead to an underestimation of the selectivity (when the PES contains only exact selectivity values, which is the case for these experiments). 
However, it can lead to a large overestimation of the selectivity.
Here, \statShort{IP(id,pv)} gives a median and max Q-error of $1.6$ and $1.1\cdot 10^5$, where independence gives  $32.7$ and $8142$. Therefore, the median Q-error of implication assumptions is $20\times$ lower than for independence assumptions, but the max Q-error is $13\times$ larger.

For both approaches, the set $s1=$\{\hasPropConstr{id0}{note}{CONTAINS}{('producer')}, \hasPropConstr{id0}{role}{=}{'actor'}\} caused the largest overestimation. 
A total of $1436536$ graph ids satisfy the first constraint, $12670688$ graph ids satisfy the second constraint, but only $13$ graph ids satisfy both constraints.

\subsection{Estimation Error Experiments for CTs}
\label{sec:experiments_CT_error}

PETs can obtain selectivity estimates for a specific subclass of \queryPatterns{} (e.g. only for star patterns, or only for all property constraints referring to the same query id).
In order to obtain a selectivity estimate for every possible \queryPattern{}, a CT will combine all PEs into a single selectivity estimate for the \queryPattern{}.

Section \ref{sec:exp_CT_labTopConstrs} focuses first on the estimation error introduced by combining PEs consisting of only labeled topological constraints. 
After that, Section \ref{sec:exp_CT_QueryPatternConstrs} focuses on the estimation error introduced by combing all PEs.

\subsubsection{Combining Labeled Topological Constraints}
\label{sec:exp_CT_labTopConstrs}
For this section, we use all the subqueries, that have at most six edge patterns, from all JOB queries. For each such subquery, the property constraints have been removed.
This allows us to focus on the error that is introduced during combination of PEs consisting of labeled topological constraints.

In order to focus on the estimation error caused by the combination techniques, this section uses only PETs that add PEs that are exact (e.g. labeled topological synopses).

\vspace{1mm}
\noindent \textbf{Conditional Independence Assumptions.}
\label{sec:condIndep_topOnly}
The CT \ctShort{condIndep} requires a specific sorting strategy. 
Table \ref{tab:condIndep_ss_qError} shows the median (and max) q-error of the selectivity estimates for different combinations of sorting strategies and PETs. Those results clearly show that a primary sort on the selectivity value of the PEs in descending order or a primary sort on the number of conjunct of the PEs in ascending order should be avoided. 
Those sorting strategies have the effect that essentially all partial estimates involving more than one constraint are ignored.
All other sort strategies perform much better and show that adding more PETs, that use more advanced statistics, lead to better estimation accuracy.

The median estimation time is between $1$ and $15$ milliseconds for the different sort strategies and different PETs, and the maximum estimation time lies between $10$ and $100$ milliseconds (due to page limit restrictions, we omit the full table). 
The estimation time increases slightly when more PETs are used.
The estimation times for different sort strategies are comparable. Only the strategies based on maximum overlap take a bit more time (compared to other sorting strategies), which is required to find the overlap between the different PEs.

\begin{figure}
\centering
\includegraphics[width=\linewidth]{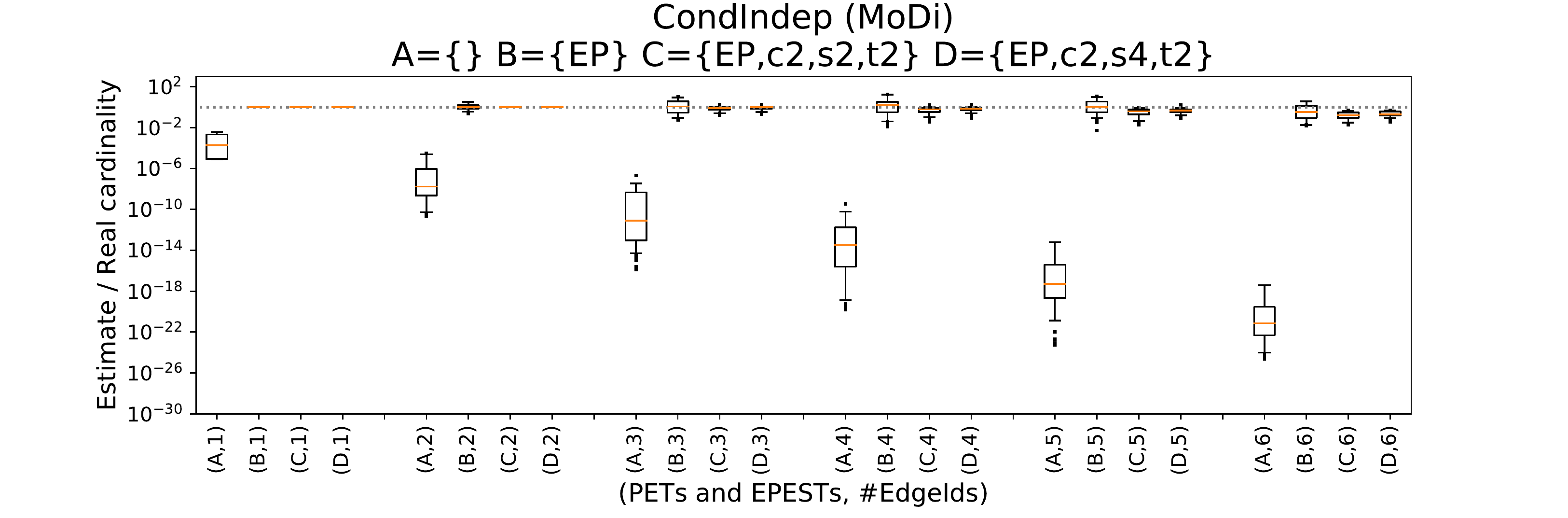}
\includegraphics[width=\linewidth]{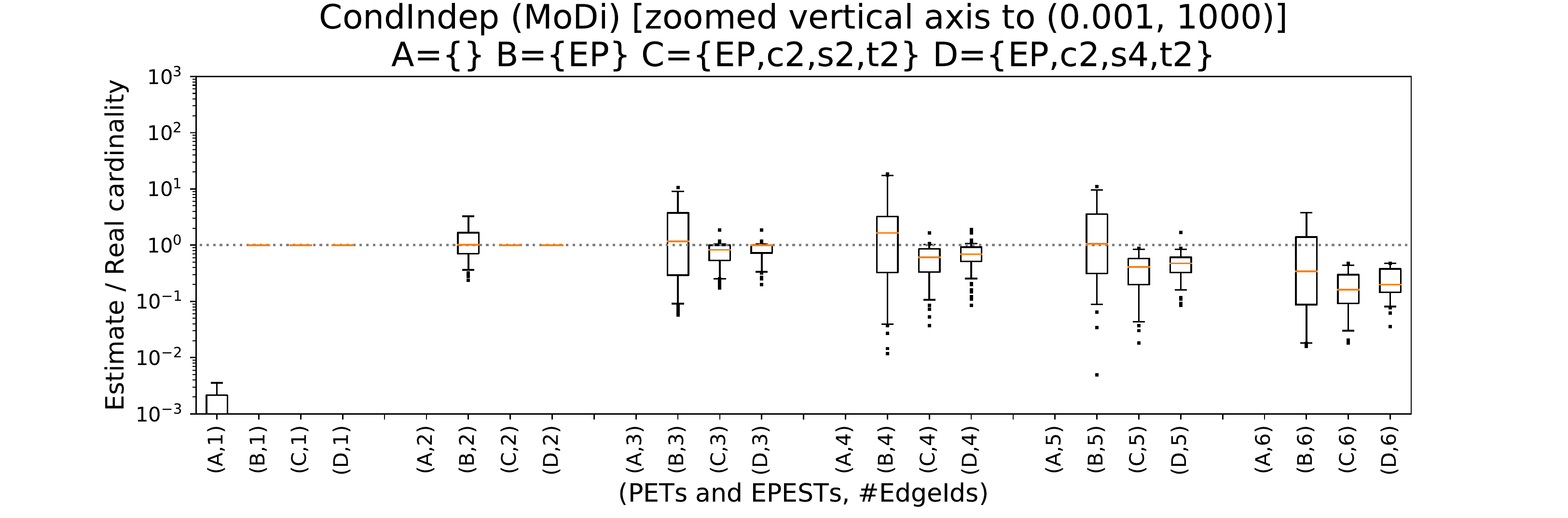}
\caption{Accuracy experiments for the CT \ctShort{condIndep} on all subqueries (up to six edge patterns) without property constraints.}
\label{fig:condIndep_topOnly_analysis}
\end{figure}

Figure \ref{fig:condIndep_topOnly_analysis} shows the boxplots of the results for the sorting strategy \sortStratShort{MoDi}, where the subqueries are partitioned on the number of edge patterns. 
This shows very accurate estimates for \{\statShort{EP, c2, s2, t2}\}, which have mean and max Q-error of $1.0$ and $54.7$ and becomes even better for labeled structural synopses for larger patterns.

\begin{table}
\centering
\caption{The median (and max) q-error of the selectivity estimates using CT \ctShort{CondIndep} with different sort strategies and different PETs.}
\label{tab:condIndep_ss_qError}
{\scriptsize
\begin{tabular}{| l || l | l | l | l |}
\hline
 & \textbf{\{\}} & \textbf{\{EP\}} & \textbf{\{EP,c2,s2,t2\}} & \textbf{\{EP,c2,s4,t2\}} \\ \hline \hline
\sortStratShort{SaNd} & \redCell{4.54e+09 (4.29e+24)} & \orangeCell{1.96 (204.48)} & \greenCell{1.0 (67.64)} & \greenCell{1.0 (19.86)} \\ \hline
\sortStratShort{Sd} & \redCell{4.54e+09 (4.29e+24}) & \redCell{\makecell{4.54e+09\\\hspace{2.5mm}(4.29e+24)}} & \redCell{\makecell{4.54e+09\\\hspace{2.5mm}(4.29e+24)}} & \redCell{\makecell{4.54e+09\\\hspace{2.5mm}(4.29e+24)}} \\ \hline
\sortStratShort{NdSa} & \redCell{4.54e+09 (4.29e+24)} & \orangeCell{1.96 (204.48)} & \greenCell{1.0 (67.64)} & \greenCell{1.0 (24.75)} \\ \hline
\sortStratShort{NdSd} & \redCell{4.54e+09 (4.29e+24)} & \orangeCell{1.96 (204.48)} & \orangeCell{1.0 (624.75)} & \greenCell{1.0 (56.16)} \\ \hline
\sortStratShort{NaSd} & \redCell{4.54e+09 (4.29e+24)} & \redCell{\makecell{4.54e+09\\\hspace{2.5mm}(4.29e+24)}} & \redCell{\makecell{4.54e+09\\\hspace{2.5mm}(4.29e+24)}} & \redCell{\makecell{4.54e+09\\\hspace{2.5mm}(4.29e+24)}} \\ \hline
\sortStratShort{NaSa} & \redCell{4.54e+09 (4.29e+24)} & \redCell{\makecell{4.54e+09\\\hspace{2.5mm}(4.29e+24)}} & \redCell{\makecell{4.54e+09\\\hspace{2.5mm}(4.29e+24)}} & \redCell{\makecell{4.54e+09\\\hspace{2.5mm}(4.29e+24)}} \\ \hline
\sortStratShort{Di} & \redCell{4.54e+09 (4.29e+24)} & \orangeCell{1.96 (204.48)} & \greenCell{1.0 (15.59)} & \greenCell{1.0 (13.81)} \\ \hline
\sortStratShort{MoNd} & \redCell{4.54e+09 (4.29e+24)} & \orangeCell{1.96 (204.48)} & \orangeCell{1.0 (108.01)} & \greenCell{1.0 (43.68)} \\ \hline
\sortStratShort{MoDi} & \redCell{4.54e+09 (4.29e+24)} & \orangeCell{1.96 (204.48)} & \greenCell{1.0 (54.72)} & \greenCell{1.0 (28.23)} \\ \hline
\end{tabular}
}
\end{table}

\vspace{1mm}
\noindent \textbf{Maximum Entropy.}
The combination technique based on the maximum entropy principle is computational expensive (exponential in the number of constraints). 
Therefore, a forced partitioning of the constraints is used, as described by Markl et al. \cite{markl2007consistent}. 
The number for the maximum partition size (mps) is given as a parameter.

The \ctShort{maxEnt} technique is performed for every partition and assumes independence between the different partitions to obtain the final selectivity estimate.

Notice that the effect of partitioning the constraints is that all PEs are lost that have constraints that belong to different partitions. Therefore, it is expected that larger mps values lead to selectivity estimates with higher accuracy, but require larger estimation times.

\begin{figure}
\centering
\includegraphics[width=\linewidth]{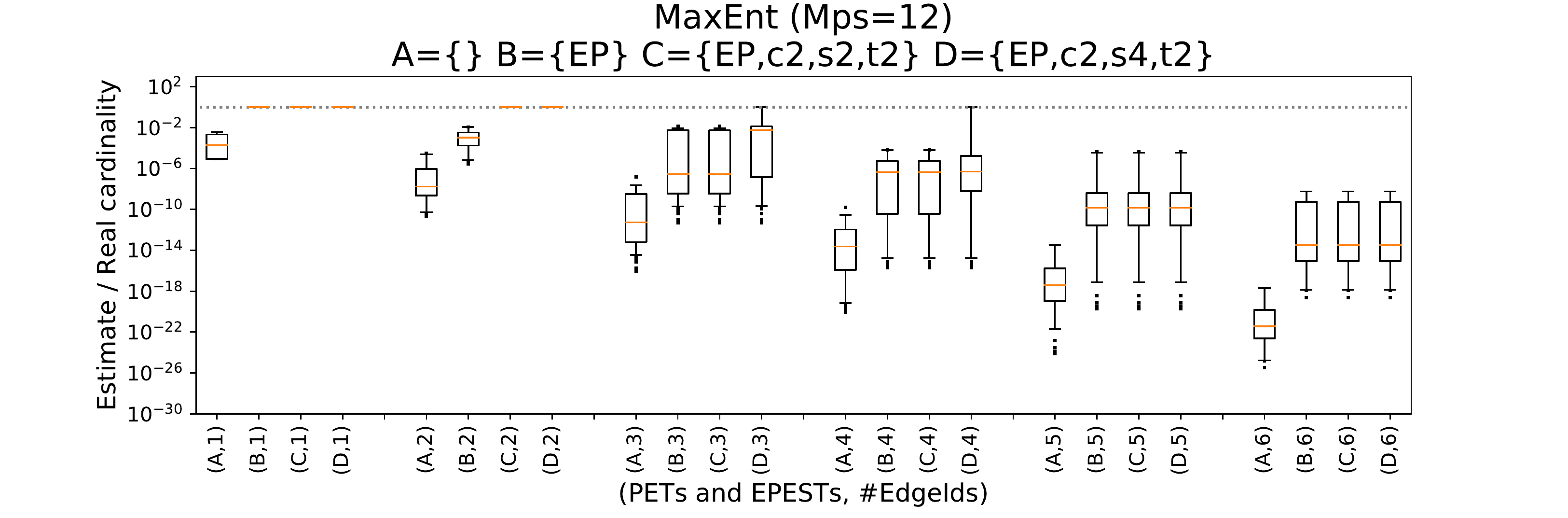}
\caption{Accuracy experiments for the CT \ctShort{maxEnt} with mps=12 on all subqueries (up to six edge patterns) without property constraints.}
\label{fig:maxEnt_topOnly_analysis}
\end{figure}

\ctShort{MaxEnt} using mps values larger than $8$ made the estimation time of some subqueries larger than a second. 
Experiments (omitted due to page limit) show indeed that larger mps values improve estimation accuracy and increase estimation times. 
More PETs, that use more advanced statistics, lead to better estimation accuracy on average. 
However, it did not show a big improvement on the worst case errors. 
Figure \ref{fig:maxEnt_topOnly_analysis} also shows this phenomenon, where for large subqueries, adding more advanced PETs does not help estimation accuracy. 
The PEs obtain by those PETs typically include many constraints, which could not all fit into a single partition. 
Therefore, due to forced partitioning, those PEs are lost.

Comparing the results of the CTs \ctShort{condIndep} and \ctShort{maxEnt}, shows that \ctShort{condIndep} leads to more accurate estimates and requires lower estimation times, which makes \ctShort{condIndep} preferable over \ctShort{maxEnt}.

%
%

\begin{figure}
\centering
\includegraphics[width=\linewidth]{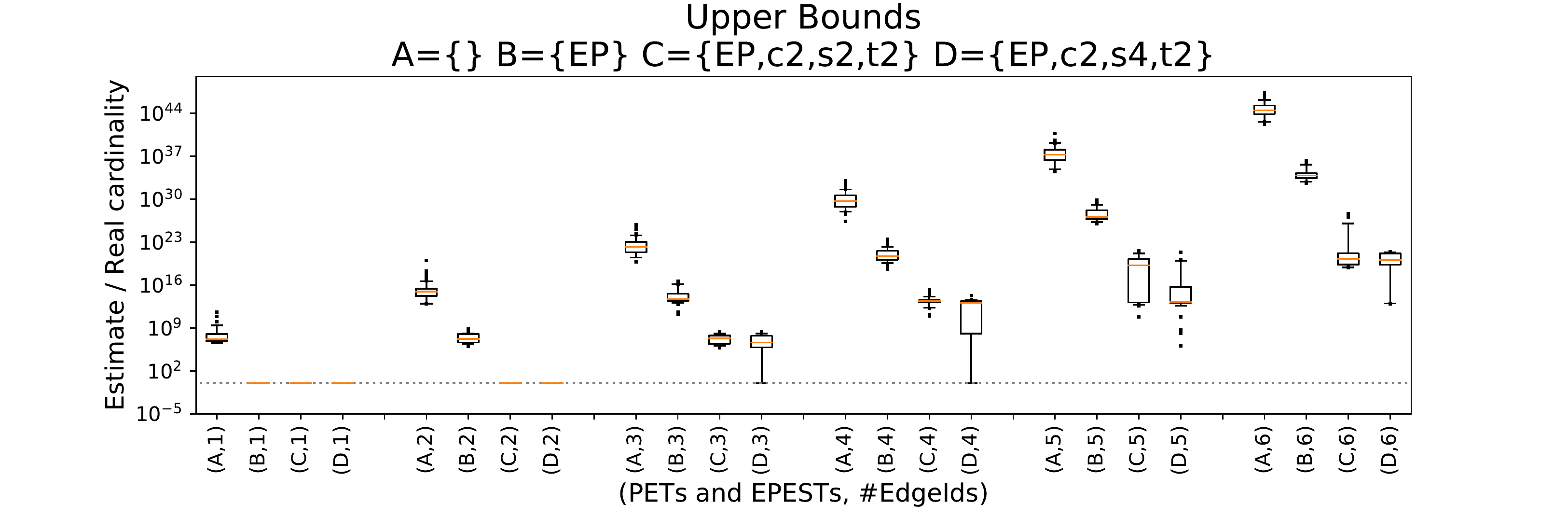}
\caption{Accuracy experiments for the CT \ctShort{upper bound} on all subqueries (up to six edge patterns) without property constraints.}
\label{fig:upperBounds_topOnly_analysis}
\end{figure}

\vspace{1mm}
\noindent \textbf{Upper Bound.}
The combination approach based on lower and upper \ctShort{bounds} gives the guarantee that the real cardinality is in between the lower and upper bound (when the PEs did not introduce any error). 
Whenever the PES does not contain a PE that includes all constraints for the query, then the lower bound was $0$. Therefore, our focus is on the upper bound. 
Figure \ref{fig:upperBounds_topOnly_analysis} shows how the upper bounds improve when more advanced statistics become available.

Table \ref{tab:upperBound_qError_estTime} shows the median (and max) q-errors and estimation times. The q-error clearly decrease with the amount of detailed statistics that become available. However, the actual errors are very large.
Obtaining all independent set of PEs by considering all subsets required time that is exponential in the number of PEs in the PES. This caused the estimation time of some subqueries to take more than $10$ seconds to obtain its upper bound, which makes it impractical for applications of cardinality estimation like query planning.

\begin{table}
\centering
\caption{The median (and max) q-error and estimation time (in ms) using the CT based on exact upper bounds with different PETs.}
\label{tab:upperBound_qError_estTime}
\begin{tabular}{| p{0.12\linewidth} | p{0.14\linewidth} | p{0.14\linewidth} | p{0.17\linewidth} | p{0.17\linewidth} |}
\hline
 & \textbf{\{\}} & \textbf{\{EP\}} & \textbf{\{EP,c2,s2,t2\}} & \textbf{\{EP,c2,s4,t2\}} \\ \hline \hline
\textbf{Q-Error} & \redCell{1.45e+21 (2.09e+47)} & \redCell{2.75e+13 (1.80e+36)} & \redCell{2.07e+06 (4.29e+27)} & \redCell{1.0 (2.85e+21)} \\ \hline
\textbf{Est time (ms)} & \redCell{14.7 (4.83e+03)} & \redCell{28.5 (1.29e+04)} & \redCell{42.8 (2.67e+04)} & \redCell{43.0 (3.63e+04)}  \\ \hline
\end{tabular}
\end{table}

\subsubsection{Combining Query Pattern Constraints}
\label{sec:exp_CT_QueryPatternConstrs}
For this section we use all the subqueries, that have at most six edge pattern, from all JOB queries (same as last section). For each such subquery, all constraints are considered (different from last section, where the property constraints were removed).

From last section, we observe that the the CT \ctShort{CondIndep} outperforms the CTs \ctShort{MaxEnt} and \ctShort{upper bounds} on both estimation time and estimation accuracy.
Therefore, this section focuses on the CT \ctShort{CondIndep}.

\begin{figure}
\centering
\includegraphics[width=\linewidth]{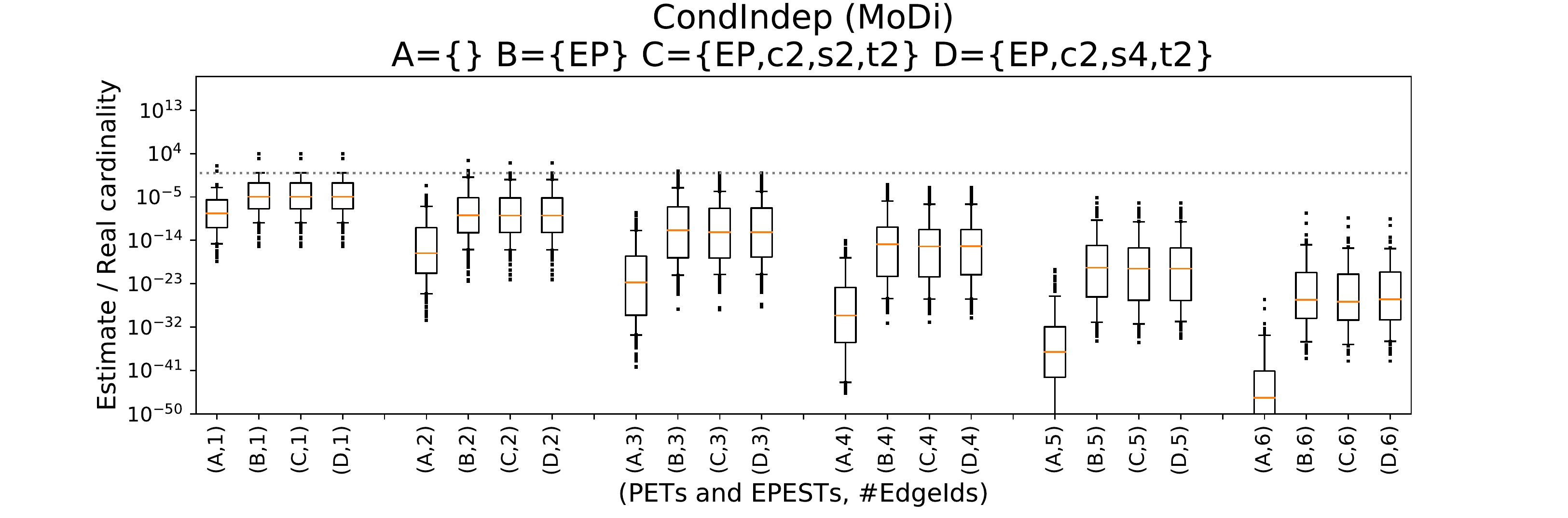}
\caption{Same experiment as in Figure \ref{fig:condIndep_topOnly_analysis}, but now all subqueries keep their property constraints.}
\label{fig:withPropConstrs_labTopSyn_PETs}
\end{figure}

\vspace{1mm}
\noindent \textbf{Labeled Topological Synopses}
\label{sec:labTopSyn_withPropConsts}
First we repeat the same experiment as in Figure \ref{fig:condIndep_topOnly_analysis}, but now all subqueries keep their property constraints.
The results are shown in Figure \ref{fig:withPropConstrs_labTopSyn_PETs}.
The labeled topological synopsis for edge patterns improves the estimation accuracy. However, adding labeled topological synopses for chain, source star and target star patterns do not further improve the accuracy.

Notice that none of the PETs include PEs for combination of property constraints or combination of property and labeled topological constraints.
This means that the error is dominated by the correlation between those combinations of constraints.


\vspace{1mm}
\noindent \textbf{Implied Constraints Assumptions.}
\label{sec:IP_exp_withPropConsts}
Using \statShort{IP} EPESTs can add PEs that contain combinations of property and labeled topological constraints. Therefore, the CT \ctShort{condIndep} will not perform independence assumptions between those constraints as was done in the previous section.

\begin{figure}
\centering
\includegraphics[width=\linewidth]{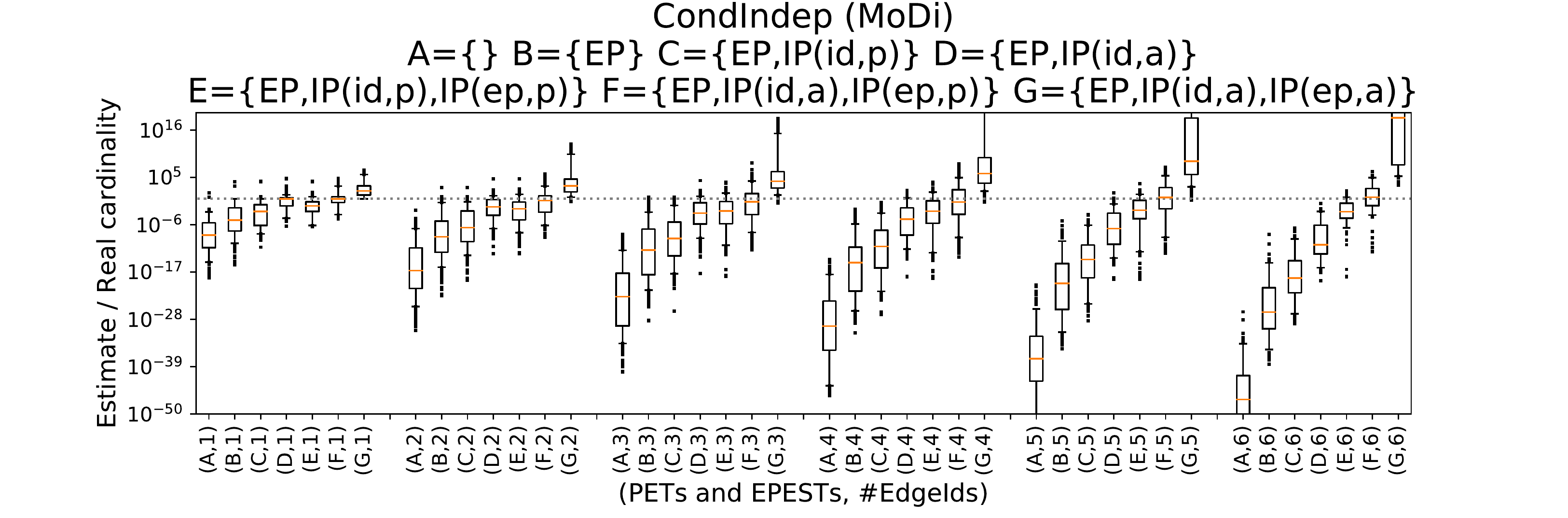}
\caption{Estimation accuracy experiments for different \statShort{IP} EPESTs.}
\label{fig:withPropConstrs_EP_IP}
\end{figure}

Figure \ref{fig:withPropConstrs_EP_IP} shows the accuracy results of experiments with different \statShort{IP} EPESTs when only the PET \statShort{EP} is used.
For this scenario, \statShort{IP(id,a)} with \statShort{IP(ep,p)} performs the best.
Here, \{\statShort{EP, IP(id,a), IP(id,p)}\} has a median and max Q-error of $101$ and $3.5\cdot 10^{13}$, where \{\statShort{EP}\} has $1.1\cdot 10^{11}$ and $3.0\cdot 10^{38}$. The maximum Q-error of independence assumptions is $10^{24}\times$ larger than for implication assumptions.
%
%
%

The IP type \statShort{IP(ep,a)} is too strong. It essentially ignores the effect of many constraints that are actually restricting the cardinality of the subqueries. 
The other IP types are more conservative, i.e. they make less implication assumptions (as a result, the CT might have to make more independence assumptions).

\begin{table*}
\centering
\caption{The median (and max) estimation time (in ms) for the experiments from Figure \ref{fig:withPropConstrs_EP_Sampling}.}
\label{tab:estTime_sampling}
{\scriptsize
\begin{tabular}{| l | l | l | l | l | l | l |}
\hline
\textbf{PETs and EPETs} & \textbf{\#EdgeIds: 1} & \textbf{\#EdgeIds: 2} & \textbf{\#EdgeIds: 3} & \textbf{\#EdgeIds: 4} & \textbf{\#EdgeIds: 5} & \textbf{\#EdgeIds: 6} \\ \hline \hline 
\{\} & \greenCell{2.20 (36.4)}  & \greenCell{6.14 (71.3)}  & \greenCell{12.4 (126)}  & \greenCell{23.4 (206)}  & \greenCell{38.3 (297)}  & \greenCell{57.3 (271)}  \\ \hline
\{\statShort{EP}\} & \greenCell{2.86 (44.5)}  & \greenCell{8.83 (93)}  & \greenCell{19.1 (185)}  & \greenCell{36.4 (314)}  & \greenCell{62.8 (463)}  & \greenCell{91.1 (468)}  \\ \hline
\{\statShort{EP, S(id,0.001)}\} & \redCell{711 (8.61e+03)}  & \redCell{1.26e+3 (1.77e+4)}  & \redCell{1.86e03 (2.08e04)}  & \redCell{2.77e+3 (2.43e+4)}  & \redCell{3.68e+03 (2.56e+4)}  & \redCell{4.37e+03 (2.72e+04)}  \\ \hline
\makecell{\{\statShort{EP, S(id,0.001),}\\\hspace{7.5mm}\statShort{S(ep,0.001)}\}} & \redCell{1.32e+03 (1.16e+04)}  & \redCell{2.86e+3 (3.63e+4)}  & \redCell{4.68e+3 (6.28e+4)}  & \redCell{6.75e+3 (8.25e+4)}  & \redCell{8.12e+3 (9.61e+4)}  & \redCell{9.49e+3 (5.42e+4)}  \\ \hline
\{\statShort{EP,IP(id,a),IP(ep,p)}\} & \greenCell{5.45 (108)}  & \greenCell{17.7 (226)}  & \greenCell{38.0 (372)} & \greenCell{73.8 (613)}  & \yellowCell{123 (939)}  & \yellowCell{185 (901)}  \\ \hline
\end{tabular}
}
\end{table*}

\begin{figure}
\centering
\includegraphics[width=\linewidth]{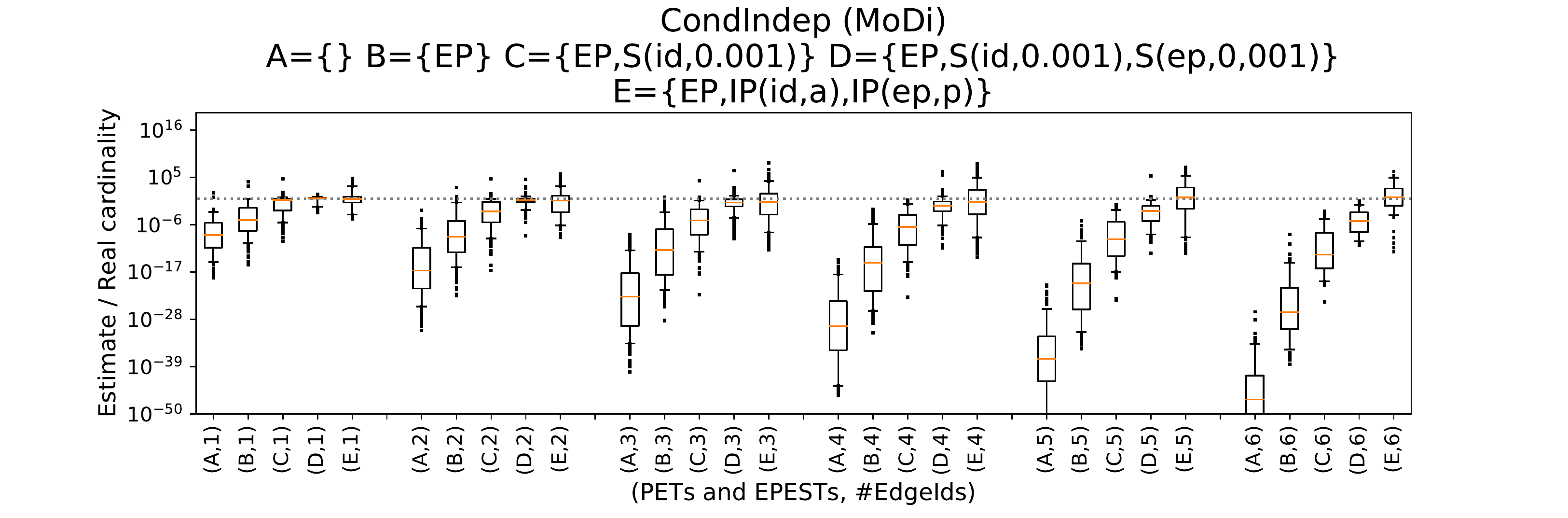}
\caption{Estimation accuracy experiment for different sampling techniques.}
\label{fig:withPropConstrs_EP_Sampling}
\end{figure}

\vspace{1mm}
\noindent \textbf{Sampling techniques}
\label{sec:sampling_exp_withPropConsts}
Sampling PETs are able to obtain PEs about combinations of property and topological constraints. 
Figure \ref{fig:withPropConstrs_EP_Sampling} shows the results of accuracy experiments for different sampling techniques. 
For the pattern parameter of sampling, we experimented with $id$ and $ep$ and for the probability we used $10^{-3}$.
For comparison reasons, we also added the best technique using \statShort{IP} EPESTs from the previous section.
Here, \{\statShort{EP,S(id,0.001),S(ep,0.001)}\} gives as median and max Q-error $6$ and $4.4\cdot 10^{12}$, where \{\statShort{EP,IP(id,a),IP(ep,p)}\} gives $101$ and $3.5\cdot 10^{13}$. Therefore, the median Q-error of sampling is $16\times$ better and max Q-error is $8\times$ better than the best approach based on implication assumptions.


Clearly, sampling techniques improve the estimation accuracy, which highlights the importance of capturing correlations between constraints referring to the same $id$ or $ep$. 
Assuming independence can lead to large underestimations. 
Assuming implications can lead to large overestimations.

The median and max estimation time (in ms) for \{\statShort{EP,S(id,0.001),S(ep,0.001)}\} is $4890$ and $9.61\cdot 10^{4}$, and for \{\statShort{EP,IP(id,a),IP(ep,p)}\} it is $36$ and $939$. 
Therefore, the median estimation time for sampling is $135\times$ higher and the max estimation time is $102\times$ higher.
This raises the question whether or not the large estimation time of sampling techniques is worth the improvement in estimation accuracy.
Table \ref{tab:estTime_sampling} shows the estimation times of the experiments in Figure \ref{fig:withPropConstrs_EP_Sampling}.

\section{Conclusions}
\label{sec:concusionFutureWork}

Many cardinality estimation techniques have been proposed in the literature which require specific statistics or indexes to be available and use a certain set of simplifying assumptions.
The framework introduced in this document makes it possible to compare, and even combine, different techniques with the goal of producing superior solutions.

The framework consists of obtaining estimates for subqueries (partial estimates) using techniques (partial estimate techniques) that have specific prerequisites, e.g. specialized statistics or indexes. 
All partial estimates are stored in a partial estimate set (PES).
This set can be extended and made complete, such that every constraint from the \queryPattern{} is available in at least one partial estimate in the PES.
Finally, all partial estimates in the PES are combined using a general combine technique (CT). A CT is a general technique that uses simplifying assumption to combine partial estimates and therefore can be applied for any complete PES, i.e. it does not require any form of statistics.

Extensive experiments show that synopsis consisting of the cardinality of small labeled topological patterns makes it possible to obtain accurate estimates for \queryPatterns{} that do not include property constraints.
For \queryPatterns{} that include property constraints, capturing correlations between property constraints and labeled topological constraints is essential. When this information is insufficiently available, then,  making implication assumptions typically performs better than making independence assumptions.

\section{Future work}

\subsection{Research Perspective}
The framework proposed in this document allows new PETs, EPESTs and CTs to be developed in isolation and finally used in combination with other existing techniques.
Future work consists of proposing new PETs that are able to capture most important correlations between property and labeled topological constraints, while keeping the estimation time low. 
For example, is it possible to find subclasses of \queryPatterns{} where sampling can be efficient and accurate?

Another type of future work is to improve upon the combination techniques.
For example, defining metrics to obtain an optimal sorting strategy for CT \ctShort{condIndep}.
The CT \ctShort{MaxEnt} might be more valuable as an EPEST.

Instead of only obtaining a cardinality estimate, it would also be very useful to obtain some sort of error guarantee, e.g. there is a 95\% change that the actual cardinality is between $[100, 250]$.

Extensions w.r.t. the query class needs to be considered in the future, e.g. estimate the cardinality for UCRPQs\cite{bonifati2018querying}. UCRPQs is an extension of  \queryPatterns{} that allows path navigation of arbitrary length, which are specified using regular expressions.

\subsection{Practitioners Perspective}
In order to benefit from recently developed cardinality estimation techniques, it is possible to implement our general cardinality estimation framework, i.e. Algorithm \ref{alg:generalSelEstProcess}. 
This makes it possible to implement current cardinality estimation techniques as PETs. 
Our framework then allows to combine different techniques by selecting the PETs of interest (depending on trade-offs, e.g. estimation time and estimation accuracy). For each query, a decision about what combination of techniques to be used can be made.
For example, complex analytical queries that are expected to run for a long time might afford to spend more time in the query optimizer, therefore using PETs with higher estimation times might be allowed.

For the CT, \ctShort{condIndep} is recommended since it outperformed the other CTs in both estimation accuracy and estimation time.

Considering a system with a classical setup for cardinality estimation, e.g. some (multidimensional) histograms and some form of sampling. Then those implementations can be transformed into PETs, which allows them to be used in combination with each other.

Experimental evaluation over the JOB dataset showed that instead of using the PET \statShort{SysR} (which uses the uniform distribution and inclusion assumptions), it is recommended to use \statShort{labTopSyn(EP)} in combination with \ctShort{condIndep} (which uses conditional independence assumptions).

\bibliographystyle{IEEEtran}
\bibliography{IEEEabrv,ref}

\begin{thebibliography}{10}
\providecommand{\url}[1]{#1}
\csname url@samestyle\endcsname
\providecommand{\newblock}{\relax}
\providecommand{\bibinfo}[2]{#2}
\providecommand{\BIBentrySTDinterwordspacing}{\spaceskip=0pt\relax}
\providecommand{\BIBentryALTinterwordstretchfactor}{4}
\providecommand{\BIBentryALTinterwordspacing}{\spaceskip=\fontdimen2\font plus
\BIBentryALTinterwordstretchfactor\fontdimen3\font minus
  \fontdimen4\font\relax}
\providecommand{\BIBforeignlanguage}[2]{{%
\expandafter\ifx\csname l@#1\endcsname\relax
\typeout{** WARNING: IEEEtran.bst: No hyphenation pattern has been}%
\typeout{** loaded for the language `#1'. Using the pattern for}%
\typeout{** the default language instead.}%
\else
\language=\csname l@#1\endcsname
\fi
#2}}
\providecommand{\BIBdecl}{\relax}
\BIBdecl

\bibitem{leis2015good}
V.~Leis, A.~Gubichev, A.~Mirchev, P.~Boncz, A.~Kemper, and T.~Neumann, ``How
  good are query optimizers, really?'' \emph{Proceedings of the VLDB
  Endowment}, vol.~9, no.~3, pp. 204--215, 2015.

\bibitem{leis2018query}
V.~Leis, B.~Radke, A.~Gubichev, A.~Mirchev, P.~Boncz, A.~Kemper, and
  T.~Neumann, ``Query optimization through the looking glass, and what we found
  running the join order benchmark,'' \emph{The VLDB Journal}, vol.~27, no.~5,
  pp. 643--668, 2018.

\bibitem{GQLstandard}
``{GQL Standard},'' \url{https://www.gqlstandards.org}, accessed: 2021-07-26.

\bibitem{bonifati2018querying}
A.~Bonifati, G.~Fletcher, H.~Voigt, and N.~Yakovets, ``Querying graphs,''
  \emph{Synthesis Lectures on Data Management}, vol.~10, no.~3, pp. 1--184,
  2018.

\bibitem{BonifatiMT20}
A.~Bonifati, W.~Martens, and T.~Timm, ``An analytical study of large {SPARQL}
  query logs,'' \emph{{VLDB} J.}, vol.~29, no. 2-3, pp. 655--679, 2020.

\bibitem{ParkKBKHH20}
Y.~Park, S.~Ko, S.~S. Bhowmick, K.~Kim, K.~Hong, and W.~Han, ``{G-CARE:} {A}
  framework for performance benchmarking of cardinality estimation techniques
  for subgraph matching,'' in \emph{{SIGMOD}}, 2020, pp. 1099--1114.

\bibitem{angles2017foundations}
R.~Angles, M.~Arenas, P.~Barcel{\'o}, A.~Hogan, J.~Reutter, and D.~Vrgo{\v{c}},
  ``Foundations of modern query languages for graph databases,'' \emph{ACM
  Computing Surveys (CSUR)}, vol.~50, no.~5, pp. 1--40, 2017.

\bibitem{van2016pgql}
O.~van Rest, S.~Hong, J.~Kim, X.~Meng, and H.~Chafi, ``Pgql: a property graph
  query language,'' in \emph{Proceedings of the Fourth International Workshop
  on Graph Data Management Experiences and Systems}, 2016, pp. 1--6.

\bibitem{angles2018g}
R.~Angles, M.~Arenas, P.~Barcel{\'o}, P.~Boncz, G.~Fletcher, C.~Gutierrez,
  T.~Lindaaker, M.~Paradies, S.~Plantikow, J.~Sequeda \emph{et~al.}, ``G-core:
  A core for future graph query languages,'' in \emph{Proceedings of the 2018
  International Conference on Management of Data}, 2018, pp. 1421--1432.

\bibitem{selinger1979access}
P.~G. Selinger, M.~M. Astrahan, D.~D. Chamberlin, R.~A. Lorie, and T.~G. Price,
  ``Access path selection in a relational database management system,'' in
  \emph{Proceedings of the 1979 ACM SIGMOD international conference on
  Management of data}, 1979, pp. 23--34.

\bibitem{cormode2012synopses}
G.~Cormode, M.~Garofalakis, P.~J. Haas, and C.~Jermaine, ``Synopses for massive
  data: Samples, histograms, wavelets, sketches,'' \emph{Foundations and Trends
  in Databases}, vol.~4, no. 1--3, pp. 1--294, 2012.

\bibitem{poosala1996improved}
V.~Poosala, P.~J. Haas, Y.~E. Ioannidis, and E.~J. Shekita, ``Improved
  histograms for selectivity estimation of range predicates,'' \emph{ACM Sigmod
  Record}, vol.~25, no.~2, pp. 294--305, 1996.

\bibitem{ioannidis2003history}
Y.~Ioannidis, ``The history of histograms (abridged),'' in \emph{Proceedings
  2003 VLDB Conference}.\hskip 1em plus 0.5em minus 0.4em\relax Elsevier, 2003,
  pp. 19--30.

\bibitem{jagadish1999substring}
H.~Jagadish, R.~T. Ng, and D.~Srivastava, ``Substring selectivity estimation,''
  in \emph{Proceedings of the eighteenth ACM SIGMOD-SIGACT-SIGART symposium on
  Principles of database systems}, 1999, pp. 249--260.

\bibitem{aboulnaga2001estimating}
A.~Aboulnaga, A.~R. Alameldeen, and J.~F. Naughton, ``Estimating the
  selectivity of xml path expressions for internet scale applications,'' in
  \emph{VLDB}, vol.~1.\hskip 1em plus 0.5em minus 0.4em\relax Citeseer, 2001,
  pp. 591--600.

\bibitem{cai2019pessimistic}
W.~Cai, M.~Balazinska, and D.~Suciu, ``Pessimistic cardinality estimation:
  Tighter upper bounds for intermediate join cardinalities,'' in
  \emph{Proceedings of the 2019 International Conference on Management of
  Data}, 2019, pp. 18--35.

\bibitem{neumann2011characteristic}
T.~Neumann and G.~Moerkotte, ``Characteristic sets: Accurate cardinality
  estimation for rdf queries with multiple joins,'' in \emph{2011 IEEE 27th
  International Conference on Data Engineering}.\hskip 1em plus 0.5em minus
  0.4em\relax IEEE, 2011, pp. 984--994.

\bibitem{poosala1997selectivity}
V.~Poosala and Y.~E. Ioannidis, ``Selectivity estimation without the attribute
  value independence assumption,'' in \emph{VLDB}, vol.~97, 1997, pp. 486--495.

\bibitem{bruno2001stholes}
N.~Bruno, S.~Chaudhuri, and L.~Gravano, ``Stholes: a multidimensional
  workload-aware histogram,'' in \emph{Proceedings of the 2001 ACM SIGMOD
  international conference on Management of data}, 2001, pp. 211--222.

\bibitem{li2016wander}
F.~Li, B.~Wu, K.~Yi, and Z.~Zhao, ``Wander join: Online aggregation via random
  walks,'' in \emph{Proceedings of the 2016 International Conference on
  Management of Data}, 2016, pp. 615--629.

\bibitem{world2014rdf}
W.~W.~W. Consortium \emph{et~al.}, ``Rdf 1.1 concepts and abstract syntax,''
  2014.

\bibitem{schreiber2014rdf}
A.~T. Schreiber and Y.~Raimond, ``Rdf 1.1 primer,'' 2014.

\bibitem{bruno2002exploiting}
N.~Bruno and S.~Chaudhuri, ``Exploiting statistics on query expressions for
  optimization,'' in \emph{Proceedings of the 2002 ACM SIGMOD international
  conference on Management of data}, 2002, pp. 263--274.

\bibitem{markl2007consistent}
V.~Markl, P.~J. Haas, M.~Kutsch, N.~Megiddo, U.~Srivastava, and T.~M. Tran,
  ``Consistent selectivity estimation via maximum entropy,'' \emph{The VLDB
  journal}, vol.~16, no.~1, pp. 55--76, 2007.

\end{thebibliography}
\vspace{-1cm}
\begin{IEEEbiography}[{\includegraphics[width=1in,height=1.25in,clip,keepaspectratio]{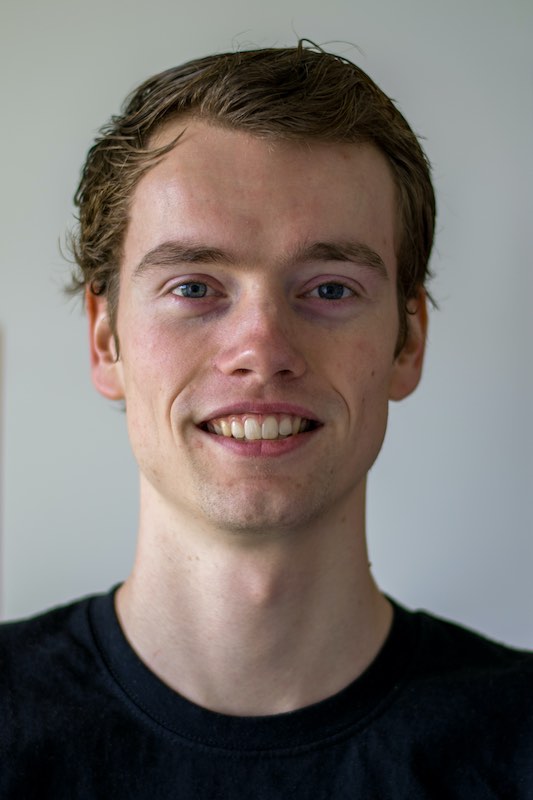}}]{Wilco van Leeuwen}
is a PhD student in the department of Mathematics and Computer Science at Eindhoven University of Technology. His current research interest is on graph query languages, query planning, cardinality estimation and approximate query evaluation.
\end{IEEEbiography}


\begin{IEEEbiography}[{\includegraphics[width=1in,height=1.25in,clip,keepaspectratio]{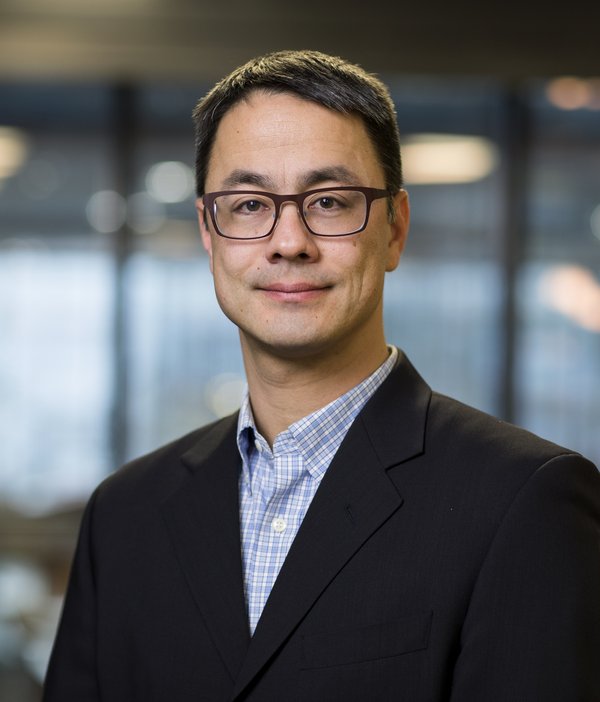}}]{George Fletcher}
is a full professor of computer science at Technische Universiteit Eindhoven where he is chair of the Database Group.  
He defended a PhD at Indiana University Bloomington in 2007.  
His research interests span query language design and engineering, foundations of databases, and data integration.  
His current focus is on management of massive graphs such as social networks and linked open data.  
\end{IEEEbiography}


\begin{IEEEbiography}[{\includegraphics[width=1in,height=1.25in,clip,keepaspectratio]{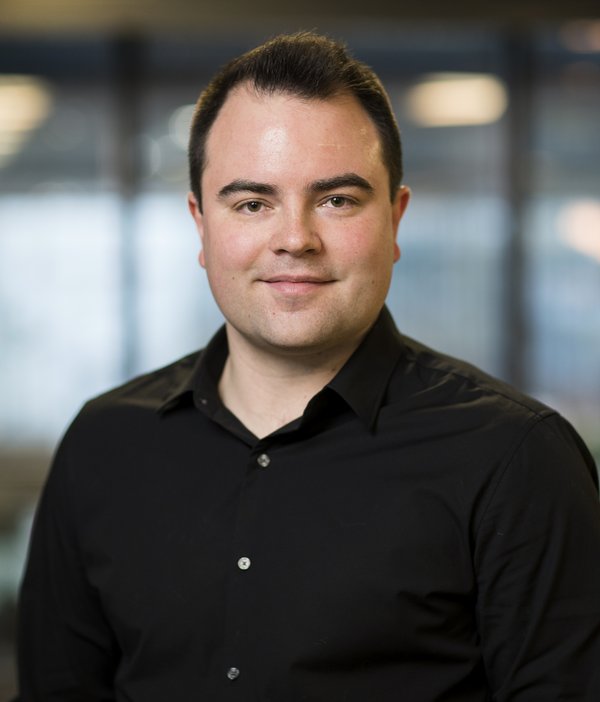}}]{Nikolay Yakovets}
is an assistant professor of computer
science at Technische Universiteit Eindhoven. 
He obtained his PhD from Lassonde School of Engineering at York University in 2017.
His current focus is on design and implementation of core database technologies, management of massive graph data, and efficient processing of queries on graphs.
\end{IEEEbiography}

\end{document}